\documentclass[a4paper,11pt,preprintnumbers]{article}

\pdfoutput=1

\usepackage{jheppub}
\usepackage{amsmath}
\usepackage{xspace}
\usepackage{hyperref}
\usepackage{mdwlist}

\usepackage{braket}
\allowdisplaybreaks 
\usepackage{color}
\usepackage{mathrsfs}
\newcommand\f{\frac}
\newcommand\as{\alpha_s}
\newcommand{\ba}{\begin{eqnarray}}
\newcommand{\ea}{\end{eqnarray}}
\def\bea#1\eea{\begin{align}#1\end{align}}
\newcommand{\bef}{\begin{figure*}[t]\centering}
\newcommand{\eef}{\end{figure*}}
\newcommand{\be}{\begin{equation}}
\newcommand{\ee}{\end{equation}}
\newcommand{\nn}{\nonumber}

\def\OMIT#1{{}}

\newcommand{\ffh}{f\!f\!H}

\newcommand{\lea}{\Big\langle\!\!\Big\langle}
\newcommand{\rea}{\Big\rangle\!\!\Big\rangle}

\newcommand{\rnd}{\text{{\bf rnd}}}

\begin{document}

\preprint{\begin{flushright}
JLAB-THY-21-3339
\\ LA-UR-21-22574
\end{flushright}}


\title{Leading jets and energy loss}

\author[a]{Duff Neill,}
\author[b]{Felix Ringer,}
\author[c]{Nobuo Sato}

\affiliation[a]{Theoretical Division, MS B283, Los Alamos National Laboratory, Los Alamos, NM 87545, USA}
\affiliation[b]{Nuclear Science Division, Lawrence Berkeley National Laboratory, Berkeley, CA 94720, USA}
\affiliation[c]{Theory Center, Jefferson Laboratory, Newport News, Virginia 23606, USA}
                                         
\emailAdd{duff.neill@gmail.com}
\emailAdd{fmringer@lbl.gov}
\emailAdd{nsato@jlab.org}


\abstract{The formation and evolution of leading jets can be described by jet functions which satisfy non-linear DGLAP-type evolution equations. Different than for inclusive jets, the leading jet functions constitute normalized probability densities for the leading jet to carry a longitudinal momentum fraction relative to the initial fragmenting parton. We present a parton shower algorithm which allows for the calculation of leading-jet cross sections where logarithms of the jet radius and threshold logarithms are resummed to next-to-leading logarithmic (NLL$'$) accuracy. By calculating the mean of the leading jet distribution, we are able to quantify the average out-of-jet radiation, the so-called jet energy loss. When an additional reference scale is measured, we are able to determine the energy loss of leading jets at the cross section level which is identical to parton energy loss at leading-logarithmic accuracy. We identify several suitable cross sections for an extraction of the jet energy loss and we present numerical results for leading subjets at the LHC. In addition, we consider hemisphere and event-wide leading jets in electron-positron annihilation similar to measurements performed at LEP. Besides the average energy loss, we also consider its variance and other statistical quantities such as the KL divergence which quantifies the difference between quark and gluon jet energy loss. We expect that our results will be particularly relevant for quantifying the energy loss of quark and gluon jets that propagate through hot or cold nuclear matter.}

\maketitle

\newpage

\section{Introduction \label{sec:intro}}

Highly energetic jets play a major role at high-energy collider experiments such as the Large Hadron Collider (LHC) and the Relativistic Heavy Ion Collider (RHIC), as well as the the future Electron-Ion Collider (EIC)~\cite{AbdulKhalek:2021gbh}. In the past years significant progress has been made in performing high-precision calculations for exclusive and inclusive jet production as well as jet substructure observables. Aside from being a means to constrain parton distribution functions (PDFs) of the proton~\cite{Hou:2019efy,Accardi:2016qay,Ball:2017nwa,Alekhin:2018pai,Bailey:2020ooq}, an integral part of searches for new physics~\cite{Larkoski:2017jix,Asquith:2018igt,Marzani:2019hun}, and a sensitive probe of the strong coupling constant~\cite{Aaboud:2018hie,Khachatryan:2014waa}, another fundamental concern in these studies is how exactly energy is distributed into the states registered in the detector. These states can be considered at different levels of resolution, from the irreducible individual hadrons, to large radius jets which may or may not have multi-prong substructure~\cite{Dasgupta:2013ihk,Dasgupta:2015lxh,Larkoski:2015kga}. The ability to resolve the final state of a collision at multiple scales is critical in being able to test our understanding of the dynamics that lead to these states.

\begin{figure}[t]
\vspace*{.7cm}
\centering
\includegraphics[width=12cm]{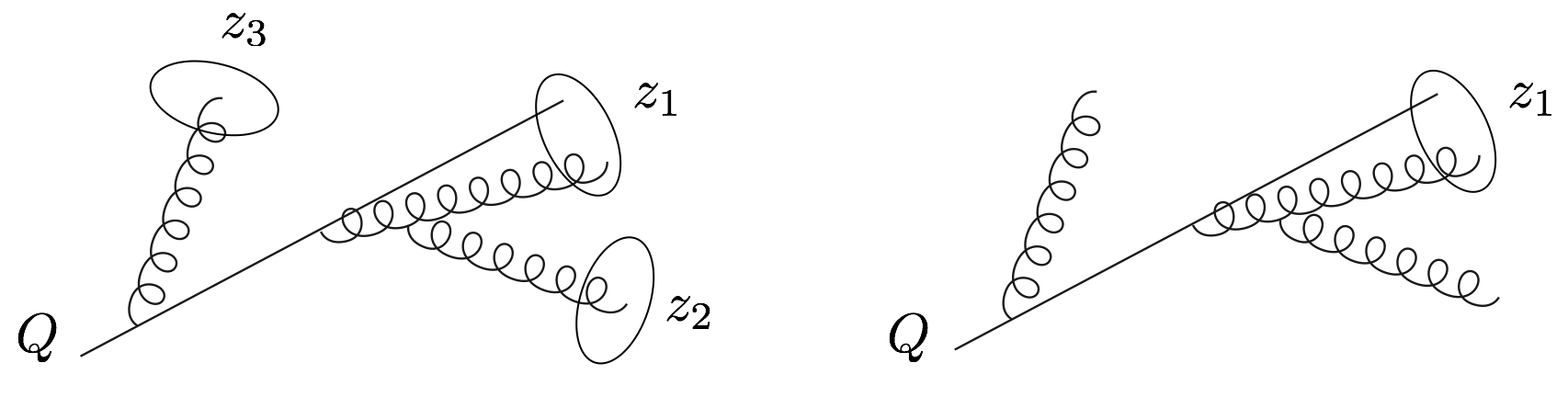}
\caption{Illustration of inclusive jets (left) and the leading jet (right) which originate from an initial fragmenting quark at LL accuracy. The $z_i$ indicate the longitudinal momentum fractions of the jets relative to the scale $Q$ of the initial quark.~\label{fig:inclusive_leading}}
\end{figure}

For example, the inclusive jet cross section $pp\to {\rm jet}+X$ has been calculated to next-to-next-to leading order (NNLO)~\cite{Currie:2016bfm,Czakon:2019tmo}, and is an important observable to constrain the gluon PDF. An inclusive jet sample is obtained by measuring the transverse momentum $p_T$ of all the jets in a given rapidity range. The factorization in QCD can be formulated in terms of hard-scattering functions and (semi-)inclusive jet functions~\cite{Catani:2013oma,Dasgupta:2014yra,Kaufmann:2015hma,Kang:2016mcy,Dai:2016hzf}. The formation and evolution of jets described by the inclusive jet function is illustrated in the left panel of Fig.~\ref{fig:inclusive_leading}. Here all jets are taken into account that are produced by the QCD fragmentation process and which are identified with a given jet algorithm. The jet functions satisfy DGLAP evolution equations which allow for the resummation of logarithms of the jet radius $R$. One can tune $R$ to capture various stages of the shower, and eventually as $R\rightarrow0$, jet production would merge into the traditional observable of inclusive hadron production~\cite{Collins:1989gx}.

However, inclusive jet production forms only one part of the set of observables one can probe in a fragmentation process, where one wishes to know the dynamical means by which the object of concern (the initial quark or gluon) is randomly broken up. Asking more differential questions about the fragmentation process, or probing more exclusive observables, can reveal the underlying mechanism of fragmentation based on general considerations of probability theory alone~\cite{Derrida_1987}. In QCD scattering, the object we are concerned with is the total momentum in the underlying hard process, and how the resulting fragments are possibly labeled according to polarization or flavor composition. While in vacuum QCD, the underlying dynamical process of fragmentation can be claimed to be qualitatively and even quantitatively understood, the propagation of partons through a strongly interacting medium has required a more careful theoretical treatment.

Thus it is critical to move beyond the consideration of inclusive jets, and in this work we focus on leading jet production. That is, we consider the cross section when only the leading jet is measured in a given rapidity interval per event (analogous to the largest fragment considered in~\cite{Derrida_1987}). The corresponding leading jet function only takes into account the formation and evolution of most energetic jet resulting from an active parton which is illustrated in the right panel of Fig.~\ref{fig:inclusive_leading}. The renormalization group (RG) equation turns out to be a non-linear DGLAP-type evolution equation which was first introduced in~\cite{Dasgupta:2014yra} at leading-order (LO) and leading-logarithmic (LL) accuracy using a generating functional approach. In~\cite{Scott:2019wlk}, these results were extended by including the full fixed order jet function at next-to-leading order (NLO). In addition, the jet functions were incorporated using a complete factorization formula at NLO which was obtained within Soft Collinear Effective Theory (SCET)~\cite{Bauer:2000ew,Bauer:2000yr,Bauer:2001yt, Bauer:2002nz, Beneke:2002ph}. Here we further extend the work of Ref.~\cite{Scott:2019wlk} by evolving the entire NLO jet function using a parton shower Monte Carlo approach and we include the resummation of threshold logarithms~\cite{Sterman:1986aj,Catani:1989ne} which dominate the cross section when the momentum fraction carried by the leading jet relative to the initial parton approaches unity. In addition, we focus specifically on cross sections where an additional reference scale $Q$ is measured such that we can directly measure the momentum fraction of the leading jet relative to $Q$. Vital to our approach is that the leading jet functions constitute normalized probability densities, even outside the Sudakov region. Thus the leading jet is a (theoretically) well-defined object of the event, whose evolution we can track. The probability distribution allows us to calculate the mean and variance of this distribution. The mean corresponds to the average energy contained inside the leading jet relative to the fragmenting parton $i$ which we denote by $\langle z_{1i}\rangle$. Correspondingly, $\langle z_{i,{\rm loss}}\rangle=1-\langle z_{1i}\rangle$ is the average out-of-jet radiation or the leading jet energy loss.

Given that leading jets form a well-defined object distributed probabilistically by the fragmentation process, they allow for a well-defined notion of jet energy loss at the jet function and cross section level. We identify the following three criteria that allow for a meaningful definition of the energy loss distribution and its average:

\begin{itemize}
    \item Different than inclusive jets, the leading jet constitutes a well-defined object which has lost energy relative to the initial parton due to out-of-jet emissions. The corresponding jet functions are normalized probability densities which allow for a perturbative evaluation of the (average) energy loss. We will also discuss a possible extension of the present work to leading hadrons which also allows for a well-defined, but nonperturbative definition of energy loss. We stress that it is not possible to construct the corresponding probability density for inclusive jets since the number of inclusive jets is not fixed but generated dynamically event-by-event through the QCD fragmentation process.
    
    \item To quantify the lost energy, we not only need to know the energy of the leading jet but also a reference scale $Q$ with respect to which we define the energy loss. We consider different observables where the reference scale is given for example by the center-of-mass (CM) $Q=\sqrt{s}$ in $e^+e^-$ collisions. Other examples include jet substructure measurements, Semi-Inclusive Deep-Inelastic Scattering (SIDIS) or photon/$Z$-jet correlations.
    
    \item Lastly, we require that the measured jet energy loss at the cross section level agrees with the (average) parton energy loss at LL accuracy. Higher order effects give corrections to this direct relation which, however, are calculable order-by-order in perturbation theory. The analogy between parton and jet energy loss here is similar to the identification of the variable Bjorken $x_B$ in Deep Inelastic Scattering (DIS) and the parton momentum fraction $x$ at LL accuracy.
\end{itemize}

The concept of jet or parton energy loss has played an important role in theoretical calculations of jet quenching in heavy-ion collisions~\cite{Gyulassy:1993hr,Baier:1996sk,Zakharov:1996fv,Gyulassy:2000er,Wiedemann:2000za,Wang:2001ifa,Arnold:2002ja,Qiu:2004da,CasalderreySolana:2011us,Ovanesyan:2011xy,Armesto:2011ht,CasalderreySolana:2012ef,Mehtar-Tani:2013pia,Burke:2013yra,Qiu:2019sfj,Caucal:2019uvr,Andres:2020vxs,Vaidya:2020cyi}. Typically, the notion of energy loss is defined in the soft gluon approximation where flavor-changing processes are suppressed, and calculations are performed in the lowest non-trivial fixed order or resummed expansions. Ideally, one would like to measure the energy of the parton before and after the interaction with the quark gluon plasma (QGP). The difference is the energy loss due to vacuum and medium-induced emissions and allows for the extraction of properties of the QGP. However, the concept of an energetic parton that exits the hard interaction, losing radiation only due to soft emissions, that then emerges from the scattering is tenuous even in the pure vacuum evolution case. The work presented here using leading jets provides the closest connection to this idealized scenario of energy loss measurements. In the vacuum, the parton/jet energy loss is calculable perturbatively, and nonperturbative effects may be modeled via shape functions. In the medium the average energy loss can be determined experimentally and compared to theoretical model calculations, and then tied to the underlying physics of the QGP.

Beyond simply calculating the average energy loss (which corresponds to the mean of the leading jet distribution), since we have the full probability distribution at hand, we also consider for the first time the variance of the jet energy loss that characterizes event-by-event fluctuations. We analyze in particular the change with the jet radius parameter $R$ for these moments of the leading jet distributions, and we explore differences between the quark and gluon energy loss. Moreover, we introduce additional statistical quantities to help understand the differences between leading quark and gluon jets and their fragmentation processes. For example, we compute the Shannon entropy and KL divergence. The latter quantifies the difference between the quark and gluon leading jet distributions. In addition, we present receiver operator characteristic (ROC) curves for different values of the leading jet radius. This represents a first step toward assessing the potential impact of this observable for quark/gluon jet tagging.

A reliable quantitative understanding of leading jets in the high-energy collider experiments is also necessary for the calibration of the jet energy using $Z/\gamma$-tagged jets~\cite{Chatrchyan:2011ds,Aad:2014bia}. See also Refs.~\cite{Chatrchyan:2012gt,Sirunyan:2017qhf,Aaboud:2018anc}. In addition, it can be advantageous from an experimental point of view to measure additional quantities such as jet substructure observables on the leading and first subleading jet in the event instead of an inclusive jet sample. See for example Refs.~\cite{Aaboud:2017qwh,Aad:2019vyi,Sirunyan:2018xdh}. A reliable evaluation of the quark/gluon fractions requires corresponding factorization formulas analogous to those developed in this work.

One of the main novelties of our work is the development of a Monte Carlo parton shower framework which solves the non-linear DGLAP-type evolution equations of leading jets, while including the complete (to NLL$'$) threshold resummed hard-scattering and jet functions. We find full agreement with analytical results for inclusive jets, validating that the parton shower framework which allows for a systematic extension to leading jet cross sections. In addition, we present numerical results for leading partons where we run the shower down to the nonperturbative scale $\sim 1$~GeV which is a first step toward understanding the fragmentation spectrum of leading hadrons. We leave more detailed studies of leading hadrons for future work. While the shower introduced here should be considered as a single or few purpose Monte Carlo event generator~\cite{Sjostrand:2014zea,Bahr:2008pv,Gleisberg:2008ta}, we expect that it allows for systematic extensions to other observables consistent with analytical results obtained within QCD factorization. See Refs.~\cite{Nagy:2012bt,Hoche:2015sya,Alioli:2015toa,Dasgupta:2018nvj,Bewick:2019rbu,Forshaw:2020wrq} for recent developments of parton shower algorithms.

The remainder of this work is organized as follows. In section~\ref{sec:2}, we introduce the main theoretical concepts of leading (and subleading) jets and compare them to inclusive jet production. We discuss the evolution equations of leading jets and factorization formulas at fixed order and in the threshold limit. In addition, we discuss the connection between leading and subleading jets and inclusive single-, di- and tri-jet functions. In section~\ref{sec:3}, we discuss the setup of the parton shower Monte Carlo framework and we present first numerical results. We discuss the resummation at LL accuracy and the extension beyond LL by including (threshold resummed) hard and jet functions in the shower algorithm. In section~\ref{sec:4}, we derive the threshold resummed hard and jet functions for $e^+e^-$ hemisphere leading jets and leading subjets. We discuss how nonperturbative effects can be included in the threshold limit which is phenomenologically relevant for leading jets and present numerical results for both processes at the cross section level. In section~\ref{sec:5} we calculate the average leading jet energy loss and the variance at NLO by taking moments of the leading jet function. We present numerical results for the mean and variance of the leading jet/energy loss distribution and focus in particular on quark/gluon differences. In addition, we present numerical results for the Shannon entropy and the KL divergence. In section~\ref{sec:6}, we study the discrimination power of leading (sub)jets for quark/gluon jet tagging. In section~\ref{sec:7}, we discuss further applications of our framework such as event-wide leading jets in $e^+e^-$ collisions similar to existing data from LEP. In addition, we consider leading jets in SIDIS and photon-jet correlations in proton-proton collisions. Both processes also allow us to perform jet energy loss measurements. Lastly, we present results for leading partons at the nonperturbative scale. We conclude in section~\ref{sec:8}.

\section{Fixed order, evolution and factorization \label{sec:2}}

We start by reviewing the NLO jet function, its evolution and the factorization formula for inclusive jets $pp\to {\rm jet}+X$ in section~\ref{sec:review_inclusive}. Correspondingly, we discuss the leading jet cross sections $pp\to{\rm jet}_1+X$ in terms of their NLO jet function, and evolution equations in section~\ref{sec:lead_evolution}. In addition, we introduce the relevant jet functions for subleading jets.  Subsequently, we discuss the structure of relevant factorization formulas in section~\ref{sec:lead_factorization}. In section~\ref{sec:leadingfrominclusive}, we extend the inclusive jet function for single jets to di- and tri-jet functions and discuss their relation to leading and subleading jet functions. Here we refer to Ref.~\cite{Derrida_1987} where these relations were proposed in the context of statistical properties of randomly broken objects and spin glasses.

\subsection{Review of inclusive jet production~\label{sec:review_inclusive}}

The inclusive jet function was calculated in Refs.~\cite{Kaufmann:2015hma,Kang:2016mcy,Dai:2016hzf} for $k_T$-type algorithms and in Ref.~\cite{Kang:2017mda} for cone algorithms. See also~\cite{Dai:2018ywt,Li:2018xuv} for the extension to massive quarks. We can write the inclusive jet function $J_i(z,Q R,\mu)$ for $i=q,g$ in terms of the momentum fraction of the inclusive jets $z$, the jet radius $R$, the large reference scale $Q$ and the renormalization scale $\mu$. For inclusive jet production in proton-proton collisions the large reference scale is given by the transverse momentum of the initial fragmenting parton $Q=\hat p_T=p_T/z$. Here $p_T$ denotes the transverse momentum of the final observed jet. We note that $\hat p_T$ is not an observable quantity in proton-proton collisions which is why we convolve the jet function with a hard-scattering function in the factorization formula of the cross section. The range of the convolution integral is determined by the allowed range of $\hat p_T$ which is determined by the jet's transverse momentum and rapidity $p_T,\eta$ and the CM energy $\sqrt{s}$, see Eqs.~(\ref{eq:inclusivefactorization}) and~(\ref{eq:inclusivefactorization1}) below. In subsequent sections we will focus on cross sections where we have access to the initial scale $Q$ which is necessary to define the lost energy of a leading jet as mentioned in the Introduction. Therefore, we keep the general notation here and denote the hard reference scale by $Q$. For quarks and gluons we find the following results at NLO
\begin{align}\label{eq:inclusiveJfunction}
 J_q(z,Q R,\mu) = 
 & \, 
 \delta(1-z)+\f{\as}{2\pi}\left(\ln\left(\f{\mu^2}{Q^2 R^2}\right)-2\ln z\right)\left[P_{qq}(z)+P_{gq}(z)\right]
 \nn \\&
 -\f{\as}{2\pi}\Bigg[C_F\left[2(1+z^2)\left(\f{\ln(1-z)}{1-z}\right)_++(1-z)\right]
 \nn\\&
-\delta(1-z) C_F\left(\f{13}{2}-\f{2\pi^2}{3}\right)+2 P_{gq}(z) \ln(1-z)+C_F z\Bigg]
\\
J_g(z,Q R,\mu) = 
& \, 
\delta(1-z) + \f{\as}{2\pi}\left(\ln\left(\f{\mu^2}{Q^2 R^2}\right)-2\ln z\right)\left[P_{gg}(z)+2 N_f P_{qg}(z)\right]
\nn \\&
-\frac{\alpha_{s}}{2 \pi}\Bigg[\frac{4 C_{A}\left(1-z+z^{2}\right)^{2}}{z}\left(\frac{\ln (1-z)}{1-z}\right)_{+}-\delta(1-z)\bigg(C_{A}\left(\frac{67}{9}-\frac{2 \pi^{2}}{3}\right)
\nn\\&
-T_{F} N_{f}\left(\frac{23}{9}\right)\bigg)+4 N_{f}\left(P_{q g}(z) \ln (1-z)+T_{F} z(1-z)\right)\Bigg]\,.
\end{align}
The inclusive jet function measures the momentum fraction $z$ of all jets that are produced from the initial parton. The evolution equations of the inclusive jet functions are the standard timelike DGLAP evolution equations~\cite{Catani:2013oma,Dasgupta:2014yra,Kang:2016mcy,Dai:2016hzf}
\be
\mu \f{{\rm d}}{{\rm d}\mu}J_i=\frac{\as}{2\pi}\sum_j P_{ji}\otimes J_{j}\,,
\ee
similar to fragmentation functions. Here $P_{ji}(z)$ denote the time-like Altarelli-Parisi splitting functions which allow for a perturbative power expansion in terms of the QCD strong coupling constant
\be
P_{ji}(z)=P_{ji}^{(0)}(z)+\frac{\as}{2\pi}P_{ji}^{(1)}(z)+\ldots
\ee
The inclusive jet functions are then evolved from their characteristic scale $\mu\sim Q R$, which removes logarithms of the jet radius in Eq.~(\ref{eq:inclusiveJfunction}), to the hard scale $\mu\sim Q$. This DGLAP evolution resums large logarithmic corrections of the jet radius to all orders. The factorization formula for the inclusive jet cross section $pp\to{\rm jet}+X$ differential in the transverse momentum $p_T$ and rapidity $\eta$ of the jet is given by
\ba\label{eq:inclusivefactorization}
\frac{{\rm d}\sigma_{pp\to{\rm jet}+X}}{{\rm d}p_T {\rm d}\eta}=\sum_{ijk}f_i(x_i,\mu)\,\otimes\, f_j(x_j,\mu)
\otimes\, H_{ijk}(x_i,x_j,p_T/z,\eta,\mu)\otimes J_k(z,p_T R,\mu)\,.\;\;\;\;\;\;
\ea
Here $f_{i,j}$ denote the parton distribution functions (PDFs) for the two incoming protons and $H_{ijk}$ are the hard-scattering functions $ij\to k$ similar to inclusive hadron production. They are known analytically at NLO~\cite{Aversa:1988vb,Jager:2002xm}. For hadron production, the same factorization formula as in Eq.~(\ref{eq:inclusivefactorization}) can be used except that the perturbative jet functions are replaced by fragmentation functions. The factorization in Eq.~(\ref{eq:inclusivefactorization}) holds up to power corrections ${\cal O}(R^2)$ which are usually found to be small even for large values of $R$~\cite{Mukherjee:2012uz,Scott:2019wlk}. The first two symbols $\otimes$ in Eq.~(\ref{eq:inclusivefactorization}) denote appropriate integrals over the momentum fractions $x_{i,j}$. The third integral denoted by $\otimes$ is a convolution integral in terms of the momentum fraction $z$. To make this structure more explicit, we rewrite Eq.~(\ref{eq:inclusivefactorization}) as
\ba\label{eq:inclusivefactorization1}
\frac{{\rm d}\sigma_{pp\to{\rm jet}+X}}{{\rm d}p_T {\rm d}\eta}=\sum_{i} \int_{z_0}^1\frac{{\rm d}z}{z} \ffh_i(z,p_T, \eta,\mu)\, J_i\bigg(\frac{z_0}{z},p_T R,\mu\bigg)\,,
\ea
where we absorbed the PDFs and corresponding integrations over the variables $x_{i,j}$ in the new hard functions $\ffh_i$. Here the lower integration limit is given by $z_0=2p_T/\sqrt{s}\cosh\eta$. Note that in Eqs.~(\ref{eq:inclusivefactorization}) or~(\ref{eq:inclusivefactorization1}) we do not have access to the partonic transverse momentum $\hat p_T$ which appeared in the jet functions in Eq.~(\ref{eq:inclusiveJfunction}). Therefore, we write both the hard and jet function in terms of the observed jet transverse momentum $p_T$ instead of the initial partonic transverse momentum using the relation $\hat p_T=p_T/z$ as in Ref.~\cite{Kang:2016mcy}. This is valid as long as we are not in the regime where $z\ll 1$ which would require an additional resummation of small-$z$ logarithms. See Refs.~\cite{Neill:2020tzl,Neill:2020bwv} as well as earlier work in Refs.~\cite{Mueller:1981ex,Bassetto:1982ma,Dokshitzer:2005bf,Kom:2012hd,Anderle:2016czy}

We end this section by summarizing some key features of the inclusive jet functions. Similar to PDFs and fragmentation functions, the inclusive jet functions constitute number densities in the sense that an integral over the jet function (first Mellin moment) yields the event averaged number of jets $\langle N_{\rm jet}\rangle$ which originate from the fragmenting parton. This number is generated dynamically through the QCD fragmentation process and depends on the jet radius, the jet algorithm and the scale $Q$. Analogously, the same integral over fragmentation functions gives the average number of hadrons or the hadron multiplicity. The first moment (second Mellin moment) is related to momentum conservation in the sense that the initial scale $Q$ has to be recovered in the inclusive jet sample that is produced resulting from the initial parton. Of course, in practice not the entire inclusive jet sample resulting from a highly energetic quark or gluon is reconstructed by the experiment. The limited range of detectors is taken into account in the factorization formula in Eq.~(\ref{eq:inclusivefactorization}) which depends on the jet rapidity $\eta$. At the level of the jet function, we thus have the following two sum rules
\ba
&&\int_0^1{\rm d}z \, J_i(z,Q R,\mu)=\langle N_{i,{\rm jets}} \rangle \,,\label{eq:sum1}\\
&&\int_0^1{\rm d}z \, z\, J_i(z,Q R,\mu)=1\,,\label{eq:sum2}
\ea
which hold for quarks and gluons separately. Note that the quantity $\langle N_{i,{\rm jets}}\rangle$ introduced here is related to the entropy of a jet~\cite{Neill:2018uqw,Hagiwara:2017uaz}. In order to evaluate the first Mellin moment analytically, the resummation of small-$z$ logarithms is required as mentioned above. Related experimental results can be found in Ref.~\cite{Chekanov:2002ux} where the subjet multiplicity was measured. The momentum sum rule for an inclusive jet sample in Eq.~(\ref{eq:sum2}) is illustrated on the left side of Fig.~\ref{fig:inclusive_leading} where three jets are reconstructed resulting from an initial fragmenting quark. In this case, the momentum fractions of the three jets have to add up to unity $z_1+z_2+z_3=1$. An important aspect is that the momentum sum rule in Eq.~(\ref{eq:sum2}) is conserved by the time-like DGLAP evolution of the jet functions which follows from
\be\label{eq:conservation1}
\mu\frac{{\rm d}}{{\rm d}\mu}\int_0^1 {\rm d}z\, z\, J_i(z,Q R,\mu)\sim \sum_{j}\int_0^1 {\rm d}z\, z\, P_{ji}(z)=0 \,.
\ee
We note that both sum rules in Eqs.~(\ref{eq:sum1}) and~(\ref{eq:sum2}) only hold if the inclusive jet functions in Eq.~(\ref{eq:inclusiveJfunction}) are written in terms of the partonic momentum $Q=\hat p_T$.

\subsection{Leading and subleading jet functions and their evolution~\label{sec:lead_evolution}}

We are now going to introduce the jet functions for leading and subleading jets analogous to the inclusive case discussed in the previous section. The LO and LL resummation for leading jets was first introduced in Ref.~\cite{Dasgupta:2014yra}. The ${\cal O}(\alpha_s)$ correction of the leading and subleading jet functions was first discussed in Ref.~\cite{Scott:2019wlk}. Here we denote the leading jet function by ${\cal J}_{i}(z_1,Q R,\mu)$ which depends on the momentum fraction $z_1$ of the leading jet relative to the initial scale $Q$ of the fragmenting quark or gluon, see the right panel of Fig.~\ref{fig:inclusive_leading}. Analogously, the jet function that describes the formation of the leading and the first subleading jet is given by ${\cal J}_{i}(z_1,z_2,Q R,\mu)$. It depends on the momentum fractions $z_1$ and $z_2$ of the leading and first subleading jet, respectively. It is a more differential version of the leading jet function. Note that throughout this work we write jet and hard-scattering functions associated with leading and subleading jets in script font to distinguish them from the corresponding inclusive jet quantities.

We start with the fixed order calculation of the leading and subleading jet function. At LO there is just one parton and we trivially find
\begin{align}
    {\cal J}_{i}^{(0)}(z_1,Q R,\mu) & =\, \delta(1-z_1) \,,\\
    {\cal J}_{i}^{(0)}(z_1,z_2,Q R,\mu) & =\, \delta(1-z_1)\,\delta(z_2) \,.
\end{align}
At NLO there are at two partons which can be clustered into a single jet or two separate jets depending on their distance and the jet algorithm. If both partons are clustered into the same jet, the contribution is the same as for inclusive jets and it is proportional $\sim\delta(1-z_1)$. If the two partons are clustered into separate jets, we only take into account the jet which contains the more energetic parton. Instead, for inclusive jets we always take into account both jets independent of how energetic they are. It turns out that at NLO we can write the leading jet functions in terms of the inclusive ones by including a theta function which requires $z_1>1/2$. We thus find for the leading and subleading jet functions up to NLO
\begin{align}\label{eq:leadJfunction}
{\cal J}_{i}(z_1,Q R,\mu)&=\,\Theta(z_1>1/2)\,J_i(z_1,Q R,\mu)\,, \\
{\cal J}_{i}(z_1,z_2,Q R,\mu)&=\,\delta(1-z_1-z_2)\,\Theta(z_1>1/2)\,J_i(z_1,Q R,\mu)~\label{eq:subleadJfunction}\,.
\end{align}
Note that the corresponding jet functions ${\cal J}_{i}(z_1,\ldots,z_n,Q R,\mu)$ which take into account the dynamics down to the $n$-th leading jet can be constructed in a similar way. We also note that the leading jet function is obtained from the subleading jet function upon integration
\begin{equation}
    \int_0^1{\rm d}z_2\,{\cal J}_{i}(z_1,z_2,Q R,\mu)={\cal J}_{i}(z_1,Q R,\mu) \,.
\end{equation}
The leading jet function at NLO is only non-zero for $z>1/2$ where it agrees with the inclusive jet function. At next-to-next-to-leading order (NNLO), we need to consider three particles in the final state which gives a lower bound for the leading jet function of $z>1/3$. In general, at N$^n$LO, the minimal non-zero value of the leading jet function is thus given by $1/(n+1)$.

\begin{figure}[t]
\vspace*{.7cm}
\centering
\includegraphics[width=10cm]{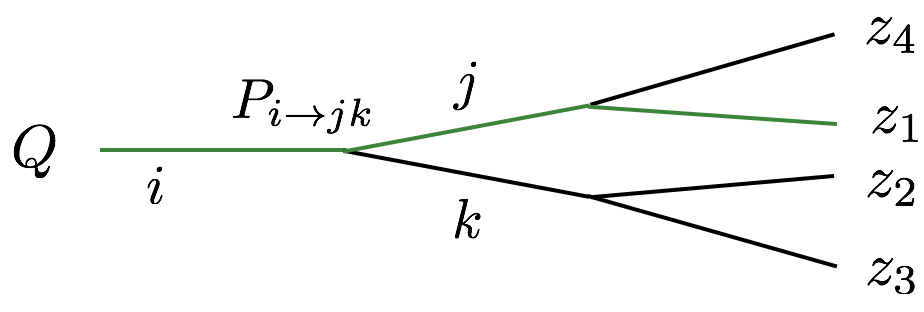}
\caption{Illustration of an exemplary branching tree: Starting from parton $i$ at the scale $Q$, four partons/jets are produced which carry a momentum fraction $z_n$ relative to the initial scale. In order to determine the correct path (green) that leads to the leading jet, we need to know the momentum fractions and leading jet functions of each branch at a given splitting $i\to jk$. This results in a non-linear evolution equation for observables involving leading jets. Instead, for inclusive jets we only need to know the final momentum fractions $z$ of each jet irrespective of the they compare to the rest of the event since we sum over all contributions.~\label{fig:splitting_example}}
\end{figure}

Next, we consider the evolution equation of the leading jet functions. Different than for inclusive jets, the evolution equations for leading and subleading jets are non-linear. At NLO there are only two partons and it is sufficient to follow the leading parton at the $1\to 2$ splitting. If there are subsequent splitting processes as well, one needs to know the value of the leading jet function at each branching point. This is illustrated in Fig.~\ref{fig:splitting_example} which shows an exemplary branching tree, where the green line leads to the leading jet which carries a momentum fraction $z_1$. In order to obtain the correct path, we need to known at each branching point the value of the branching fraction $z$ of the splitting $i\to jk$ as well as the leading jet functions ${\cal J}_{j}(z_{j1})$ and ${\cal J}_{k}(z_{k1})$ of each branch. This feature of leading jets makes the evolution equations non-linear. Instead, for inclusive jets we only need to sum over all possible paths of the branching tree and we obtain the usual linear DGLAP evolution equations, see section~\ref{sec:review_inclusive} above. The measurement of leading jets requires knowledge of all the other jets when the evolution terminates. Only if we know all other jets we can determine which jet is the leading jet. Instead, for an inclusive measurement we simply sum over all possible jets at the end which does not require simultaneous knowledge about the other jets that are produced.

We can write the non-linear evolution equation for the leading jet functions as\footnote{Note that we write $z_{i1}$ instead of $z_1$ in case the association of the momentum fraction to a particular jet function may be ambiguous or if the flavor dependence is relevant.}
\begin{align}\label{eq:nonlinearevol}
\mu\frac{{\rm d}}{{\rm d}\mu}{\cal J}_{i}(z_{1i},Q R,\mu)=\,&\f12 \sum_{jk}\int {\rm d}z\, {\rm d}z_{j1} {\rm d}z_{k1} \f{\alpha_s(\mu)}{\pi}P_{i\to jk}(z)\, {\cal J}_{j}(z_{j1},Q R,\mu)\, {\cal J}_{k}(z_{k1},Q R,\mu) \nn\\
&\times \delta(z_{i1}-\text{max}\left\{z z_{j1},(1-z)z_{k1} \right\})
\end{align}
Here we follow Ref.~\cite{Elder:2017bkd} and use the notation $P_{i\to jk}(z)=P_{ji}^{(0)}(z)$ for the LO Altarelli-Parisi splitting functions and we regulate both endpoints $1/(1-z)_+$ and $1/z_+$. We thus have $P_{q\to qg}(z)=P_{q\to gq}(1-z)$ and $P_{g\to gg}(z)=P_{g\to gg}(1-z)$. We can then check explicitly that the leading jet functions in Eq.~(\ref{eq:leadJfunction}) satisfy the non-linear evolution equations. Similar non-linear evolution equation were obtained in the context of fractal jet substructure observables~\cite{Elder:2017bkd}, the jet charge~\cite{Krohn:2012fg,Waalewijn:2012sv}, track functions~\cite{Chang:2013rca,Chen:2020vvp} and di-hadron fragmentation functions~\cite{deFlorian:2003cg,Ceccopieri:2007ip,Metz:2016swz}.

We also note that the evolution equations here are different than for the central subjets or the jet shape with the winner-take-all axis (WTA)~\cite{Bertolini:2013iqa} which was considered in Refs.~\cite{Neill:2016vbi,Kang:2017mda,Neill:2018wtk}. Even though the fixed order expressions of the jet functions contain a similar term $\sim\Theta(z>1/2)P_{ji}(z)$, the use of the WTA axis leads to a linear DGLAP-type evolution equation with modified kernels.

Similar to the evolution equation for the leading jet functions in to Eq.~(\ref{eq:nonlinearevol}), we can write down an evolution equation for the subleading jet functions. We have
\begin{align}\label{eq:nonlinearevolsubleadingjet}
&\mu\frac{{\rm d}}{{\rm d}\mu}{\cal J}_{i}(z_{i1},z_{i2},Q R,\mu)=
\nn\\&\quad
\f12 \sum_{jk}\int {\rm d}z\, {\rm d}z_{j1} {\rm d}z_{k1}{\rm d}z_{j2} {\rm d}z_{k2} \f{\alpha_s(\mu)}{\pi}P_{i\to jk}(z)\,
{\cal J}_{j}(z_{j1},z_{j2},Q R,\mu)\, {\cal J}_{k}(z_{k1},z_{k2},Q R,\mu) 
\nn\\ &\times 
\big[\Theta(z z_{j1}-(1-z)z_{k1})\, \delta(z_{i1}- z z_{j1}) \,\delta(z_{i2}-\text{max}\{z z_{j2},(1-z)z_{k1}\})
\nn\\& \quad \;\, +
\Theta((1-z)z_{k1}-z z_{j1})\, \delta(z_{i1}- (1-z)z_{k1}) \,\delta(z_{i2}-\text{max}\{(1-z) z_{k2},z z_{j1}\})\big]\,.
\end{align}
Different than for inclusive jet functions in Eq.~(\ref{eq:sum1}), we find that the leading jet functions are normalized to unity (first Mellin moment). The first moment (second Mellin moment) corresponds to the average energy or transverse momentum fraction $\langle z_{1}\rangle$ which is contained in the leading jet. Recall that for inclusive jets, the second Mellin moment is unity due to momentum conservation, see Eq.~(\ref{eq:sum2}). We thus find the following expressions for the leading jet functions
\begin{align}
\int_0^1 {\rm d}z_{i1} \, {\cal J}_{i}(z_{i1},Q R,\mu)&=\,1 \,,~\label{eq:sum3}\\
\int_0^1 {\rm d}z_{i1}\, z_{i1} \, {\cal J}_{i}(z_{i1},Q R,\mu)&=\,\langle z_{i1}\rangle \,.~\label{eq:sum4}
\end{align}
In order to interpret the leading jet functions as probability densities for the leading jet to carry the longitudinal momentum fraction $z_1$, they need to be $i)$ normalized to unity and $ii)$ positive for all values of $z_1$. The first requirement is satisfied as can be seen from Eq.~(\ref{eq:sum3}). For example, we can check this result using the NLO expression. In addition, this normalization is conserved under the non-linear evolution, i.e., we have
\begin{equation}\label{eq:conservation2}
\mu\frac{{\rm d}}{{\rm d}\mu}\int_0^1 {\rm d}z_{1}\, {\cal J}_i(z_{1},Q R,\mu) = 0\,,
\end{equation}
which follows from Eq.~(\ref{eq:nonlinearevol}). The conservation of the first Mellin moment for leading jets is analogous to the conservation of the second Mellin moment for inclusive jets, see Eq.~(\ref{eq:conservation1}) and only holds when the jet function is written in terms of $Q=\hat p_T$ instead of the transverse momentum of the observed jet. The second requirement that the leading jet functions need to be positive for all values of $z_{1}$ can be violated at finite perturbative accuracy. For example, at NLO the leading jet function can become negative especially for large values of $z_{1}$ and small values of the jet radius $R$. However, we observe that this problem can be solved by including threshold resummation which we discuss in more detail in section~\ref{sec:4} below. Therefore, we can indeed treat the leading jet functions as probability densities which allow us to calculate the expectation value or the average energy contained in the leading jet $\langle z_{i1}\rangle$ as in Eq.~(\ref{eq:sum4}) above. We note that probability densities are rather unusual in this context since we typically obtain number densities as for inclusive jets or inclusive hadrons. Similar sum rules as in Eqs.~(\ref{eq:sum3}) and~(\ref{eq:sum4}) hold for leading hadrons which we discuss in more detail in section~\ref{sec:hadrons}. In addition, analogous sum rules as in Eqs.~(\ref{eq:sum3}) and~(\ref{eq:sum4}) hold for subleading jets and the corresponding jet functions can be treated as probability densities.

Having defined the average energy contained in the leading jet $\langle z_{i1}\rangle$, we can now define the average energy loss of a leading jet which is given by all the energy which is not contained in the leading jet
\begin{equation}\label{eq:zloss}
    \langle z_{i,{\rm loss}}\rangle = 1-\langle z_{i1}\rangle \,.
\end{equation}
We discuss $\langle z_{i,{\rm loss}}\rangle$ and other statistical quantities that quantify the probability distribution of jet energy loss in section~\ref{sec:5} for both the jet functions and at the level of cross sections.

We also would like to stress that the notion of a leading or subleading jet does not directly correspond to the leading jet in the entire event in proton-proton collisions which is not possible due to the finite detector coverage and the incoming particle beams. Instead, we calculate the leading jet in a given rapidity interval. This  dependence is taken into account in factorization formulas which we discuss in the next section similar to inclusive jets.

We can now extend Eq.~(\ref{eq:sum4}) to subleading jets. For example, we can use the subleading jet function to calculate the the average energy which is contained in the first subleading jet as
\begin{align}
    \int_0^1{\rm d}z_{i1}\, {\cal J}_i(z_{i1},z_{i2},Q R,\mu)&=\, {\cal J}_{i2}(z_{i2},Q R,\mu) \,,
    \\
   \int_0^1 {\rm d}z_{i2}\, z_{i2}\, {\cal J}_{i2}(z_{i2},QR,\mu)&=\,\langle z_{i2}\rangle \,.
\end{align}
Here, ${\cal J}_{i2}$ is the jet function of the second leading jet which is indicated by the additional subscript 2. It constitutes the probability density for finding the first subleading jet with momentum fraction $z_{i2}$ and we can thus calculate the average quantity $\langle z_{i2}\rangle$. Note that the full subleading jet function is needed for the evolution in Eq.~(\ref{eq:nonlinearevolsubleadingjet}). However, the information of the leading jet momentum fraction $z_{i1}$ can be integrated out after the evolution and we can calculate the probability density for the first subleading jet. Analogous equations hold for the $n$-th leading jet
\begin{align}\label{eq:njet}
    \int_0^1{\rm d}z_{i1}\ldots {\rm d}z_{i(n-1)}\, {\cal J}_i(z_{i1},\ldots ,z_{in},Q R,\mu)&=\, {\cal J}_{in}(z_{in},QR,\mu) \,,
    \\
   \int_0^1 {\rm d}z_{in}\, z_{in}\, {\cal J}_{in}(z_{in},QR,\mu)&=\,\langle z_{in}\rangle \,.
\end{align}
The average values of the energy fractions contained in the subleading jets is also relevant for studies of the jet energy loss. They contain the information of how the lost energy of the leading jet $\langle z_{i,{\rm loss}}\rangle$, as defined in Eq.~(\ref{eq:zloss}), is distributed amongst the subleading jets. We have
\begin{equation}
    \langle z_{i,{\rm loss}}\rangle = \sum_{n\geq 2}
    \langle z_{in}\rangle\,.
\end{equation}
We note that the non-linear evolution equations for leading and subleading jets in Eqs.~(\ref{eq:nonlinearevol}) and~(\ref{eq:nonlinearevolsubleadingjet}), respectively, are difficult to solve analytically. In Ref.~\cite{Dasgupta:2014yra,Scott:2019wlk} an iterative approach was used starting with the LO jet function as the initial condition of the evolution. However, this approach is impractical when the whole NLO jet function is evolved and when additional threshold logarithms are resummed to all orders which is discussed in more detail in section~\ref{sec:4} below. In addition, in order to calculate the (sub)leading jet spectrum at small-$z$ many iterations would need to be computed analytically. Ref.~\cite{Dasgupta:2014yra} also explored a strictly leading order parton shower method. However, we introduce a new Monte Carlo approach to solve the non-linear evolution equations for leading and subleading jets which we discuss in detail in section~\ref{sec:3}, that allows for the incorporation of threshold resummation effects and jet function contributions, moving beyond the LO process. Different than general purpose Monte Carlo event generators our approach is a single purpose (or few purpose) parton shower which is designed to specifically solve the above evolution equations. It allows for a well defined perturbative accuracy similar to analytical calculations and it allows for systematic improvements to yet higher perturbative accuracy.

Before introducing the parton shower algorithm, we first discuss relevant factorization formulas for leading and subleading jets in the next section.

\subsection{Factorization~\label{sec:lead_factorization}}

Different than for inclusive jets, see Eq.~(\ref{eq:inclusivefactorization}), the form of the factorization formula for leading jets $pp\to{\rm jet}_1+X$ depends on the perturbative accuracy we are working at. The structure of the hard functions changes and additional jet functions need to be included as the perturbative accuracy is increased. Here we consider the cross section differential only in transverse momentum of the leading jet $p_{T1}$ for $\eta=0$. The extension to $\eta\neq 0$ is straightforward since only the hard functions depend on the rapidity and the overall theta functions which enforce momentum conservation. Following Ref.~\cite{Scott:2019wlk}, we can write the leading jet cross section at LO and LL accuracy as
\begin{align}\label{eq:pp_leadingjet1}
\frac{{\rm d}\sigma_{pp\to{\rm jet}_1+X}^{(0)}}{{\rm d}p_{T1}}=&\sum_{ij}\int {\rm d}\hat p_{Ti}\,{\rm d}\hat p_{Tj}\,\int{\rm d}z_{i1}\,{\rm d}z_{j1}\; f\!f{\cal H}^{(0)}_{ij}(\hat p_{Ti},\hat p_{Tj},\mu)
\nn\\
&\times \, {\cal J}_{i}(z_{i1},\hat p_{Ti} R,\mu)\,{\cal J}_{j}(z_{j1},\hat p_{Tj}R,\mu)
\nn\\
&\times \,\delta(p_{T1}-{\rm max}\{z_{i1}\hat p_{Ti},z_{j1} \hat p_{Tj}\})\,\Theta(2\hat p_{Ti}/\sqrt{s}<1)\, \Theta(2\hat p_{Tj}/\sqrt{s}<1)\,,
\end{align}
where we sum over all contributing channels $ij$. Note that we denote the appropriate LO hard-scattering functions by $f\!f{\cal H}^{(0)}_{ij}$ whereas the inclusive hard functions are written as $ffH_{i}$. The hard functions $f\!f{\cal H}_{ij}^{(0)}$ describe the production of two partons with momenta $\hat p_{Ti,j}$ in a hard-scattering event which are back-to-back at LO in the transverse plane. In principle we could also write the LO hard function only as a function of $\hat p_{Ti}=\hat p_{Tj}$. The hard functions also include the initial state parton distribution functions and appropriate integrals over the momentum fractions of the incoming partons. The jet functions ${\cal J}_{i,j}$ take into account the formation and evolution of the two leading jets which originate from partons $i$ and $j$, respectively. The delta function in the last line of Eq.~(\ref{eq:pp_leadingjet1}) then picks one of the two leading jets from partons $i,j$. The one with the larger transverse momentum is measured and denoted by $p_{T1}$. The two theta functions in the last line ensure momentum conservation. We can also rewrite Eq.~(\ref{eq:pp_leadingjet1}) in terms of the measured jet transverse momenta instead of the partonic quantities $p_{Ti}=z_{i1}\hat p_{Ti}$ which is valid as long as we are sufficiently far away from the region where $z_{i1}\ll 1$. We find \begin{align}\label{eq:pp_leadingjet2}
\frac{{\rm d}\sigma_{pp\to{\rm jet}_1+X}^{(0)}}{{\rm d}p_{T1}}=&\sum_{ij}\int {\rm d} p_{Ti}\,{\rm d} p_{Tj}\,\int\frac{{\rm d} z_{i1}}{z_{i1}}\,\frac{{\rm d}z_{j1}}{z_{j1}}\; f\!f{\cal H}^{(0)}_{ij}(p_{Ti}/z_{i1},p_{Tj}/z_{j1},\mu)
\nn\\
&\times \, {\cal J}_{i}(z_{i1},p_{Ti} R,\mu)\,{\cal J}_{j}(z_{j1},p_{Tj}R,\mu)
\nn\\
&\times \,\delta(p_{T1}-{\rm max}\{p_{Ti}, p_{Tj}\})\,\Theta(2p_{Ti}/\sqrt{s}<z_{i1})\,\Theta(2p_{Tj}/\sqrt{s}<z_{j1})\,,
\end{align}
which is closer to the factorization formula for inclusive jets in Eq.~(\ref{eq:inclusivefactorization}). In either case the evolution equations only resum logarithms of the jet radius. Next, we consider the leading jet cross section with NLO hard-scattering functions. At NLO a third hard parton can be radiated which requires us to introduce an additional jet function. The final jet with the highest transverse momentum can result from the fragmentation process of either of the three hard partons produced in the hard-scattering process. We thus have
\begin{align}\label{eq:pp_leadingjetNLO}
\frac{{\rm d}\sigma_{pp\to{\rm jet}_1+X}^{(1)}}{{\rm d}p_{T1}}=&\sum_{ijk}\int {\rm d} \hat p_{Ti}\,{\rm d}\hat p_{Tj}\,{\rm d} \hat p_{Tk}\,\int {\rm d}z_{i1}\,{\rm d}z_{j1}\,{\rm d}z_{k1}\; f\!f{\cal H}^{(1)}_{ijk}(\hat p_{Ti},\hat p_{Tj},\hat p_{Tk},\mu)
\nn\\
&\times \, {\cal J}_{i}(z_{i1},\hat p_{Ti} R,\mu)\,{\cal J}_{j}(z_{j1},\hat p_{Tj}R,\mu)\,{\cal J}_{k}(z_{k1},\hat p_{Tk}R,\mu)
\nn\\
&\times \,\delta(p_{T1}-{\rm max}\{z_{i1} \hat p_{Ti},z_{j1} \hat p_{Tj},z_{k1} \hat p_{Tk}\})
\nn\\
&\times \, \Theta(2\hat p_{Ti}/\sqrt{s}<1)\,\Theta(2\hat p_{Tj}/\sqrt{s}<1)\,\Theta(2\hat p_{Tk}/\sqrt{s}<1)\,,
\end{align}
where $ffH^{(1)}_{ijk}$ denotes the NLO hard function with three hard partons $ijk$ which subsequently fragment into jets. This result can be generalized to higher orders. 

Next, we consider the cross section where we not only measure the transverse momentum of the leading jet but also the first subleading jet. We denote the transverse momentum of the fist subleading jet by $p_{T2}$. The result obtained here can also be extended to the measurement of further subleading jets. At LO and LL, we find the following result
\begin{align}\label{eq:pp_subleadingjet1}
\frac{{\rm d}\sigma_{pp\to{\rm jet}_1+{\rm jet}_2+X}^{(0)}}{{\rm d}p_{T1}\,{\rm d}p_{T2}}=&\sum_{ij}\int {\rm d}\hat p_{Ti}\,{\rm d}\hat p_{Tj}\,\int{\rm d}z_{i1}\,{\rm d}z_{i2}\,{\rm d}z_{j1}\,{\rm d}z_{j2}\; f\!f{\cal H}^{(0)}_{ij}(\hat p_{Ti},\hat p_{Tj},\mu)
\nn\\
&\times \, {\cal J}_{i}(z_{i1},z_{i2},\hat p_{Ti} R,\mu)\,{\cal J}_{j}(z_{j1},z_{j2},\hat p_{Tj}R,\mu)
\nn\\
&\times \,\delta(p_{T1}-{\rm max}\{z_{i1}\hat p_{Ti},z_{j1} \hat p_{Tj}\})
\nn\\
&\times\,
\delta(p_{T2}-{\rm max}\{{\rm min}\{z_{i1}\hat p_{Ti},z_{j1} \hat p_{Tj}\},z_{i2}\hat p_{Ti}, z_{j2}\hat p_{Tj}\})
\nn\\
&\times \,
\Theta(2\hat p_{Ti}/\sqrt{s}<1)\, \Theta(2\hat p_{Tj}/\sqrt{s}<1)\,.
\end{align}
Here the leading jet $p_{T1}$ is given by the transverse momentum of the leading jet originating either from parton $i$ or $j$ as above. The transverse momentum of the first subleading jet $p_{T2}$ is given by the smaller one of the leading jets from partons $i$ and $j$ or one of the subleading jets. The theta functions in the last line do not change as they are written in terms of the initial partonic transverse momentum. The factorization for leading and subleading jets in Eq.~(\ref{eq:pp_subleadingjet1}) can be generalized to higher perturbative accuracy similar to Eq.~(\ref{eq:pp_leadingjetNLO}).

It is instructive to consider how the factorization for inclusive jets is recovered by summing over the leading jet and all subleading jets. After carrying out the sum over all jets, the factorization structure simplifies significantly and it has the same structure to all orders. To make this connection more explicit, we work with a factorization formula as in Eq.~(\ref{eq:pp_leadingjet1}). The delta function in the last line of that equation, which specifies the measurement, needs to be replaced by a sum over delta functions that measure the transverse momentum of all jets. Schematically, for two initial partons we have
\begin{equation}\label{eq:leadinginclusivedelta}
    \delta(p_T-z_{i1}\hat p_{Ti})+\delta(p_T-z_{j1}\hat p_{Tj})+\,\text{other subleading jets} \,,
\end{equation}
where the first two terms which are written out explicitly correspond to the two leading jets originating from partons $i$ and $j$. As an example, we consider only the first delta function. We find that we can rewrite the corresponding contribution to the inclusive cross section as
\begin{align}\label{eq:pp_leadinginclusive1}
&\int {\rm d}\hat p_{Ti}\,{\rm d}\hat p_{Tj}\,\int{\rm d}z_{i1}\,{\rm d}z_{j1}\; f\!f{\cal H}^{(0)}_{ij}(\hat p_{Ti},\hat p_{Tj},\mu) \, {\cal J}_i(z_{i1},\hat p_{Ti} R,\mu)\,{\cal J}_j(z_{j1},\hat p_{Tj}R,\mu)
\nn\\&
\times \,\delta(p_{T}-z_{i1}\hat p_{Ti})\,\Theta(2\hat p_{Ti}/\sqrt{s}<1)\,\Theta(2\hat p_{Tj}/\sqrt{s}<1)
\nn\\
=& \int^1_{z_{i1}^{0}}\frac{{\rm d}z_{i1}}{z_{i1}}{\cal J}_i(z_{i1},p_{T}R,\mu)\int {\rm d}\hat p_{Tj}\int {\rm d}z_{j1}\,J_j(z_{j1},\hat p_{Tj}R,\mu)\, f\!f{\cal H}_{ij}^{(0)}(p_T/z_{i1},\hat p_{Tj},\mu) \,\Theta(2\hat p_{Tj}/\sqrt{s}<1)
\nn\\
=& \int^1_{z_{i1}^{0}}\frac{{\rm d}z_{i1}}{z_{i1}}{\cal J}_i(z_{i1},p_{T}R,\mu) \, f\!f{\cal H}'^{\,(0)}_{i}\bigg(\frac{z_{i1}^{0}}{z_{i1}},\mu\bigg)\,,
\end{align}
where the lower integration limit here is given by $z_{i1}^{0}=2p_T/\sqrt{s}$ as expected from Eq.~(\ref{eq:inclusivefactorization}) for $\eta=0$. The jet function ${\cal J}_i$ in the second and third line is written in terms of the observed jet $p_T$ instead of the parton transverse momentum similar to Eq.~(\ref{eq:pp_leadingjet2}). After performing the integral over $z_j$ and $\hat p_{Tj}$ and implicitly defining $f\!f{\cal H}'_i$, the last equation has the typical convolution structure as it is found for inclusive jets. Similar steps hold for the other jets that are produced and eventually we find
\begin{equation}
   \sum_{\rm all\, jets}\sum_i {\cal J}_i\otimes f\!f{\cal H}'_i= \sum_i J_i\otimes f\!fH_i \,.
\end{equation}
The right hand side of the equation is written in terms of the inclusive jet and hard functions of Eq.~(\ref{eq:inclusivefactorization}). Note that if we consider the inclusive cross section where only the two leading jets are measured for every event, it would approximate the inclusive jet cross section at high $p_T$ but differ at low $p_T$ where subleading jets are important. The factorization of the corresponding cross section would involve the subleading jet functions as in Eq.~(\ref{eq:pp_subleadingjet1}). These kind of differences may also be relevant in the context of assessing QCD scale uncertainties of jet cross sections, see for example~\cite{Currie:2017ctp,Bellm:2019yyh}.

We also note that a factorization formula similar to Eq.~(\ref{eq:pp_subleadingjet1}) is relevant for the precise calculation of jet substructure observables in order to compare to recent measurements from ATLAS~\cite{Aaboud:2017qwh,Aad:2019vyi} and CMS~\cite{Sirunyan:2018xdh}. In this case the jet substructure measurements were performed only on the leading and the first subleading jet. In addition, a requirement $p_{T1}/p_{T2}<1.5$ (ATLAS) or $(p_{T1}-p_{T2})/(p_{T1}+p_{T2})<0.3$ (CMS) was imposed on the ratio of the transverse momenta of the two leading jets. These constraints affect the quark/gluon fractions which are not be properly taken into account when a factorization formula for inclusive jets is used instead.


\subsection{Leading and subleading jets from inclusive single-, di- and tri-jet functions~\label{sec:leadingfrominclusive}}

The (single-)inclusive jet function $J_i(z)$ was already introduced in section~\ref{sec:review_inclusive}. Analogously, we can introduce the inclusive di- and tri-jet functions which we denote by
\begin{equation}
    J_i(z_1,z_2)\,,\quad J_i(z_1,z_2,z_3)\,.
\end{equation}
The inclusive di-jet function is similar to di-hadron fragmentation functions considered for example in Refs.~\cite{deFlorian:2003cg,Ceccopieri:2007ip,Metz:2016swz}. Analogously, we denote the inclusive $n$-jet function by $J_i(z_1,\ldots,z_n)$. Different than for the leading and subleading jet functions no ordering of the momentum fractions $z_i$ is implied. For example, at NLO we can write the di- and tri-jet functions in terms of the (single-)inclusive jet function
\begin{align}
    J_i(z_1,z_2) &=\, \delta(1-z_1-z_2)\, J_i(z_1,QR,\mu)\,,
    ~\label{eq:dijetnlo}\\
    J_i(z_1,z_2,z_3) &=\, \delta(1-z_1-z_2)\,\delta(z_3)\, J_i(z_1,QR,\mu)\,.
\end{align}
Note that the NLO di-jet function in Eq.~(\ref{eq:dijetnlo}) is similar to the NLO subleading jet function in Eq.~(\ref{eq:subleadJfunction}) but without the theta function $\Theta(z_1>1/2)$. Higher order results for the di- and tri-jet functions can be obtained with the parton shower which will be introduced in the next section. In the following, we leave the arguments $Q R,\mu$ of the jet functions implicit. Following Ref.~\cite{Derrida_1987}, we should be able to compute the leading and subleading jet functions from the inclusive $n$-jet functions. If the observed jet has a momentum fraction $z>1/2$ it is the largest jet. We thus have
\begin{equation}
    {\cal J}_{i}(z) = {\cal J}_{i1}(z) = J_{i}(z)\,, \;\; \text{for}\;\; z>1/2 \,.
\end{equation}
In the range of $1/3<z<1/2$ the inclusive jet function is given by the sum of the leading and subleading jet functions.
\begin{equation}
    {\cal J}_{i1}(z)+{\cal J}_{i2}(z) = J_{i}(z)\,, \;\; \text{for}\;\; 1/3<z<1/2 \,.
\end{equation}
Analogous to the notation introduced in Eq.~(\ref{eq:njet}) we include a subscript $n$ to denote the jet function which depends only on the momentum fraction of the $n$-th leading jet. For example, we have
\begin{equation}
    {\cal J}_{i2}(z_2)=\int_0^1{\rm d}z_1\, {\cal J}_{i}(z_1,z_2) \,.
\end{equation}
More generally, we have
\begin{equation}
    {\cal J}_{i1}(z)+\ldots +{\cal J}_{in}(z) = J_{i}(z)\,, \;\; \text{for}\;\; 1/(n+1)<z<1/n \,.
\end{equation}
Here we limit ourselves to confirm the observation made in Ref.~\cite{Derrida_1987} in the range of $1/3<z<1/2$. For lower values of $z$ a similar analysis is possible as outlined in Ref.~\cite{Derrida_1987}. We have
\begin{equation}
    {\cal J}_{i2}(z_2)=\int_{z_2}^1{\rm d}z_1\, J_i(z_1,z_2)\,,\;\; \text{for}\;\; 1/3<z<1/2 \,.
\end{equation}
And the leading jet function is given by
\begin{equation}
    {\cal J}_{i1}(z_1)= J_i(z)-\int_{z_2}^1{\rm d}z_1\, J_i(z_1,z_2) \,,\;\; \text{for}\;\; 1/3<z<1/2 \,.
\end{equation}
These relations were derived in~\cite{Derrida_1987} from general considerations. In a general fragmentation process, one would need to know the whole set of leading, sub-leading, or sub-sub-leading fragment functions, etc., or the whole suite of inclusive multi-fragment distribution functions to completely characterize the fragmentation process, unless a more dynamical rule for generating these functions can be found. In the jet case, the parton shower constitutes such a dynamical rule, at least to leading logarithmic accuracy.


\section{Monte Carlo setup~\label{sec:3}}
 
In this section, we introduce the new Monte Carlo setup which solves the non-linear evolution equations of leading and subleading jets including the threshold resummed hard and jet functions. We start by presenting the Monte Carlo setup at LO/LL accuracy in section~\ref{sec:MC_LL} following Ref.~\cite{Dasgupta:2014yra}. For inclusive jets we compare the Monte Carlo result to an analytical solution of the DGLAP evolution equations in Mellin space where the contour integral of the inverse transformation is performed numerically. In section~\ref{sec:MC_LLp}, we then discuss in detail how the Monte Carlo code can be extended beyond LL accuracy by including the (threshold resummed) hard and jet functions which brings the perturbative accuracy of the shower to the same level as analytical calculations of inclusive jets.

\subsection{The parton shower at leading log~\label{sec:MC_LL}}

\begin{figure}[t]
\vspace*{.7cm}
\centering
\includegraphics[width=\textwidth]{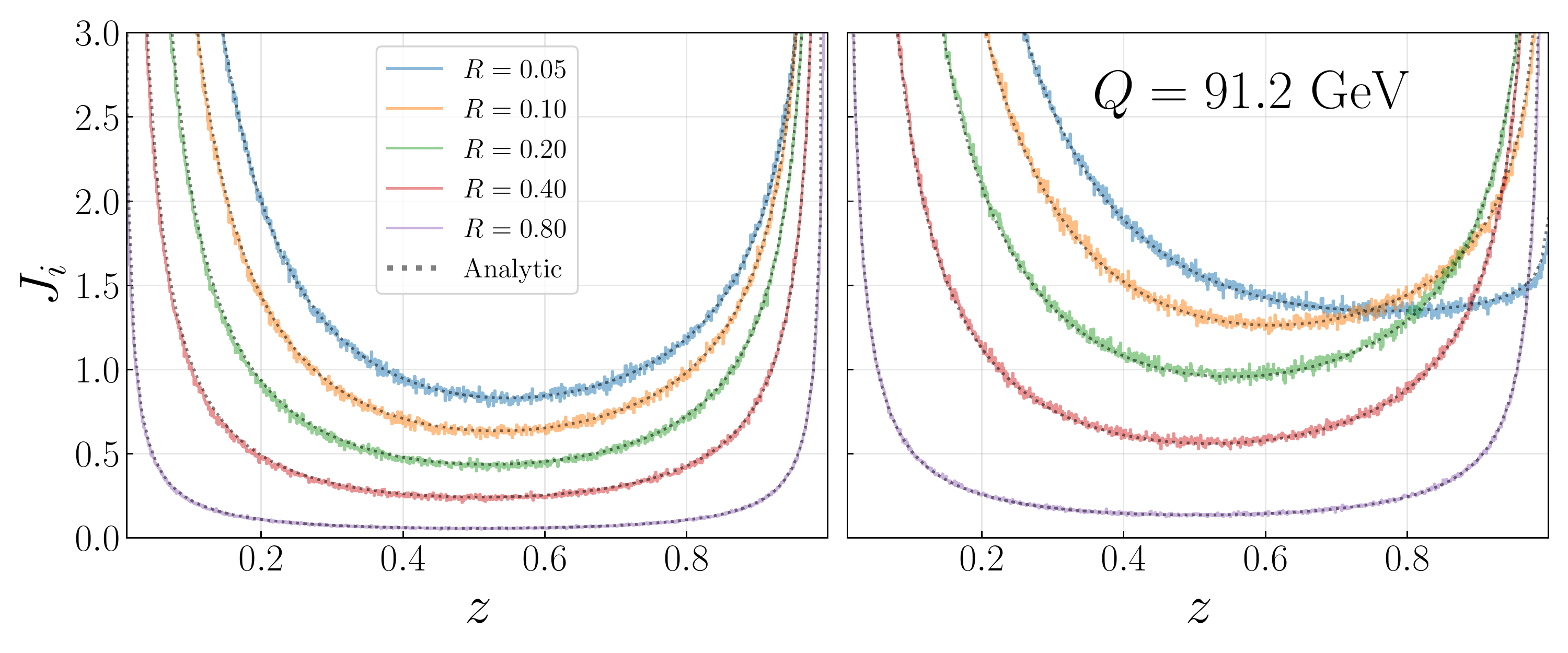}
\caption{LO/LL inclusive jet functions for quarks (left panel) and gluons (right panel). We show the results obtained analytically in Mellin space where the contour integral of the inverse transformation is performed numerically and the result from the parton shower. As an example, we choose $Q=100$~GeV and five exemplary values of the jet radius.~\label{fig:Comparison_MC_Mellin}}
\end{figure} 

At LO/LL order we follow the parton shower setup of Ref.~\cite{Dasgupta:2014yra}. See also Ref.~\cite{Lai:2020byl}. We have at any time a set of partons $S$ that represents the state of the shower:
\begin{align}
S=\{\{z_1,f_1\},\{z_2,f_2\},....,\{z_n,f_n\}\}\,.    
\end{align}
Here, $z_i$ is the energy fraction carried by the $i$-th parton, and $f_i=q,\bar{q},$ or  $g$ is its flavor. We choose a uniformly distributed random number $\rnd \in [0,1]$ and solve for a time step $\Delta t$ according to
\begin{equation}\label{eq:Deltat}
    \rnd=\exp\Bigg[-\Delta t \sum_{i\in S}\int\limits_\epsilon^{1-\epsilon}{\rm d}z\, P_{f_i}(z)\Bigg]\,.
\end{equation}
The $P_{f_i}$ denote the final state summed splitting functions for a quark to split to a quark and gluon ($f_i=q$ or $\bar{q}$) or a gluon to split to anything ($f_i=g$). Here $\epsilon$ is a small parameter which cuts off the integral at both endpoints. The sum in the exponent in Eq.~(\ref{eq:Deltat}) runs over all the generated partons in the event. We then advance the Monte Carlo time $t\to t+\Delta t$ and solve for ${\cal R}$ in
\begin{align}\label{eq:MCtime}
    t(Q,{\cal R})&= \int\limits_{Q}^{Q{\cal R}}\frac{{\rm d}t'}{t'}\frac{\alpha_s(t'e^{-K/\beta_0})}{\pi}\,,\\
    K&=\left(\frac{67}{9}-\frac{\pi^{2}}{3}\right) C_{A}-\frac{20}{9} T_{F} N_{f}\,.
\end{align}
That way the MC time and ${\cal R}=\tan(R/2)$ are ordered variables and the shower terminates when $t>t^{\rm max}=t(Q,R)$. For strict LO/LL comparisons, we take $K=0$, otherwise, $K$ is a factor which makes the threshold contribution consistent with the two-loop cusp, using the so-called CMW trick~\cite{Catani:1990rr}. The minimum angle ${\cal R}$ is set by the jet radius $R$, which is where the shower algorithm terminates. We then pick a parton to split with a probability given by its relative contribution to the decay rate of Eq.~\eqref{eq:Deltat}. The momentum fraction $z$ of the splitting of the chosen parton is sampled according to the appropriate DGLAP splitting functions, and in the case of a splitting gluon, the flavor is assigned based on the relative contribution of the final state particles to the gluon's phase space.

We now histogram the inclusive momentum fraction $z$ of the partons produced by the shower relative to the initial quark or gluon. The result is the LO evolved jet function. To establish the consistency of the parton shower result and the analytical Mellin space evolution of Ref.~\cite{Kang:2016mcy}, we compare the two results in Fig.~\ref{fig:Comparison_MC_Mellin}. As an example, we show the results for $Q=91.2$~GeV and five exemplary jet radii $R=0.05 - 0.8$. Here we use the LO expression for the running coupling constant and $\alpha_s(M_Z)=0.1187$. We observe that the two results agree exactly for the five jet radii and over the entire range of $z$. 

\subsection{Extension beyond leading-logarithmic accuracy~\label{sec:MC_LLp}}

In order to improve the accuracy of the shower, it is helpful to examine the factorization theorem for inclusive jet production. We write
\begin{align}\label{eq:incl_jets_resum}
\frac{{\rm d}\sigma}{{\rm d}z}&=\sum_{i,j}\int_{z}^{1}\frac{{\rm d}z'}{z'}\int_{z'}^{1}\frac{{\rm d}z''}{z''}J_j\Big(\frac{z}{z'},QR\Big)U_{ji}\Big(\frac{z'}{z''};QR,Q\Big)C_i(z'',Q) \,.   
\end{align}
Here $C_i$ represents a hard-scattering/coefficient function of a particular process. We can then describe how the parton shower generates $U$. First, we introduce:
\begin{itemize}
\item {\bf Notation:} Let $\lea f(S)\rea_R^i$ denote the shower average on \emph{partonic} states $S$ of some function $f$ on those states at the angular scale $R$, when the shower was initiated by the parton of flavor $i$. $S=\{\{z_1,f_1\},\{z_2,f_2\},....,\{z_n,f_n\}\}$ is the final state of the shower given above, which is the stochastic variable being averaged over.
\end{itemize}
Then the DGLAP evolution kernel $U(x;R;Q)$ to leading logarithmic accuracy, evolved from a hard scale $Q$ down to $QR$, is given by
  \begin{align}\label{eq:shower_to_DGLAP}
    U_{ij}(x;R) = \lea\delta_{f_k j}\delta\Big(x-z_k\Big)_{k\in S}\rea_{R}^{i}\,.
  \end{align}
  Where we have introduced the energy fraction $z_k$ of a parton in the event. In practical algorithmic terms, what Eq. \eqref{eq:shower_to_DGLAP} actually means is that we start with a histogram $H^{ij}_{x}$, indexed by the momentum fraction $x\in[0,1]$, and labeled by the  initial and final partons $i,j$. Each $x$ falls into some bin of size $\Delta(x)\ll 1$ (which can also depend on where we are in the distribution). Initialize $H_{x}=0$. Then:
  \begin{enumerate}
  \item Run shower to get the set of momenta $S$ at angular scale $R$.
  \item For each $k\in S$, add $\Delta^{-1}(x)$ to the bin in $H^{i f_k}_{x}$ at position $x$ where $x-\Delta(x)/2<z_k<x+\Delta(x)/2$.
  \end{enumerate}
  After a sufficient number of events, divide all bins in $H_x$ by the number of events. Then $H_{x}=U(x;R)$ in the limit $\Delta(x)\rightarrow 0$. Note that this effectively approximates the delta function in Eq. \eqref{eq:shower_to_DGLAP} by the step function:
  \begin{align}
    \delta(x-z)&\sim\frac{1}{\Delta(x)}\Theta\Big((x+\Delta(x)/2)-z\Big)\Theta\Big(z-(x-\Delta(x)/2)\Big) \,.
  \end{align}
Going back to the resummed factorization formula of Eq.~\eqref{eq:incl_jets_resum}, after a change of variables, we can combine the jet and coefficient function into a single function:
\begin{align}
CJ_{ji}(z,QR,Q)&=\int_{z}^{1}\frac{{\rm d}z'}{z'}J_j\Big(\frac{z}{z'},QR\Big)C_i(z',Q)\,,\\
\frac{{\rm d}\sigma}{{\rm d}z}&=\sum_{i,j}\int_{z}^{1}\frac{{\rm d}z'}{z'}CJ_{ji}\Big(z',QR,Q\Big)U_{ji}\Big(\frac{z}{z'},QR,Q\Big)\,.
\end{align}
Substituting in Eq. \eqref{eq:shower_to_DGLAP}, we then get
\begin{align}
\frac{{\rm d}\sigma}{{\rm d}z}&=\sum_{i,j}\int_{z}^{1}{\rm d}z'\, CJ_{ji}\Big(z',QR,Q\Big)\lea\delta_{f_k j}\delta\Big(z-z'\times z_k\Big)_{k\in S}\rea_{R}^{i}\,.
\end{align}
We wish to implement this last integral completely stochastically in the shower. We proceed as follows. First we introduce the cumulant function and its functional inverse
\begin{align}
  \mathcal{C\!J}_{ij}(x)&=\int_{x}^{1}{\rm d}z\, CJ_{ji}\Big(z,QR,Q\Big)\,,\\
  \mathcal{C\!J}^{-1}_{ij}\Big(\mathcal{C\!J}_{ij}(x)\Big)&=x\,.
\end{align}
Then we change variables
\begin{align}
  z &= \mathcal{C\!J}^{-1}_{ij}(\sigma)\,,\\
  {\rm d}z &= \frac{{\rm d}\sigma}{CJ_{ij}(\mathcal{C\!J}^{-1}(\sigma))}\,,\\
\frac{{\rm d}\sigma}{{\rm d}z}&=\sum_{i,j}\int_{\sigma_{min}}^{\sigma_{max}}{\rm d}\sigma\lea\delta_{f_k j}\delta\Big(z-z_k\times \mathcal{C\!J}^{-1}_{ij}(\sigma)\Big)_{k\in S}\rea_{R}^{i}\,.
\end{align}
If we can normalize $CJ(x)$ on some interval, so $0<\mathcal{C\!J}(x)<1$, then we can take $\sigma$ to be a random number between 0 and 1, and thus perform this final integral via a Monte Carlo sampling\footnote{We note that for the NLO coefficient and jet functions in the inclusive case, the function $\mathcal{CJ}(x)$ is not normalizable, due to soft singularities as $x\rightarrow 0$. However, we avoid this issue by only using the threshold resummed expressions for these functions, which is appropriate for leading jet production.}
\begin{align}
  \frac{{\rm d}\sigma}{{\rm d}z}&\propto\sum_{i,j}\Bigg[\lea\delta_{f_k j}\delta\Big(z-z_k\times \mathcal{C\!J}^{-1}_{ij}(\rnd)\Big)_{k\in S}\rea_{R}^i\Bigg]\,.
\end{align}
Here $[\cdot]$ is the final averaging process over the random variable $\rnd\in[0,1]$ uniformly distributed on the unit interval. The $\propto$ arises from normalization issues. Given a coefficient and jet function, we can compute the stochastic averaging as follows: we start with a histogram $H^{ij}_{x}$, indexed by the momentum fraction $x\in[0,1]$, and labeled by the initial and final parton flavors $i,j$. Each $x$ falls into some bin of size $\Delta(x)\ll 1$. Initialize $H_{x}^{ij}=0$. Then:
  \begin{enumerate}
  \item Run shower starting with initial parton $i$ to get the set of momenta $S$ at angular scale $R$.
  \item For each $k\in S$, draw a random number uniformly between $0$ and $1$, {\bf rnd}, and calculate $\mathcal{C\!J}^{-1}_{if_k}(\text{\bf rnd})$ (recall that $f_k$ is the flavor of the $k$-th parton in the shower), then add $\Delta^{-1}(x)$ to the bin in $H^{if_k}_{x}$ at position $x$ where $x-\Delta(x)/2 < z_k\times\mathcal{C\!J}^{-1}_{if_k}(\rnd) <x+\Delta(x)/2$.    
  \end{enumerate}
After a sufficient number of events, divide all bins in $H_x^{ij}$ by the number of events.

We can then improve the accuracy of the shower by using instead of the leading order coefficient and jet functions their threshold resummed expressions, as described in the next section. The algorithm for leading jet production with threshold corrections is then as follows:
  \begin{enumerate}
  \item Run shower starting with initial parton $i$ to get the set of momenta $S$ at angular scale $R$.
  \item For each $k\in S$, draw a random number uniformly between $0$ and $1$, {\bf rnd}, and set $x_k=z_k\times \mathcal{C\!J}^{{\rm thr}-1}_{if_k}(\text{{\bf rnd}})$ (recall that $f_k$ is the flavor of the $k$-th parton in the shower). This gives a set of jet momenta $\{x_1,...,x_n\}$. Then add $1$ to the bin in $H^{if_k}_{x}$ at position $x$ where $x-\Delta(x)/2 < \text{max}\{x_1,...,x_n\} <x+\Delta(x)/2$.    
  \end{enumerate}
After a sufficient number of events, divide all bins in $H_x^{ij}$ by the number of events. We note that the inverse function $\mathcal{CJ}^{\rm thr}$ is calculated from the threshold resummation expressions for $C$ and $J$. 

Finally, we note that in principle the NLO DGLAP evolution kernels can be included in a parton shower following Ref.~\cite{Hoche:2017hno}. We leave a combination of the NLO DGLAP evolution and the NLO/threshold resummed hard and jet functions for future work.


\section{Threshold resummation for leading jet observables~\label{sec:4}}

\begin{figure}[t]
\vspace*{.7cm}
\centering
\includegraphics[width=\textwidth]{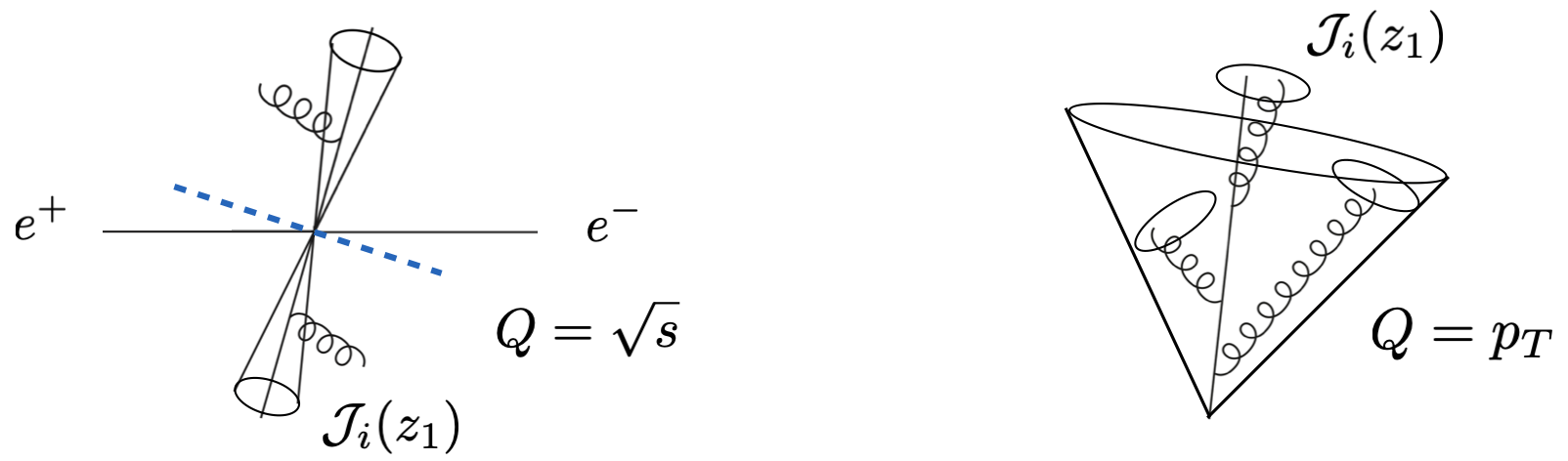}
\caption{Illustration of $e^+e^-$ hemisphere leading jets (left) and leading subjets (right).~\label{fig:hemisphere+subjets}}
\end{figure}

For leading jets the threshold region $z\to 1$~\cite{Sterman:1986aj,Catani:1989ne} is phenomenologically important. Therefore, we need to include the corresponding double logarithmic corrections to all orders at NLL$'$ accuracy.  We note that the jet at threshold is the leading jet -- up to power corrections, and so we do not need to consider the difference between the inclusive and leading jets when performing the resummation. We consider first the threshold resummation for $e^+e^-$ hemisphere jets which will be the standard reference process for the subsequent section. Second, we consider the threshold resummation for subjets in proton-proton collisions. The two processes are illustrated in Fig.~\ref{fig:hemisphere+subjets} and the corresponding threshold resummation is discussed in sections~\ref{sec:4.1} and~\ref{sec:4.2}. Different than event-wide leading jets in proton-proton or $e^+e^-$ collisions, for both processes considered here we only have one initial parton at LO/LL accuracy. Different than for fixed order calculations of leading jets, as discussed in the previous section, the structure of the factorization does not change order-by-order in the threshold region. By including the threshold resummed hard and jet functions in the Monte Carlo approach discussed above, we can obtain numerical results for leading jets. We discuss nonperturbative effects in section~\ref{sec:4.3} which can be included in the threshold region by convolving the perturbative spectrum with a shape function. We then present numerical results for both processes in section~\ref{sec:4.4}. Parton showers should naturally, at least to leading logarithmic accuracy, resum final state threshold logarithms, such as treated in this paper. We also note that in Refs.~\cite{Nagy:2016pwq,Nagy:2017ggp} initial state threshold resummation was also included in the Deductor parton shower.  

\subsection{$e^+e^-$ hemisphere jets~\label{sec:4.1}}

\begin{figure}[t]
\vspace*{.7cm}
\centering
\includegraphics[width=.9\textwidth]{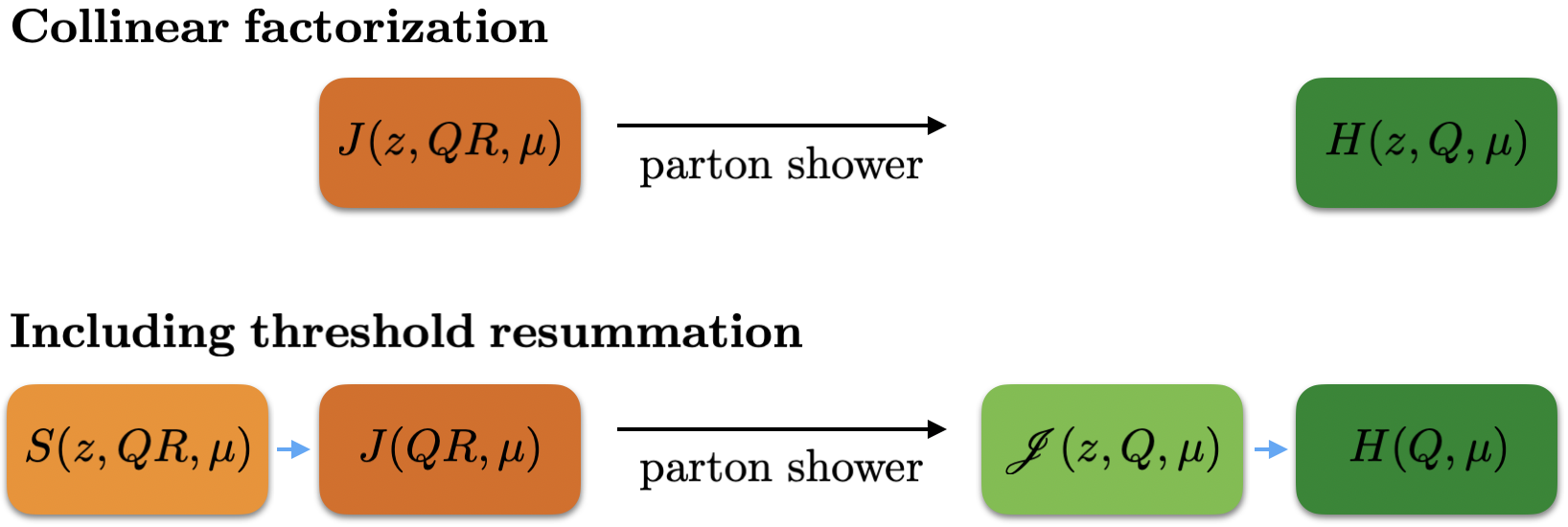}
\vspace*{.3cm}
\caption{Comparison of collinear factorization (left) and when threshold resummation is included (right). The evolution indicated by the blue arrows is carried out analytically and in both cases the parton shower resums large logarithms between the scales $QR\to Q$. Functions with the same colored box have the same characteristic scales.~\label{fig:collinear_threshold}}
\end{figure}

For $e^+e^-$ hemisphere jets\footnote{The hemispheres in $e^+e^-$ collisions can be obtained with the exclusive k$_T$ algorithm~\cite{Ellis:1993tq} or the thrust axis~\cite{Farhi:1977sg}.} there is only one initial parton at LO which is different than for $e^+e^-$ event-wide leading jets which will be discussed in section~\ref{sec:epemeventleading} below. Note that for inclusive jets this distinction is not relevant at the level of the factorization theorem since we sum over all possible jets in the final state. Per-event and per-hemisphere inclusive jets only differ in terms of their normalization. We consider the cross section ${\rm d}\sigma/{\rm d}z$ where additional angular dependencies are integrated out and $z=2E/Q$ is the hemisphere jet energy relative to half of the CM energy $Q=\sqrt{s}$. The collinear factorization using $\ln R$ resummed jet functions is analogous to the results discussed above and it's form depends on the perturbative accuracy. At LO/LL accuracy we have
\begin{equation}\label{eq:factorization_hemisphere}
    \frac{{\rm d}\sigma_{e^+e^-\to{\rm jet}_1+X}}{{\rm d}z_1} = \sum_i H_i(Q,z_1,\mu)\otimes {\cal J}_i(z_1,QR,\mu) \,.
\end{equation}
Therefore, this process is ideally suited to study jet energy loss since all three criteria listed in the Introduction are satisfied. At LL accuracy, we find a direct connection between the leading jet cross section and the average parton energy loss
\begin{equation}\label{eq:avg_hemisphere}
    \int_0^1{\rm d}z_1\,z_1\,\frac{1}{\sigma_{\rm tot}}\frac{{\rm d}\sigma_{e^+e^-\to{\rm jet}_1+X}}{{\rm d}z_1} = \langle z_1\rangle\,,
\end{equation}
see Eq.~(\ref{eq:sum4}). Here the average energy loss is a function of the energy scale $Q$ and the jet radius $R$. At higher perturbative accuracy, additional jet functions need to be introduced in Eq.~(\ref{eq:factorization_hemisphere}). However, in the threshold limit $z\to 1$, where additional emissions are soft, we can obtain a closed form of the factorization formula. We derive the threshold resummation in Mellin transform space and then perform the inverse transformation analytically. Throughout this work we adopt the following convention for the Mellin transform and it's inverse
\begin{align}
    f(N) &=\int_{0}^{1} \mathrm{d} z\, z^{N-1} f(z) \,, \\
    f(z) &=\int_{\mathcal{C}_{N}} \frac{\mathrm{d} N}{2 \pi i} z^{-N} f(N)\,.
\end{align}
In this section we derive the threshold resummed cross section for inclusive $e^+e^-$ hemisphere jets. The leading jet cross section is then obtained by including the threshold resummed hard and jet functions in the parton shower as discussed in the previous section. In the threshold region, the cross section can be refactorized as
\begin{equation}\label{eq:refactorization_epem}
    \frac{{\rm d}\sigma_{e^+e^-\to{\rm jet}+X}}{{\rm d}z} = \sum_i H_i(Q,\mu)\times \mathscr{J}_i(z,Q,\mu)\,\otimes\, S_i(z,Q R,\mu)\, \otimes\, S_i^{{\rm NGL}}(z) \times J_i(Q R,\mu) \,.
\end{equation}
Here $H_i$ is the hard function~\cite{Altarelli:1979kv}, $\mathscr{J}_i$ accounts for the recoiling soft radiation~\cite{Becher:2006qw,Becher:2010pd}, ${\cal S}_i$ is a soft-collinear function~\cite{Chien:2015cka,Becher:2015hka} taking into account soft radiation in the direction of the observed jet~\cite{Dai:2017dpc,Liu:2017pbb} and $J_i$ takes into account unmeasured collinear radiation inside the observed jet~\cite{Ellis:2010rwa,Liu:2021xzi}. In addition, $S_i^{\rm NGL}$ accounts for non-global logarithms (NGLs) due to correlations between the in- and out-of-jet region~\cite{Dasgupta:2001sh}. For our numerical results presented below we include the fit to the Monte Carlo algorithm of Ref.~\cite{Dasgupta:2001sh}. To the accuracy we are working at, the NGL factor can be included multiplicatively. For completeness, we summarize the NLO order expressions of the relevant functions in the Appendix~\ref{app:NLO_expressions}. For phenomenological applications at LEP, the sum in Eq.~(\ref{eq:refactorization_epem}) runs over quark and anti-quark channels. For our numerical results presented below we also consider leading jets initiated by gluons. In $e^+e^-$ collisions they can be obtained from a hard-scattering process with an intermediate scalar $\phi\to gg$, where $\phi$ could be an intermediate Higgs boson. The natural scales which eliminate the large logarithms of the different functions at fixed order and set the initial scale of their RG evolution are given by
\begin{equation}\label{eq:scales}
    \mu_H\sim Q\,,\quad \mu_{\mathscr{J}}\sim (1-z)^{1/2}Q\,,\quad \mu_S\sim (1-z)QR\,,\quad \mu_J\sim QR \,.
\end{equation}
We note that both within collinear factorization, Eq.~(\ref{eq:factorization_hemisphere}), and when threshold resummation is included as in Eq.~(\ref{eq:refactorization_epem}), the parton shower resums logarithms between the scales $QR\to Q$ which is illustrated in Fig.~\ref{fig:collinear_threshold}. The additional resummation of the threshold logarithms is carried out analytically as described in this section and included in the parton shower algorithm.
As an example, we consider the RG equation of the soft-collinear function $S_i$. In Mellin space the evolution equation is multiplicative and can be written as
\begin{equation}\label{eq:evoleq}
    \mu \frac{\mathrm{d}}{\mathrm{d} \mu} S_i\left(Q R /N, \mu\right)=\gamma_i^{S}\left(Q R / N, \mu\right) S_i\left(Q R / N, \mu\right) \,.
\end{equation}
The anomalous dimension is given by
\begin{equation}
    \gamma_i^{S}\left(Q R / N, \mu\right)= -\frac{\alpha_{s}}{\pi} C_i \ln \left(\frac{\mu^2 \bar{N}^2}{Q^2 R^2}\right) \,,
\end{equation}
with $\bar N=Ne^{\gamma_E}$. In general, $\gamma_i^S$ has the structure
\begin{equation}
    \gamma_i^{S}\left(Q R / N, \mu\right)= -\Gamma_i\left(\alpha_{s}\right) \ln \left(\frac{\mu^2 \bar{N}^2}{Q^2 R^2}\right)+\gamma_i^S\left(\alpha_{s}\right) \,.
\end{equation}
The cusp $\Gamma_i$ and non-cusp $\gamma_i$ anomalous dimensions allow for the following perturbative expansion
\begin{equation}
\Gamma_{i}\left(\alpha_{s}\right)=\sum_{n=0}^{\infty} \Gamma_{i}^{n}\left(\frac{\alpha_{s}}{4 \pi}\right)^{n+1} \,,\quad \gamma_i(\alpha_s)=\sum_{n=0}^{\infty}\gamma_i^n\left(\frac{\alpha_s}{4\pi}\right)^{n+1} \,.
\end{equation}
For the soft-collinear function $S_i$ the non-cusp term vanishes to NLL, $\gamma_i^{S,0}=0$. The relevant terms of the cusp anomalous dimension at NLL are given by
\begin{equation}
\Gamma_{i}^{0}=4 C_{i}\,, \quad
\Gamma_{i}^{1}=4 C_{i}\left[\left(\frac{67}{9}-\frac{\pi^{2}}{3}\right) C_{A}-\frac{20}{9} T_{F} N_{f}\right]\,.
\end{equation}
Solving the evolution equation in Eq.~(\ref{eq:evoleq}), we find
\begin{equation}\label{eq:S_evolved}
    S_i\left(Q R / N, \mu\right)=S_i\left(Q R / N, \mu_S\right) \exp \left[\int_{\mu_S}^{\mu} {\rm d}\ln \mu^{\prime} \,\gamma_i^{S}\left(Q R / N, \mu^{\prime}\right)\right] \,.
\end{equation}
Instead of performing the inverse transformation numerically using for example the minimal prescription of Ref.~\cite{Catani:1996yz}, we follow here the method introduced in Ref.~\cite{Becher:2006nr,Becher:2007ty} and perform the inverse transformation analytically. Following the notation of Ref.~\cite{Stewart:2010qs}, we introduce the functions $K_i$, $\eta_i$ and $\eta_{i,\gamma}$ as
\begin{align}
    K_{i}\left(\mu, \mu_{0}\right) &=\int_{\alpha_{s}\left(\mu_{0}\right)}^{\alpha_{s}(\mu)} \frac{\mathrm{d} \alpha}{\beta(\alpha)} \Gamma_{i}(\alpha) \int_{\alpha_{s}\left(\mu_{0}\right)}^{\alpha} \frac{\mathrm{d} \alpha^{\prime}}{\beta\left(\alpha^{\prime}\right)} \,, \\
    \eta_{i}\left(\mu, \mu_{0}\right) &=\int_{\alpha_{s}\left(\mu_{0}\right)}^{\alpha_{s}(\mu)} \frac{\mathrm{d} \alpha}{\beta(\alpha)} \Gamma_{i}(\alpha) \,,
    \\
    \eta_{i,\gamma}\left(\mu, \mu_{0}\right) &=\int_{\alpha_{s}\left(\mu_{0}\right)}^{\alpha_{s}(\mu)} \frac{\mathrm{d} \alpha}{\beta(\alpha)} \gamma_{i}(\alpha)\,. 
\end{align}
Evaluating these expressions at NLL, we find
\begin{align}
    K_i\left(\mu_{0}, \mu\right)&=-\frac{\Gamma_i^{0}}{4 \beta_{0}^{2}}\left[\frac{4 \pi}{\alpha_{s}\left(\mu_{0}\right)}\left(1-\frac{1}{r}-\ln r\right)+\left(\frac{\Gamma_i^{1}}{\Gamma_i^{0}}-\frac{\beta_{1}}{\beta_{0}}\right)(1-r+\ln r)+\frac{\beta_{1}}{2 \beta_{0}} \ln ^{2} r\right] \,,
    \\
    \eta_i\left(\mu_{0}, \mu\right)&=-\frac{\Gamma_i^{0}}{2 \beta_{0}}\left[\ln r+\frac{\alpha_{s}\left(\mu_{0}\right)}{4 \pi}\left(\frac{\Gamma_i^{1}}{\Gamma_i^{0}}-\frac{\beta_{1}}{\beta_{0}}\right)(r-1)\right] \,,
\end{align}
with $r=\alpha_s(\mu)/\alpha_s(\mu_0)$ and similarly for $\eta_{i,\gamma}$. The function $\eta_{i,\gamma}$ vanishes for the soft function $S_i$ at NLL. We can then write the evolved soft function $S_i$ in Eq.~(\ref{eq:S_evolved}) in Mellin space as
\begin{equation}\label{eq:soft_Mellin}
    S_i\left(Q R / N, \mu\right)=S_i\left(Q R / N, \mu_S\right) e^{-2K_i(\mu,\mu_S)} \bigg(\frac{\mu_S\bar N}{Q R}\bigg)^{-2\eta_i(\mu,\mu_S)}
\end{equation}
Next, we consider the Mellin inverse transformation back to $z$-space. Note that here we did not make a scale choice for $\mu_S$. Therefore, the $N$-dependence appears both in the factor $N^{-2\eta_i}$ and also the initial condition of the evolution $S_i\left(Q R / N, \mu_S\right)$ which we need to take into account at NLL$'$ accuracy. The initial condition of the evolved soft function $S_i$ is a polynomial of $L=\ln(\mu_S^2 \bar N^2/(Q^2R^2))$. With
\begin{equation}
    L^m \bigg(\frac{\mu_S\bar N}{Q R}\bigg)^{-2\eta(\mu,\mu_S)}  \; = \; (-1)^m\partial^m_\eta \bigg(\frac{\mu_S\bar N}{Q R}\bigg)^{-2\eta(\mu,\mu_S)} \,,
\end{equation}
we can then write the initial condition of the soft function in Eq.~(\ref{eq:soft_Mellin}) as 
\begin{equation}
    S_i(L,\mu_S)\; \rightarrow\; S_i(-\partial_\eta,\mu_S) \,.
\end{equation}
With this replacement, the only remaining dependence on the Mellin variable $N$ is the factor $N^{-2\eta}$. We calculate the following Mellin transformation
\begin{equation}
    \int_0^1{\rm d}z\, z^{N-1}(1-z)^{-1+2\eta}\frac{1}{\Gamma[2\eta]}=\frac{\Gamma[N]}{\Gamma[N+2\eta]} = N^{-2\eta} + {\cal O}\bigg(\frac{1}{N}\bigg) \,.
\end{equation}
In the threshold limit we can drop terms which are power suppressed as ${\cal O}(1/N)$. With this result we can now write the NLL resummed soft function in $z$-space as
\begin{equation}\label{eq:NLL1}
    S_i(z,QR,\mu)=e^{-2K_i(\mu,\mu_S)} \frac{(1-z)^{-1+2\eta_i(\mu,\mu_S)}}{\Gamma[2\eta_i(\mu,\mu_S)]} \bigg(\frac{\mu_S e^{\gamma_E}}{Q R}\bigg)^{-2\eta_i(\mu,\mu_S)} \,.
\end{equation}
In order to implement the threshold resummed jet function in the Monte Carlo code discussed in section~\ref{sec:3} above, we need the cumulant instead of the differential result. We take the cumulant to be the integral of Eq.~(\ref{eq:NLL1}) from $z$ to 1. We find
\begin{equation}\label{eq:NLL2}
    S_i^c(z,QR,\mu)=e^{-2K_i(\mu,\mu_S)} \frac{(1-z)^{2\eta_i(\mu,\mu_S)}}{\Gamma[1+2\eta_i(\mu,\mu_S)]} \bigg(\frac{\mu_S e^{\gamma_E}}{Q R}\bigg)^{-2\eta_i(\mu,\mu_S)} \,.
\end{equation}
Here the superscript $c$ indicates that $S_i^c$ is the cumulative result of the soft function. Next, we extend our result to NLL$'$ accuracy where we need to include the initial condition $S_i(-\partial_\eta,\mu_S)$ of the evolved soft function at ${\cal O}(\alpha_s)$. Taking the first and second order derivatives of Eq.~(\ref{eq:NLL2}) with respect to $\eta$, we find
\begin{align}
    -\partial_\eta &\to O_1=2\ln\bigg(\frac{(1-z)Q R}{\mu_S e^{\gamma_E}}\bigg)-2\psi_0(1+2\eta(\mu,\mu_S))\,, \nonumber\\
    (-)^2\partial_\eta^2& \to O_2= O_1^2-4\psi_1(1+2\eta(\mu,\mu_S)) \,.
\end{align}
Substituting the derivatives $(-1)^m\partial_\eta^m$ for the operators $O_m$ in $S_i(-\partial_\eta,\mu_S)$, we find the following expression for the resummed cumulant of the soft function at NLL$'$ accuracy
\begin{align}\label{eq:final}
    S_i^c\left(z,Q R,\mu\right)=&\, \bigg[1-\frac{\alpha_{s}(\mu_S)}{4 \pi} C_i\bigg[\bigg(2\ln\bigg(\frac{(1-z)Q R}{\mu_S e^{\gamma_E}}\bigg)-2\psi_0(1+2\eta_i(\mu,\mu_S))\bigg)^2
    \nonumber\\
    &-4\psi_1(1+2\eta_i(\mu,\mu_S))+\frac{\pi^{2}}{2}\bigg]\bigg]   
    \nonumber \\
    &\times e^{-2K_i(\mu,\mu_S)} \frac{(1-z)^{2\eta_i(\mu,\mu_S)}}{\Gamma[1+2\eta_i(\mu,\mu_S)]} \bigg(\frac{\mu_S e^{\gamma_E}}{Q R}\bigg)^{-2\eta_i(\mu,\mu_S)} \,.
\end{align}
Lastly, we need to calculate the convolution of the resummed soft function and the jet function $\mathscr{J}_i$. At NLL, we find
\begin{align}
    \mathscr{J}^c_i\otimes S_i^c(z,Q,R,\mu_\mathscr{J},\mu_S,\mu)=&\; e^{2K_i(\mu,\mu_{\mathscr{J}})+\eta_{i,\gamma}^{\mathscr{J}}(\mu,\mu_{\mathscr{J}})-2K_i(\mu,\mu_S)}
    \nonumber\\ &\times 
    \frac{(1-z)^{2\eta_i(\mu,\mu_S)-2\eta_i(\mu,\mu_{\mathscr{J}})}}{\Gamma[1-2\eta_i(\mu,\mu_{\mathscr{J}})+2\eta_i(\mu,\mu_{S})]} 
    \nonumber\\ &\times 
    \bigg(\frac{\mu_{\mathscr{J}} e^{\gamma_E}}{Q}\bigg)^{2\eta_i(\mu,\mu_{\mathscr{J}})} \bigg(\frac{\mu_{S} e^{\gamma_E}}{Q R}\bigg)^{-2\eta_i(\mu,\mu_{S})} \,.
\end{align}
The relevant anomalous dimensions can be found in the Appendix~\ref{app:NLO_expressions}. The extension to NLL$'$ can be obtained in analogy to Eq.~(\ref{eq:final}) above. Besides the NGLs, the remaining functions in the refactorized expression in Eq.~(\ref{eq:refactorization_epem}) are independent of $z$ and satisfy multiplicative evolution equations. Their anomalous dimensions are also listed in the Appendix~\ref{app:NLO_expressions}.

\subsection{Subjets~\label{sec:4.2}}

In proton-proton and heavy-ion collisions we do not have access to anx event-wide reference scale like in $e^+e^-$ collisions to define the energy loss of the leading jet. However, we can construct a reference scale by first identifying an inclusive jet sample with jet radius $R$. The transverse momentum of the identified jets can be used a reference scale. We then recluster the particles inside a given jet with a jet radius $r<R$ and measure the momentum of the identified leading subjet $p_T^r$ relative to the initial jet $z_r=p_{T}^{r}/p_{T}$. A calculation of the inclusive subjet distribution within collinear factorization was discussed in~\cite{Dai:2016hzf,Kang:2017mda}. See also Refs.~\cite{Apolinario:2017qay,KunnawalkamElayavalli:2020pir}. We briefly review the factorization for inclusive subjets within collinear factorization and we then extend it to include the resummation of threshold logarithms which can also be implemented in the parton shower framework introduced above. We consider the cross section
\begin{equation}\label{eq:subjets1}
\frac{{\rm d}\sigma_{pp\to {\rm j}_R ({\rm j}_r)+X}}{{\rm d}p_T\, {\rm d}\eta\, {\rm d}z_r} = \sum_{ijk}f_i(x_i,\mu)\otimes f_j(x_j,\mu)\otimes H_{ijk}(x_i,x_j,p_T/z,\eta,\mu)\otimes G_k(z,z_r,p_TR,r,\mu)\,,
\end{equation}
where $p_T$ and $\eta$ is the transverse momentum rapidity of the inclusive jet with radius $R$. See also Eq.~(\ref{eq:inclusivefactorization1}). Up to power corrections ${\cal O}(R^2)$, the cross section can be factorized in terms of parton distribution functions $f_{i,j}$, hard-scattering functions $H_{ijk}$ and a jet function $G_k$ (typically denoted by ${\cal G}_k$ e.g. in Ref.~\cite{Kang:2017mda}) which depends on the jet substructure observable $z_r$. The symbols $\otimes$ denote appropriate integrals over the involved longitudinal momentum fractions $x_{i,j}$ and the fraction $z$ contained in the observed jet with radius $R$. We change the index of the jet function to $G_i$ from here on. At NLO, the jet function for a quark is given by
\begin{align}\label{eq:G1}
G_{q}\left(z, z_{r}, p_T R,r, \mu\right)=&\, \delta(1-z) \delta\left(1-z_{r}\right)+\frac{\alpha_{s}}{2 \pi}\left\{\delta\left(1-z_{r}\right) \ln\bigg(\frac{\mu^2}{p_T^2 R^2}\bigg)\left[P_{q q}(z)+P_{g q}(z)\right]\right.
\nn\\ &\hspace*{-1.5cm}
+\delta(1-z) \ln\bigg(\frac{R^2}{r^2}\bigg)\left[P_{q q}\left(z_{r}\right)+P_{g q}\left(z_{r}\right)\right]+C_{F} \delta\left(1-z_{r}\right)\left[\delta(1-z)\left(\frac{13}{2}-\frac{2 \pi^{2}}{3}\right)\right.
\nn\\ & \hspace*{-1.5cm}
\left.\left.-2\left(1+z^{2}\right)\left(\frac{\ln (1-z)}{1-z}\right)_{+}-2 \ln (1-z) \frac{1+(1-z)^{2}}{z}-1\right]\right\} \,. \end{align}
and similarly for a gluon, see Ref.~\cite{Kang:2017mda}. Note that here $z$ denotes the momentum fraction contained in the inclusive jet with radius $R$ relative to the initial parton and $z_r$ the fraction of the observed jet contained in the subjet with radius $r$. The logarithm $\ln(\mu^2/p_T^2 R^2)$ in the first line is resummed through DGLAP evolution which is satisfied by the entire jet function $G_i$. In order to resum the logarithm $\ln(R^2/r^2)$ in the second line of Eq.~(\ref{eq:G1}), the jet function $G_i$ can be further factorized in terms of a hard matching coefficient and a jet function for the subjet $\sim \tilde H_{ij}\otimes J_j$ which was carried out in~\cite{Kang:2017mda}. Here we extend this calculation and include also the resummation of threshold logarithms which is phenomenologically important similar to the $e^+e^-$ hemisphere jets discussed above. At threshold $z_r\to 1$, the jet function $G_i$ section can be refactorized as
\begin{align}\label{eq:refactorization_subjets}
    G_i(z,z_r,p_TR,r,\mu) & = \sum_j H_{ij}(z,p_T R,\mu)\times {\cal S}_j(z_r,p_TR,\mu)\otimes S_j(z_r,p_T r,\mu) 
    \nonumber \\
    &\quad \otimes S_j^{{\rm NGL},R}(z_r) \otimes S_j^{{\rm NGL},r}(z_r) \times J_j(p_T r,\mu) \,,
\end{align}
which allows for the joint resummation of threshold logarithms and $\ln(R^2/r^2)$ similar to Eq.~(\ref{eq:refactorization_epem}) above. The hard functions ${H}_{ij}$ can be combined with the remaining functions in Eq.~(\ref{eq:subjets1}) which altogether can be considered effective quark/gluon fractions. Here ${\cal S}_j$ is a collinear-soft function -- the same as for hadon-in-jet production at threshold which was discussed at NLL$'$ accuracy in Ref.~\cite{Kaufmann:2019ksh}. See also Ref.~\cite{Procura:2011aq}. The soft-collinear function $S_j$ and the jet function $J_j$ are the same as for $e^+e^-$ hemisphere jets in Eq.~(\ref{eq:refactorization_epem}) except that they are evaluated at the subjet radius $r$. We note that the NLO expressions of the soft-collinear and collinear-soft funtion are the same up to an overall minus sign (and the different jet radii). For the subjet refactorization at threshold, we find two types of NGLs. They arise independently at the boundary of the initial jet due to a correlation of the functions ${H}_{ij}$, ${\cal S}_j$ and at the boundary of the subjet due to the correlation of the functions $S_j$, $J_j$. The fixed order expressions of all relevant functions and their anomalous dimensions can be found in the Appendix~\ref{app:NLO_expressions}. The natural scales of the different functions in the factorization at threshold in Eq.~(\ref{eq:refactorization_subjets}) are given by
\begin{equation}
    \mu_{H}\sim p_T R\,,\quad \mu_{\cal S}\sim (1-z_r)p_T R\,,\quad \mu_S\sim (1-z_r)p_Tr\,,\quad \mu_J\sim p_Tr \,.
\end{equation}
The solution of the evolution equations can be obtained in analogy to the calculation outlined for $e^+e^-$ hemisphere jets above. Similar to leading $e^+e^-$ hemisphere jets, Eq.~(\ref{eq:avg_hemisphere}), we can relate the first moment (second Mellin moment) of the leading subjet cross section to the average partonic energy loss
\begin{equation}\label{eq:avg_subjets}
    \int_0^1{\rm d}z_1\,z_1\,\frac{1}{\sigma_{\rm tot}}\frac{{\rm d}\sigma^{(0)}_{pp\to {\rm j}_R ({\rm j}_r)+X}}{{\rm d}p_T\, {\rm d}\eta\, {\rm d}z_1} = f_q\langle z_{1q}\rangle+f_g\langle z_{1g}\rangle\,.
\end{equation}
Here the result is weighted by appropriate quark/gluon fractions $f_{q,g}$ and $\sigma_{\rm tot}$ is the inclusive jet cross section (without the substructure measurement).

\subsection{Nonperturbative effects~\label{sec:4.3}}

Finally, we must deal with possible nonperturbative effects in the shower. To regulate the Landau pole, we shift the argument of the running coupling as:
\begin{align}
\alpha_s(\mu)\rightarrow\alpha_s\Big(\sqrt{\mu^2+m_{\rm reg}^2}\Big)\,.
\end{align}
We take $m_{\rm reg}\sim 500$~MeV. To model the nonperturbative contribution to the cross section, we focus on the case of leading jets in $e^+e^-$. Then we replace the soft function in Eq. \eqref{eq:refactorization_epem} as
\begin{align}
S^c_i(z,QR,\mu)&\rightarrow \int_{z}^{1}\frac{{\rm d}z'}{z'}S^c_i(z',QR,\mu)\Big|_{\rm pert}\mathcal{S}\Big(\frac{z}{z'},d\Big)\,,\\
    \mathcal{S}(z,d)&=\frac{(1-z)\text{exp}\Big(-\frac{1}{d}(1-z)\Big)}{d(d-(1+d)e^{-\frac{1}{d}})}\,,\qquad
    d\sim\frac{\Lambda}{QR}\,.
\end{align}
The function $S^c_i(z,QR,\mu)\Big|_{\rm pert}$ explicitly refers to the perturbative resummed result of Eq.~\eqref{eq:final}. Here $\mathcal{S}$ is a shape function that will shift the spectrum near $z=1$, similar to what happens in the case of event-shapes \cite{Dokshitzer:1995zt,Korchemsky:1999kt,Lee:2006nr}. The $1/R$ scaling of the nonperturbative correction was first identified in Ref. \cite{Dasgupta:2007wa}.

\subsection{Numerical results~\label{sec:4.4}}

\begin{figure}[t]
\vspace*{.7cm}
\centering
\includegraphics[width=0.6\textwidth]{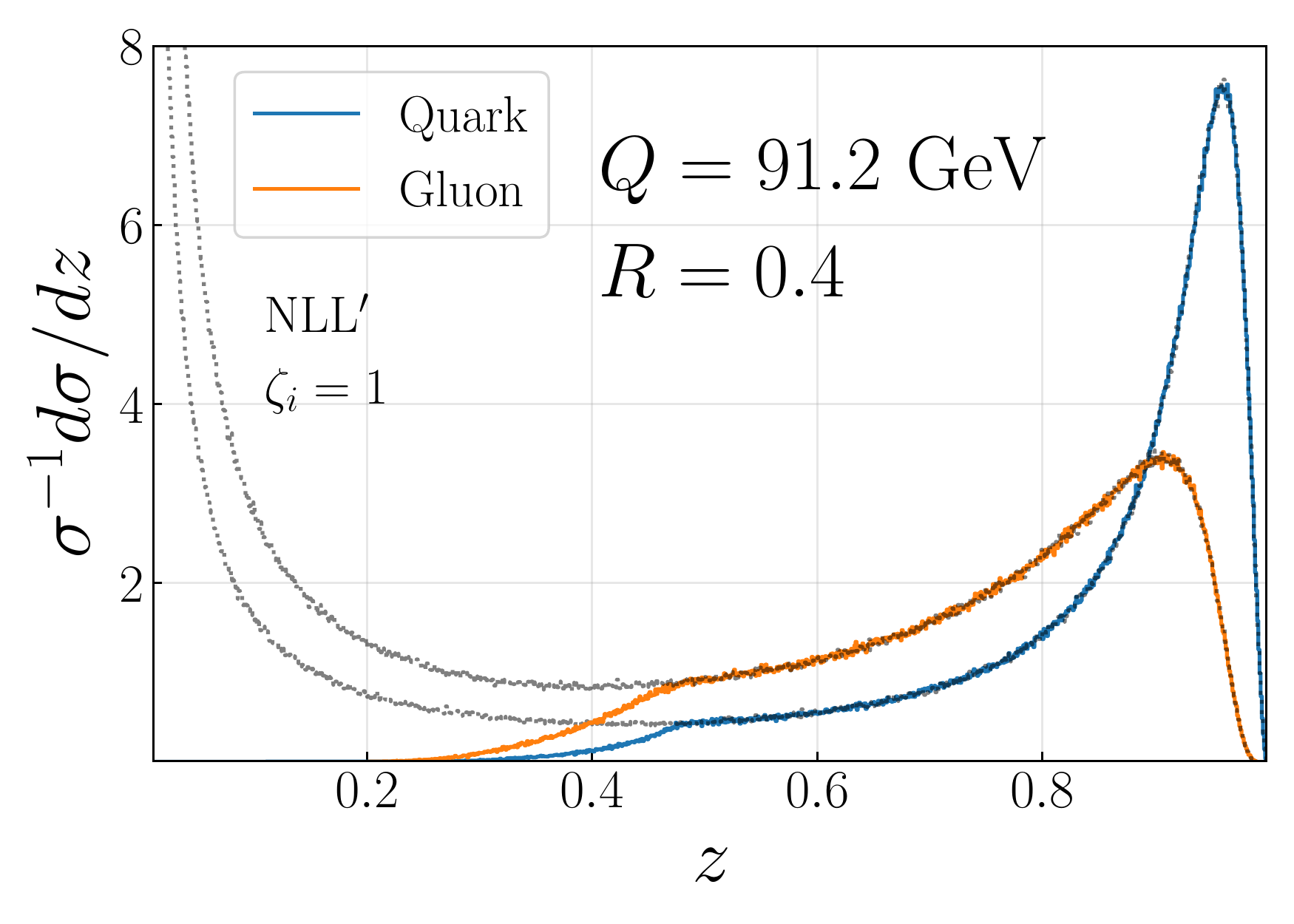}
\caption{Inclusive and leading jet spectra for quark/gluon $e^+e^-$ hemisphere jets and $\sqrt{s}=Q=91.2$~GeV.~\label{fig:inclusive_leading_NLL}}
\end{figure}

In this section we present numerical results for $e^+e^-$ hemisphere leading jets and leading subjets in proton-proton collisions. We implement the threshold resummed hard and jet functions in the Monte Carlo parton shower discussed above which allows us to calculate both the inclusive and leading jet cross section at NLL$'$ accuracy. In Fig.~\ref{fig:inclusive_leading_NLL}, we show the results for $e^+e^-$ hemisphere jets for quarks and gluons separately.\footnote{When showing our results, we vary a given scale $\mu_i\rightarrow \zeta_i\mu_i$. The range in which we vary $\zeta_i$ is given in each figure, where relevant. The scales varied are the observed jet, soft-collinear, and inclusive jet (Eq. \eqref{eq:scales}), and also the Landau-pole regularization, and the non-perturbative model parameter, and we take the envelope as a measure of uncertainty. } As an example, we choose the jet radius of $R=0.5$ and the hard scale $Q=\sqrt{s}=91.2$~GeV. The inclusive and leading jet spectra agree for $z>1/2$. For $e^+e^-$ hemisphere jets, a jet with momentum fraction $z>1/2$ is automatically the leading jet. Note that this does apply to event-wide leading jets in $e^+e^-$ collisions as discussed in section~\ref{sec:epemeventleading} below. We observe that both spectra peak at large values of $z$ which indicates that it is very likely to find a jet that carries a large momentum fraction of the initial quark or gluon. See also Ref.~\cite{Arratia:2020ssx}. The peak is less pronounced for an initial gluon than for quarks which is expected due to the different color factors. The peak structure at large values of $z$ confirms that the identified leading jet is a good proxy of the underlying parton level degrees of freedom. We note that the peak arises due to the threshold resummation. At LO/LL accuracy the numerical result diverges near the endpoint, see Fig.~\ref{fig:Comparison_MC_Mellin}. Therefore, it is phenomenologically important to include threshold resummation for leading jet measurements. Note that the suppression of the cross section for $z\to 1$ is unusual since threshold resummation is typically associated with an enhancement of the cross section~\cite{Sterman:1986aj,Catani:1989ne}. For $z<1/2$ the inclusive and leading jet spectrum differ due to the subleading jets which contribute only to the inclusive jet spectrum. The leading jet cross section drops significantly below $z=1/2$ indicating that it is very unlikely to find a leading jet that carries only a small momentum fraction $z$.

\begin{figure}[t]
\vspace*{.7cm}
\centering
\includegraphics[width=\textwidth]{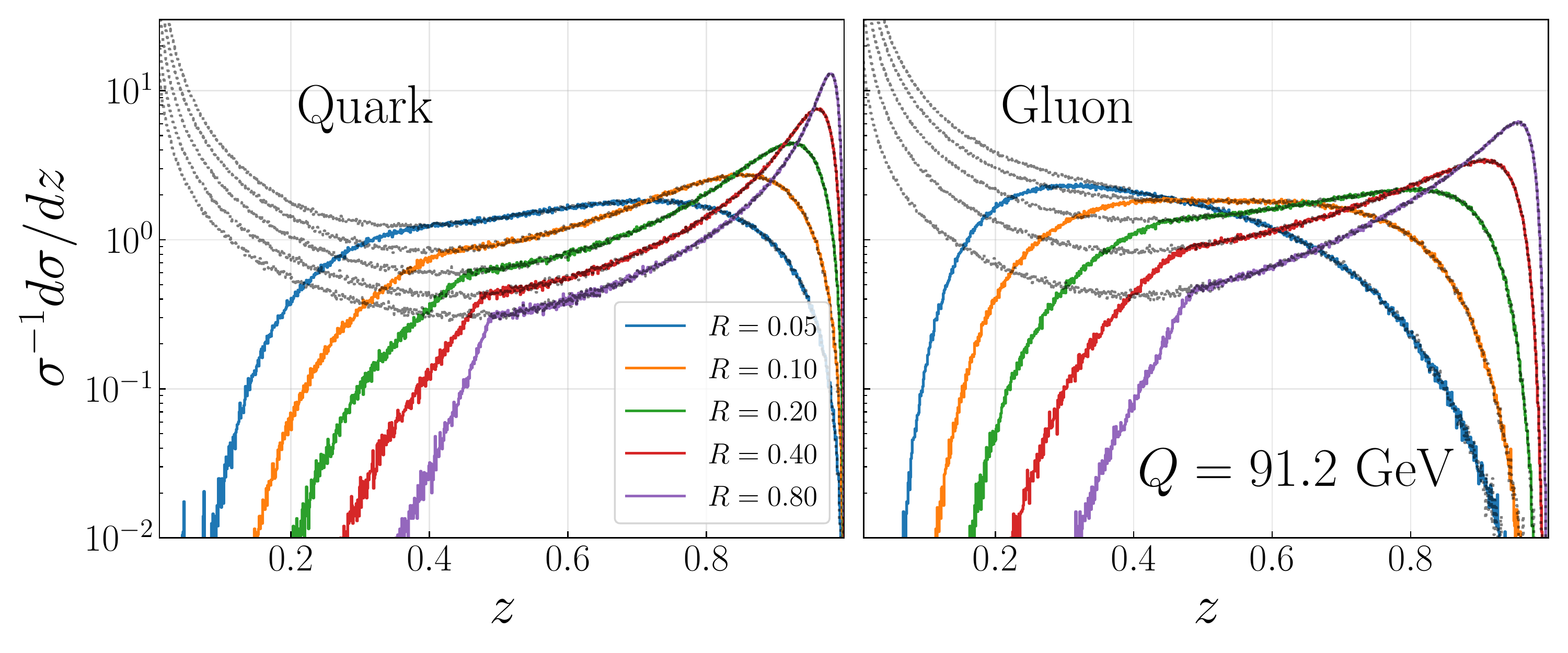}
\caption{The $e^+e^-$ hemisphere leading jet cross section for different values of the jet radius $R$. We choose the same kinematics as in Fig.~\ref{fig:inclusive_leading_NLL}.~\label{fig:lead_epem}}
\end{figure}

Another intriguing feature of the results in Fig.~\ref{fig:inclusive_leading_NLL} is the shape of the leading jet spectrum around $z=1/2$. At LO/LL accuracy there is a sharp kink or cusp which is now smeared out. In addition, we observe that the leading jet spectrum now extends to relatively small values of $z$ which is numerically quite different compared to the result at LO/LL accuracy. We would like to emphasize that the resummation here is critical to obtain the full leading jet spectrum. For all values of $R$, the tail of the distribution extends to small values of $z$. Instead, any fixed order computation at N$^n$LO would not give finite values below $z=1/(n+1)$.

Next, we study the dependence of the $e^+e^-$ hemisphere leading jet cross section on the jet radius $R$. The results for $R$ values in the range of $0.05-0.8$ are shown in Fig.~\ref{fig:lead_epem}. For small values of $R$, the leading jet can only capture a small fraction of the initial energy. Indeed, we observe that the curves shift toward smaller values of $z$ as we decrease the jet radius $R$. This shift differs significantly for quark and gluon jets. In the next section we consider the average $z$ value of these distributions which is related to the jet energy loss.

\begin{figure}[t]
\vspace*{.7cm}
\centering
\includegraphics[width=\textwidth]{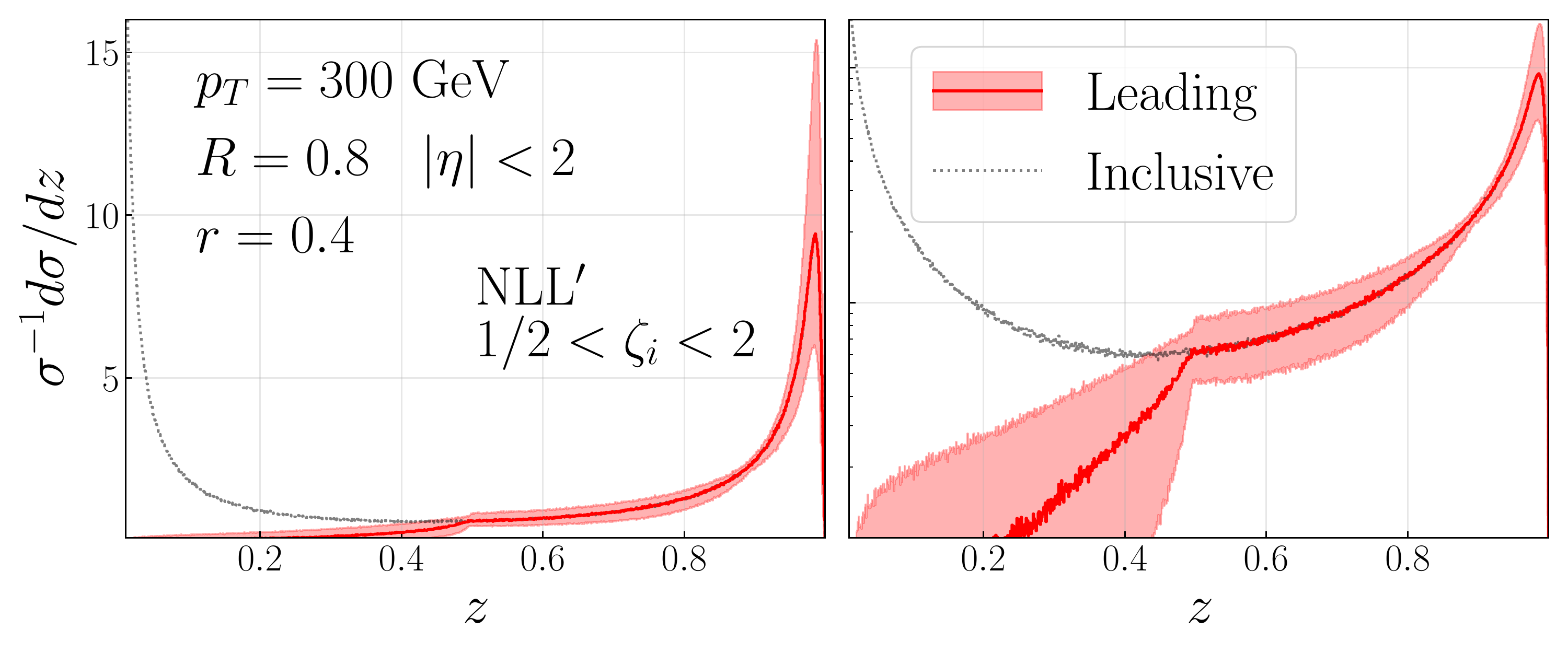}
\caption{Cross section for inclusive and leading subjets using a linear (left) and logarithmic scale (right). We show the result for $\sqrt{s}=13$~TeV proton-proton collisions and exemplary values of the jet kinematics.~\label{fig:inclusive_leading_NLL_subjets}}
\end{figure}

In Fig.~\ref{fig:inclusive_leading_NLL_subjets}, we show numerical results for inclusive and leading subjets in proton-proton collisions at $\sqrt{s}=13$~TeV. We choose exemplary jet kinematics as indicated in the figure and include appropriate quark/gluon fractions. The leading subjet cross section is the most straightforward possibility to directly quantify jet energy loss in proton-proton and heavy-ion collisions. We note that the size of the scale uncertainty band could be reduced by including higher order corrections.

\section{Quantifying jet energy loss~\label{sec:5}}

As discussed in section~\ref{sec:2}, the leading jet functions constitute probability densities that allow us to quantify the energy loss of leading jets. In this section we discuss different statistical quantities both at the level of the (threshold resummed) jet functions and and cross sections. In particular, we focus on $e^+e^-$ hemisphere leading jets. However, due to universality of the leading jet functions, the same arguments apply to subjets or other suitable observables discussed in section~\ref{sec:7} below. In section~\ref{sec:meanvariance} we start with the mean $\langle z_{i1}\rangle$ and variance $\sigma_i$ of the leading jet energy distribution. The mean is directly related to the average energy loss which we define as all the energy that is not contained in the leading jet $\langle z_{i,{\rm loss}}\rangle=1-\langle z_{i1}\rangle$. In section~\ref{sec:meanvariance_quarkgluon} we focus on differences between quark and gluon leading jets and in section~\ref{sec:entropyandKL}, we study the Shannon entropy and the KL divergence to further quantify differences between quarks and gluons.

\subsection{Mean and variance~\label{sec:meanvariance}}

We start by studying the mean and variance which are two fundamental quantities that quantify parton/jet energy loss. The mean or average energy of the initial parton which is contained inside the leading-jet $\langle z_{i1}\rangle$ is given by the first moment (second Mellin moment) of the leading jet function ${\cal J}_i$ as introduced in section~\ref{sec:lead_evolution}. We repeat the relevant equation here for convenience. For quarks and gluons, we find
\be\label{eq:average_repeated}
\int_0^1 {\rm d}z\, z\, {\cal J}_{i}(z,Q R,\mu) = \langle z_{i1}\rangle \,.
\ee
The average energy fraction of the leading jet depends on the scale $Q$ and the jet radius $R$ which we omit on the right hand side. In Ref.~\cite{Dasgupta:2014yra}, an expansion of the average out-of-jet radiation in $\as \ln R$ was performed. Here we perform the complete expansion in powers of the strong coupling constant $\alpha_s$ which requires knowledge of the entire jet function. The mean or average energy loss of the leading jet is given by $\langle z_{i,{\rm loss}}\rangle=1-\langle z_{i1}\rangle$. This relation holds to all orders. At NLO, $\langle z_{i,{\rm loss}}\rangle$ coincides with the average energy fraction contained in the first subleading jet. At higher orders, the average lost energy $\langle z_{i,{\rm loss}}\rangle$ is shared amongst the different subleading jets. We consider both cone~\cite{Akers:1994wj,Salam:2007xv} and k$_T$-type~\cite{Catani:1993hr,Dokshitzer:1997in,Wobisch:1998wt,Cacciari:2008gp} jets. We start with an NLO computation of the average momentum fraction which is contained in the leading jet at the jet function level. From Eq.~(\ref{eq:average_repeated}), we find for quarks and gluons
\begin{align}\label{eq:NLOloss}
\langle z^{k_T}_{q1} \rangle &=1+\f{\as}{2\pi}C_F \ln(1/R^2)\left(\frac{3}{8}-2\ln 2\right)+\f{\as}{2\pi}C_F\left(\f{19}{8} - \f32 \ln 2 - 4 \ln^2 2 - \f{\pi^2}{3}\right)\,, \\
\langle z^{k_T}_{g1} \rangle &=1+\f{\as}{2\pi}\ln(1/R^2)\left[C_A \left(\f{43}{96} - 2 \ln 2\right)- N_f T_F \f{7}{48}  \right]\nn\\
&+\f{\as}{2\pi}\left[C_A \left(\f{793}{288} - 4 \ln^2 2 - \f{15}{8} \ln 2 - \f{\pi^2}{3}\right) +N_f T_F \left(-\f{65}{72} + \f{3}{4} \ln 2\right)\right]\,,\\
\langle z^{\rm cone}_{q1} \rangle &=1+\f{\as}{2\pi}C_F \ln(1/R^2)\left(\frac{3}{8}-2\ln 2\right)+\f{\as}{2\pi}C_F\left(\f12 - 2\ln^2 2 + \f34 \ln 2 -\frac{\pi^2}{3} \right) \,,\\
\langle z^{\rm cone}_{g1} \rangle &=1+\f{\as}{2\pi}\ln(1/R^2)\left[C_A \left(\f{43}{96} - 2 \ln 2\right)- N_f T_F \f{7}{48}  \right]\nn\\
&+\f{\as}{2\pi}\left[C_A \left(\f{265}{576}-2\ln^2 2+\f{43}{48} \ln 2-\f{\pi^2}{3}\right) -N_f T_F \left(\f{19}{288}+\f{7}{24}\ln 2\right)\right] \,.
\end{align}
We note that the terms $\sim\ln(1/R^2)$ are the same for both k$_T$ and cone-type jets and they agree with the result in~\cite{Dasgupta:2014yra}. The remaining finite ${\cal O}(\alpha_s)$ corrections are reported here for the first time. As expected those terms depend on the jet algorithm. An important aspect of the full ${\cal O}(\alpha_s)$ result is that it leads to a finite energy loss even for $R\to 1$, where the logarithms of the jet radius vanish. At small jet radii the logarithmically enhanced terms in Eq.~(\ref{eq:NLOloss}) dominate and need to be resummed to all orders. In principle analytical results at higher orders in $\alpha_s$ could be obtained from the non-linear evolution equations in Eq.~(\ref{eq:nonlinearevol}). Instead, here we are going present numerical results using the parton shower framework which was introduced in the previous sections which also includes the resummation of threshold logarithms.

\begin{figure}[t]
\vspace*{.7cm}
\centering
\includegraphics[width=\textwidth]{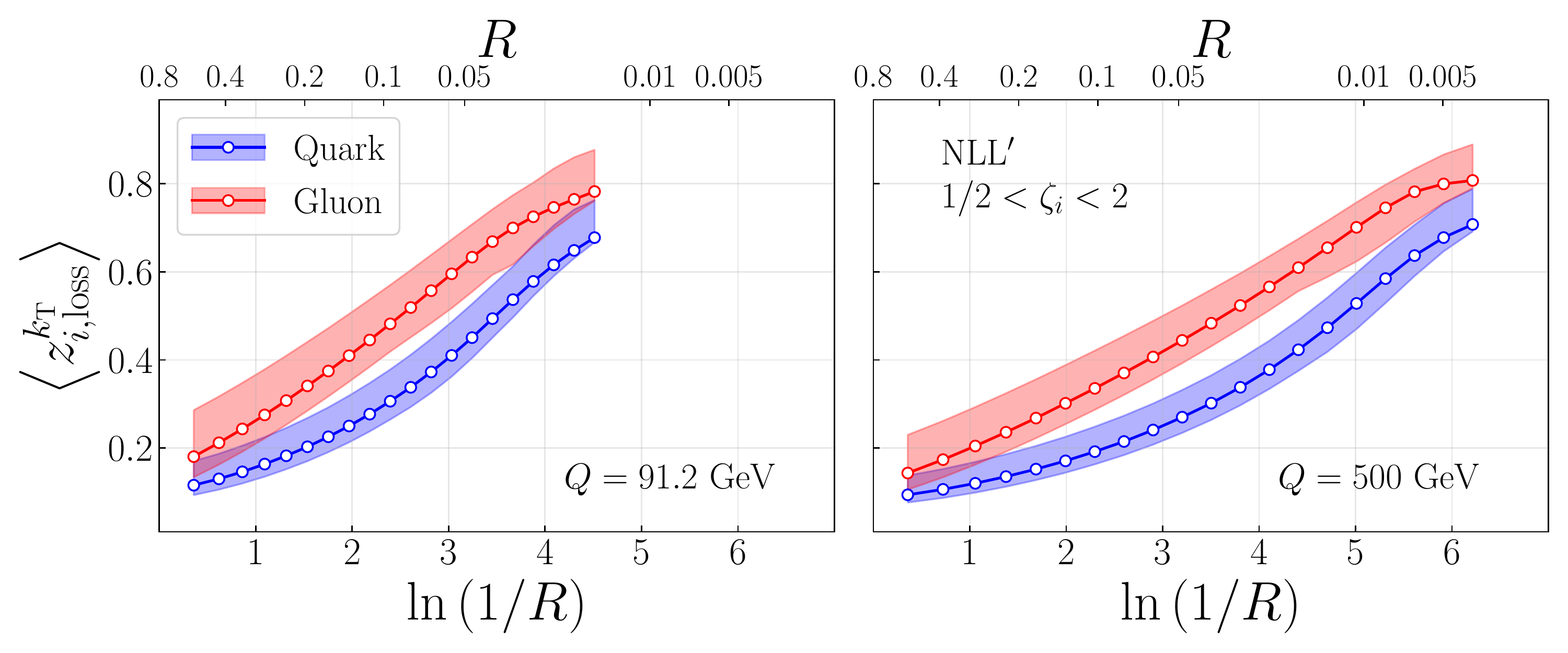}
\caption{Average energy loss of $e^+e^-$ hemisphere leading jets $\langle z_{i,{\rm loss}}^{k_T}\rangle$ for k$_T$-type jets. We separately show the quark and gluon results for an initial energy of $Q=91.2$~GeV and $500$~GeV as a function of the jet radius.~\label{fig:zloss}}
\end{figure}

We present numerical results not at the jet function level but for a cross section which can be measured in experiments, i.e. including also the (threshold resummed) hard function. We consider $e^+e^-$ hemisphere leading jets as an example. The average energy loss for quark and gluon jets is shown in Fig.~\ref{fig:zloss} using the parton shower algorithm described above including threshold resummation at NLL$'$. We plot the numerical result for $\langle z_{i,{\rm loss}}\rangle$ as a function of the jet radius on a logarithmic scale. We choose two exemplary hard scales of $Q=91.2$~GeV and $500$~GeV which set the initial parton energy. The rightmost $R$ values correspond to $QR=1$~GeV, where the average energy loss for both $Q$ values turns out to be very similar. We observe that the energy loss of gluons is larger than for quarks which is generally expected due to the different color factors and DGLAP splitting functions. In addition, the average energy loss of leading jets is larger at smaller scales $Q$. For large jet radii, the average energy loss of a quark is around 10\% and 15-20\% for an initial gluon. This observation quantifies why jets are generally considered to be excellent proxies of parton level dynamics compared to hadrons. For the small jet radii, the average energy loss increases up to around 70\% for quarks and 80\% for gluons. As expected, a relatively small jet radius can only capture a small fraction of the initial momentum of the quark or gluon and the energy loss grows significantly. We note that hadrons correspond to jets with vanishing radius where an additional nonperturbative parton-to-hadron transition needs to be taken into account which makes the average energy loss of hadrons even larger than what is shown in Fig.~\ref{fig:zloss}. We would like to stress again that such a statement is not possible for inclusive jets since any emission outside the leading jet constitutes another jet which is also taken into account when an inclusive jet sample is measured as discussed in section~\ref{sec:2} above. The result for the parton/jet energy loss presented here depends significantly on the perturbative higher order corrections which we include in the parton shower~\ref{sec:3}. In this sense, the results presented here constitute the first quantitative calculation of vacuum parton/jet energy loss which allows for a meaningful connection to experimental measurements.

\begin{figure}[t]
\vspace*{.7cm}
\centering
\includegraphics[width=\textwidth]{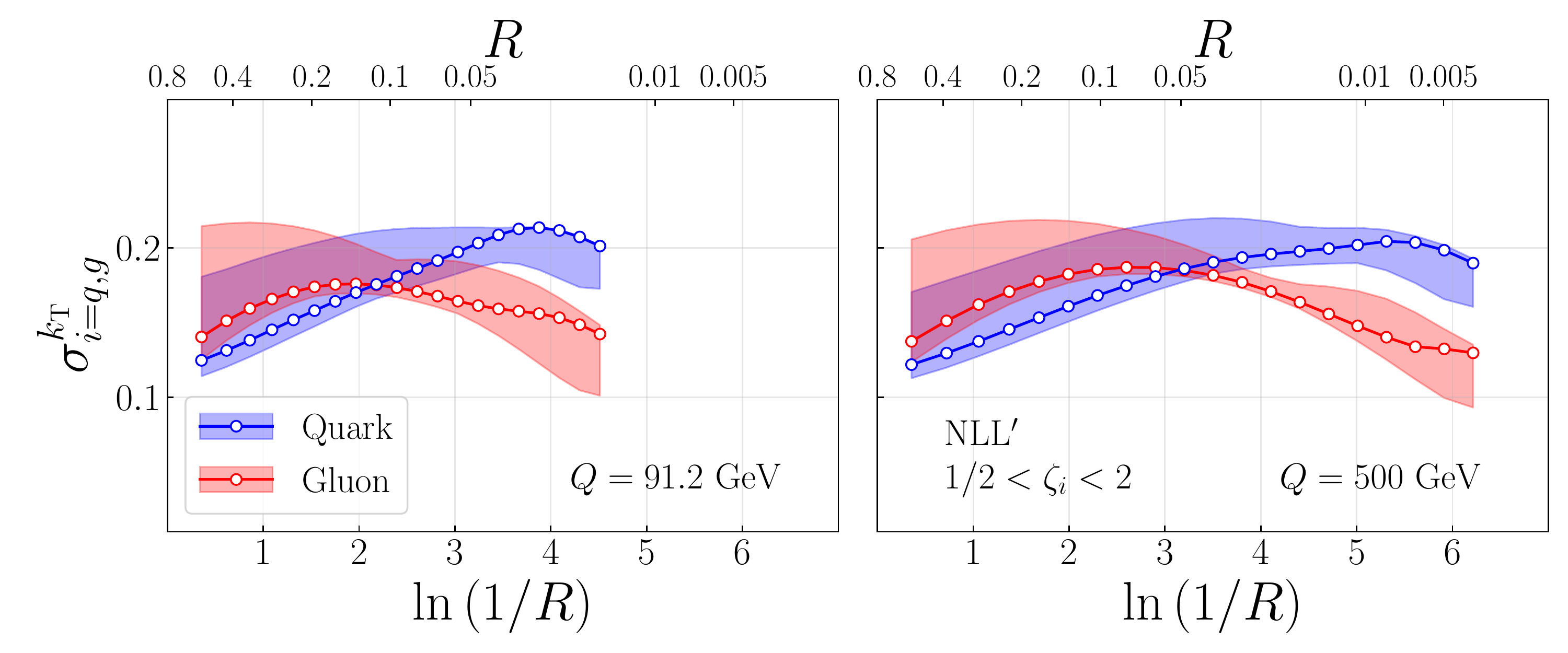}
\caption{Variance $\sigma_{i}^{k_T}$ of the energy loss of $e^+e^-$ hemisphere leading jets for quark and gluons. As in Fig.~\ref{fig:zloss}, we choose a k$_T$-type jet algorithm, two initial reference scale of $Q=91.2$~GeV and $500$~GeV and we plot the result as a function of the jet radius.~\label{fig:variance}}
\end{figure}

Besides the mean of the leading jet probability distribution, we can also calculate the variance or the fluctuations of jet energy loss. The variance for quarks and gluons is defined as
\begin{equation}
    \sigma_i = \langle (z_i-\langle z_i\rangle)^2\rangle = \langle z_i^2\rangle - \langle z_i\rangle^2 \,.
\end{equation}
Similar to the mean, the variance $\sigma_i$ depends on the jet algorithm, the jet radius $R$ and initial scale $Q$. We omit the explicit dependence of $\sigma_i$ on those quantities here for notational convenience. However, we study the dependence of $\sigma_i$ on those variables numerically below. Here $\langle z_i^2\rangle$ is the second moment (third Mellin moment) of the leading jet function. At the jet function level it is given by
\begin{equation}
    \left\langle z_{i}^2\right\rangle=\int_{0}^{1} \mathrm{d} z\, z^2 \,\mathcal{J}_{i}\left(z, Q R, \mu\right)\,.
\end{equation}
We start again with an NLO calculation of the variance at the jet function level. For quarks and gluons, we find
\begin{align}
\sigma^{k_T}_{q} &
=\f{\as}{2\pi}C_F \ln(1/R^2)\left(-\frac98+2\ln 2\right)+\f{\as}{2\pi}C_F\left(-\frac{131}{24}+2\ln^2 2  +\frac32\ln 2+\frac{\pi^2}{3}\right) \,,
\\
\sigma^{k_T}_{g} &
=\f{\as}{2\pi}\ln(1/R^2)\left[C_A \left(-\frac{551}{480}+2\ln 2\right)+ N_f T_F\frac{11}{240} \right]
\nn\\&
+\f{\as}{2\pi}\left[C_A \left(-\frac{41377}{7200}+4\ln^2 2+\frac{15}{8}\ln 2+\frac{\pi^2}{3}\right) +N_f T_F \left(\frac{1121}{1800}-\frac{3}{4}\ln 2\right)\right]\,,
\\
\sigma^{\rm cone}_{q} &
=\f{\as}{2\pi}C_F \ln(1/R^2)\left(-\frac98+2\ln 2\right)+\f{\as}{2\pi}C_F\left(-\frac{17}{8}+2\ln^2 2-\frac{9}{4}\ln 2+\frac{\pi^2}{3} \right) \,,
\\
\sigma^{\rm cone}_{g} &=
\f{\as}{2\pi}\ln(1/R^2)\left[C_A \left(-\frac{551}{480}+2\ln 2\right)+ N_f T_F\frac{11}{240} \right]
\nn\\&+
\f{\as}{2\pi}\left[C_A \left(-\frac{30247}{14400}+2\ln^2 2-\frac{551}{240}+\frac{\pi^2}{3}\right) -N_f T_F \left(-\frac{23}{7200}+\frac{11}{120}\ln 2\right)\right] \,.
\end{align}
Similar to the mean in Eq.~(\ref{eq:NLOloss}), the $\sim\ln(1/R^2)$ term is independent of the jet algorithm. Next, we study the variance using the full parton shower for $e^+e^-$ hemisphere leading jets. The results for quarks and gluons are shown in Fig.~\ref{fig:variance} as a function of the jet radius. We observe that the variance for both quark and gluon jets is in the range of $\sim 0.1-0.2$ for $Q=91.2$~GeV and $500$~GeV. For large $R$, the variance is almost independent of the scale $Q$ but the scale dependence becomes more visible toward smaller $R$. For gluons the variance peaks at intermediate values of the jet radius $R$. The leading jet distribution for gluons peaks either at large-$z$ (large $R$) or large-$z$ (small R) which leads to relatively small values of $\sigma_g^{k_T}$. Only in the intermediate $R$ region, the gluon $z$-distribution is broad which leads to the maximum of $\sigma_g^{k_T}$ that we observe in Fig.~\ref{fig:variance}. For quarks the $z$-distribution is even more peaked at large $R$ than for gluons which leads to a smaller variance. It evolves more slowly toward small-$z$ than the gluon distribution which is why the variance continues to increase toward small $R$ and eventually becomes even larger than for gluons. At small $R$ the variance for quarks has a maximum and (close to the nonperturbative region) becomes smaller again. It will be interesting to study how the mean and variance of leading jets/the energy loss is modified in heavy-ion or electron-nucleus collisions where the notion of (medium induced) parton/jet energy loss plays an important role.

We end this section by noting that there is no unique definition of energy loss. For example, we could adopt the definition that all the energy which is not contained in the first two leading jets is ``lost energy''. We can also calculate the average energy loss $\langle \tilde z_{i,{\rm loss}}\rangle$ for this alternative definition from the subleading jet function as
\begin{align}
    \langle \tilde z_{i,{\rm loss}}\rangle&=\,\int_0^1{\rm d}z_{i1}{\rm d}z_{i2}{\rm d}\tilde z_{i,{\rm loss}}\, \tilde z_{i,{\rm loss}} \, {\cal J}_{i}(z_{i1},z_{i2},\hat p_T R,\mu)\,\delta(\tilde z_{i,{\rm loss}}+z_{i1}+z_{i2}-1)
    \nn\\&=\,
    1-\langle z_{i1}\rangle-\langle z_{i2}\rangle \,,
\end{align}
and therefore
\begin{equation}
    \langle \tilde z_{i,{\rm loss}}\rangle = \sum_{n\geq 3}\langle z_{in}\rangle \,.
\end{equation}
Also other definitions are possible as long as we consider a fixed number of jets. For inclusive jets it is not possible to construct a probability density since the number of jets is not fixed but it is generated dynamically event-by-event.

\subsection{Quark/gluon energy loss~\label{sec:meanvariance_quarkgluon}}

\begin{figure}[t]
\vspace*{.7cm}
\centering
\includegraphics[width=.7\textwidth]{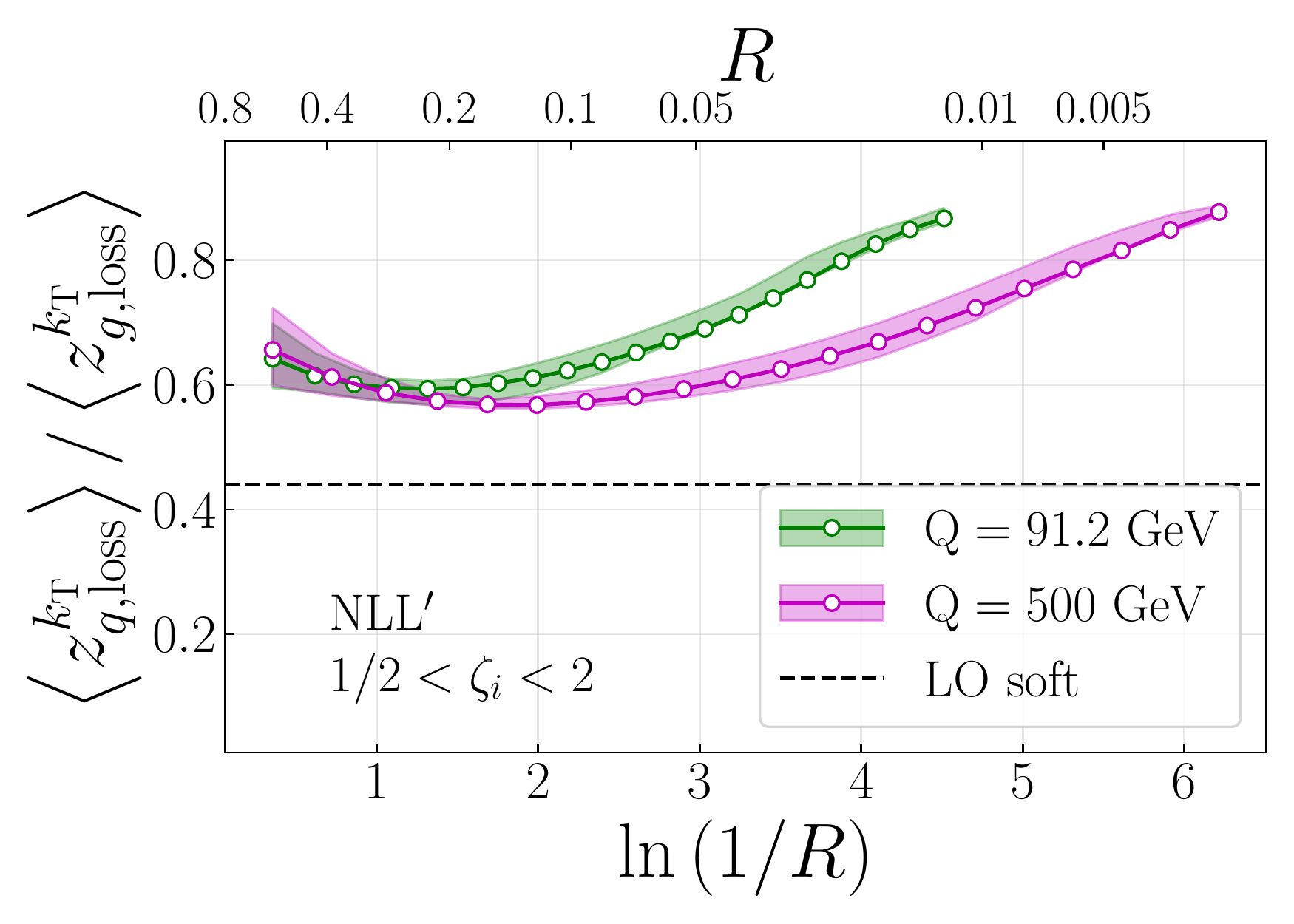}
\caption{Ratio of the average quark and gluon jet energy loss as a function of the jet radius $R$ for two different values of the energy scale $Q=91.2$~GeV and $500$~GeV. For comparison, we also show the result at LO in the soft approximation.~\label{fig:zloss_quark_gluon}}
\end{figure}

The dependence of the energy loss mechanism on the parton flavor has been discussed extensively in the literature -- in particular in the context of jet quenching studies in heavy-ion collisions. While we only focus on vacuum energy loss in this work, our results set the baseline for studies of energy loss in the nuclear medium. In this section we compare the LO estimate of the quark/gluon energy loss to the full result from the parton shower. First, we consider the LO emission spectrum in the soft gluon approximation. For an initial quark/gluon, we find
\begin{equation}\label{eq:dIdz}
    \frac{{\rm d}I_{q,g}}{{\rm d}z}\sim \frac{C_{F,A}}{1-z} \,.
\end{equation}
This relation implies that in this limit the ratio of the average jet energy loss of a quark and gluon is
\begin{equation}\label{eq:softqg}
    \frac{\langle z_{q,{\rm loss}}\rangle}{\langle z_{g,{\rm loss}}\rangle}=\frac{C_F}{C_A}\approx 0.44\,,
\end{equation}
which is sometimes referred to as Casimir scaling in the literature.

The ratio of the quark/gluon energy loss $\langle z_{q,{\rm loss}}\rangle/\langle z_{g,{\rm loss}}\rangle$ is shown in Fig.~\ref{fig:zloss_quark_gluon} as a function of the jet radius $R$ for $Q=91.2$~GeV and $500$~GeV. Here we choose again $e^+e^-$ leading jet production as a representative example. We observe that the ratio of the average energy loss of quarks and gluons is almost identical for the two different $Q$ values at large $R$ but differs at intermediate and smaller $R$. Interestingly, the ratio of the average energy loss of quarks and gluons is significantly closer to 1 compared to the LO estimate in the soft gluon approximation. Thus, quark/gluon differences of the (vacuum) energy loss are in fact significantly less pronounced than one would naively expect from the so-called Casimir scaling in Eq.~(\ref{eq:softqg}). We also observe that the ratio is relatively flat for large to intermediate values of $R$ and the curves have a minimum at intermediate values of $R$ for both values of $Q$. We explore this feature in more detail in the context of quark/gluon tagging below. It turns out that the value of the ratio of the quark/gluon average energy loss agrees for $QR=1$~GeV, which corresponds to the rightmost points in Fig.~\ref{fig:zloss_quark_gluon} and is given by $\approx 0.9$.

\subsection{Shannon entropy, KL divergence~\label{sec:entropyandKL}}

\begin{figure}[t]
\vspace*{.7cm}
\centering
\includegraphics[width=\textwidth]{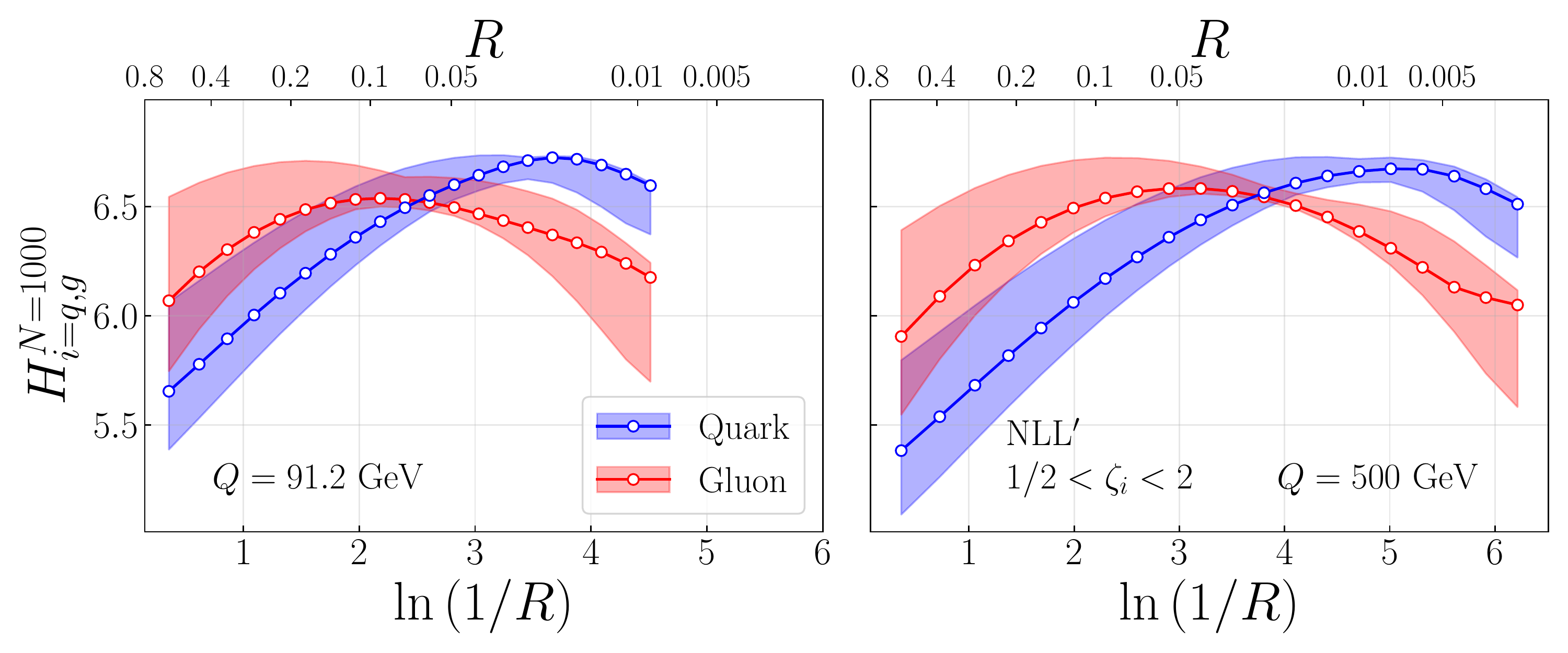}
\caption{Shannon entropy for quark and gluon jets and $Q=91.2$~GeV and $500$~GeV as a function of the jet radius $R$.~\label{fig:Shannon}}
\end{figure}

Since the leading jet cross section constitutes a probability distribution, we can compute various other statistical quantities besides the mean and variance. In this section we consider the Shannon entropy and the KL divergence. In order to quantify the average uncertainty of the leading jet/energy loss probability distribution, we consider the Shannon/information entropy. We can write the continuous version of the Shannon entropy $h_i$ at the jet function level as
\begin{equation}\label{eq:shannon}
    H_i=-\int_0^1{\rm d}z\,{\cal J}_i(z) \ln({\cal J}_i(z)) \,.
\end{equation}
In addition, we consider the KL divergence $D_{\rm KL}$ to quantify the difference between quarks and gluons
\begin{equation}\label{eq:KL}
    D_{\rm KL}({\rm Quark}||{\rm Gluon})=-\int_0^1{\rm d}z\, {\cal J}_q(z)\ln\bigg(\frac{{\cal J}_q(z)}{{\cal J}_g(z)}\bigg) \,.
\end{equation}
The KL divergence is not symmetric under ${\cal J}_q\leftrightarrow {\cal J}_g$. Nevertheless, it is a very useful measure to quantify the similarity of two probability distributions. An alternative measure would be the Jensen-Shannon divergence which is symmetric. Note that here we introduced both quantities in Eqs.~(\ref{eq:shannon}) and~(\ref{eq:KL}) for a continues variable $z$. However, experimental measurements are binned which is why we replace the integral versions of these metrics in Eqs.~(\ref{eq:shannon}) and~(\ref{eq:KL}) by their corresponding discrete versions. For our numerical results presented below we choose a binning of $N=1000$ steps in $z$. The measurement of both of these quantities will also shed new light on the energy loss mechanism in the hot or cold nuclear matter environment. Here we only consider the Shannon entropy and the KL divergence for the probability densities of the leading jet but they can be extended to subleading jets as well.

\begin{figure}[t]
\vspace*{.7cm}
\centering
\includegraphics[width=.7\textwidth]{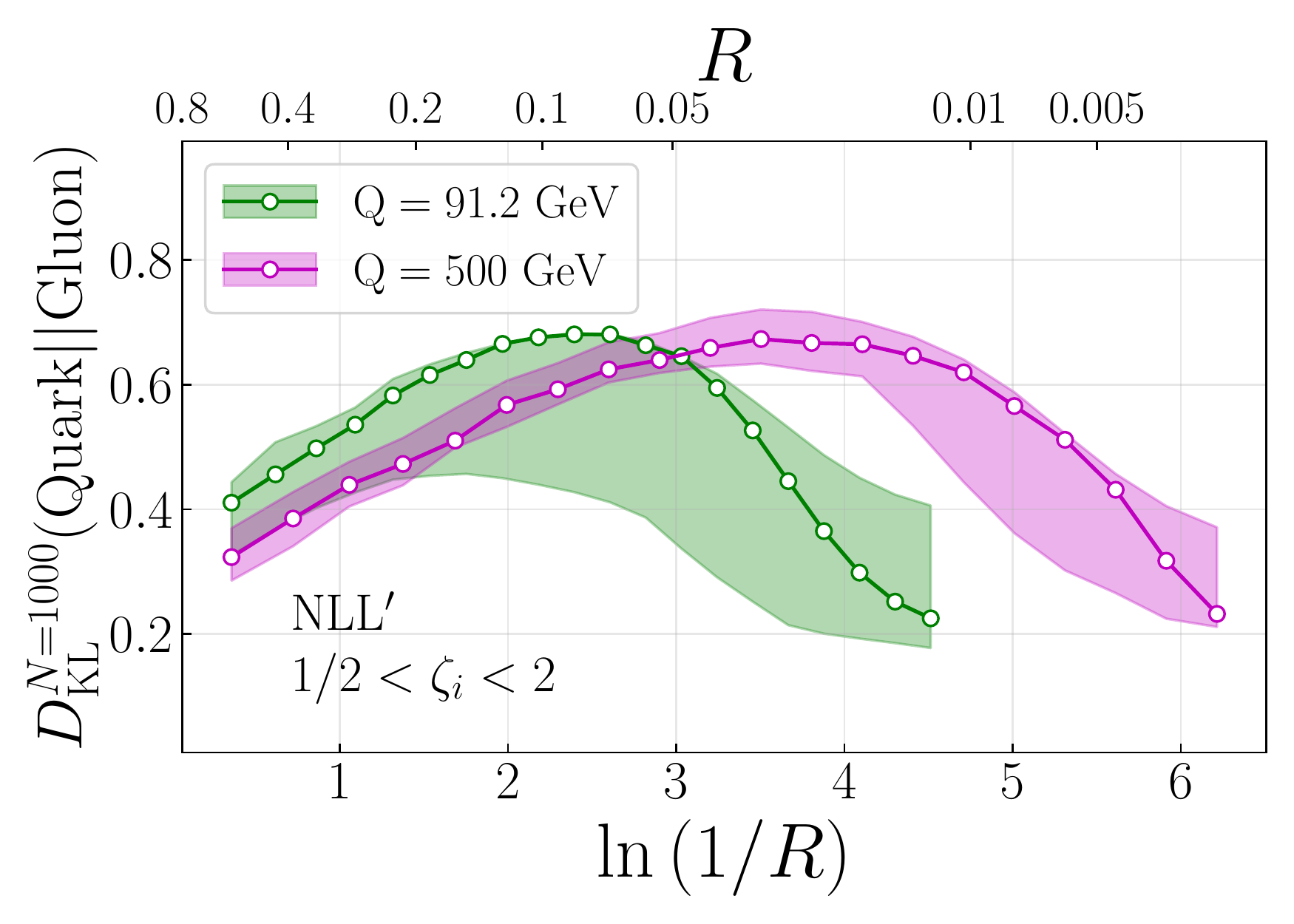}
\caption{KL divergence of the leading jet distributions as defined in Eq.~(\ref{eq:KL}) which quantifies the difference between the quark and gluon leading jet energy loss.~\label{fig:KL}}
\end{figure}

The Shannon entropy at the cross section level for $e^+e^-$ hemisphere leading jets is shown in Fig.~\ref{fig:Shannon}. It peaks at intermediate values of the leading jet radius. At large values of $R$, the uncertainty of gluon jets is larger compared to quark jets and a larger value of $Q$ leads to a lower value. At intermediate to small values of $R$ the ordering changes. The KL divergence is shown in Fig.~\ref{fig:KL}. We observe that the KL divergences peaks at intermediate values of $R$ which indicates that there is an optimal value of $R$ to distinguish quark and gluon leading jets. The maximum value depends on the scale $Q$ and is shifted toward smaller $R$ for larger values of $Q$. Interestingly, this maximum is in the perturbative regime indicating that leading (sub)jets may be a good observable for quark/gluon discrimination which is under perturbative control. We further explore the potential of leading (sub)jets as a quark/gluon discriminant in the next section. We also notice that the rightmost points of the two curves in Fig.~\ref{fig:KL} agree, which correspond to $QR=1$~GeV in both cases. This is likely due to the relative simplicity of the non-perturbative model that we are using, without tuning the it to account for differences between quark or gluon initiated jets.

\section{Quark/gluon discrimination with leading (sub)jets~\label{sec:6}}

A typical task of jet substructure observables is the discrimination between quark and gluon jets. In this section we study the tagging performance of leading (sub)jets. It is instructive to compare leading (sub)jets to established quark/gluon jet tagging techniques in the literature~\cite{Gras:2017jty}. Using the largest momentum fragment in a jet as a classifier is in the same class of observables known as ``fractal jet observables''~\cite{Elder:2017bkd}, see also Refs.~\cite{Krohn:2012fg,Kang:2021ryr}. These observables are tuned to the energy flow patterns generated in DGLAP evolution of the final state. Another set of useful classifiers is given by the so-called generalized angularities~\cite{Larkoski:2014pca} (see also~\cite{Berger:2003iw,Almeida:2008yp,Ellis:2010rwa,Larkoski:2014uqa})
\begin{equation}\label{eq:angularities}
    \lambda_\beta^\kappa = \sum_{i\in{\rm jet}} z_i^\kappa\theta_i^\beta  \,.
\end{equation}
Here the sum runs over all particles inside the jet with radius parameter $R$, and $z_i,\theta_i$ are their momentum fractions and angles with respect to the jet axis, respectively. For example, the average momentum fraction of the leading jet can be obtained from  Eq.~(\ref{eq:angularities}) as follows. First, instead of summing over individual particles in the jet, we sum over all subjets which are obtained by reclustering the initial jet with a jet radius $r<R$, see also section~\ref{sec:4.2} above. Second, we choose $\beta=0$ which is similar to jet multiplicities and the jet $p_T^D$ observable~\cite{Pandolfi:2012ima,Chatrchyan:2012sn}. Third, take the expression to the power of $1/\kappa$ and we take the limit $\kappa\to\infty$ which singles out the leading jet momentum fraction.

\begin{figure}[t]
\vspace*{.7cm}
\centering
\includegraphics[width=.9\textwidth]{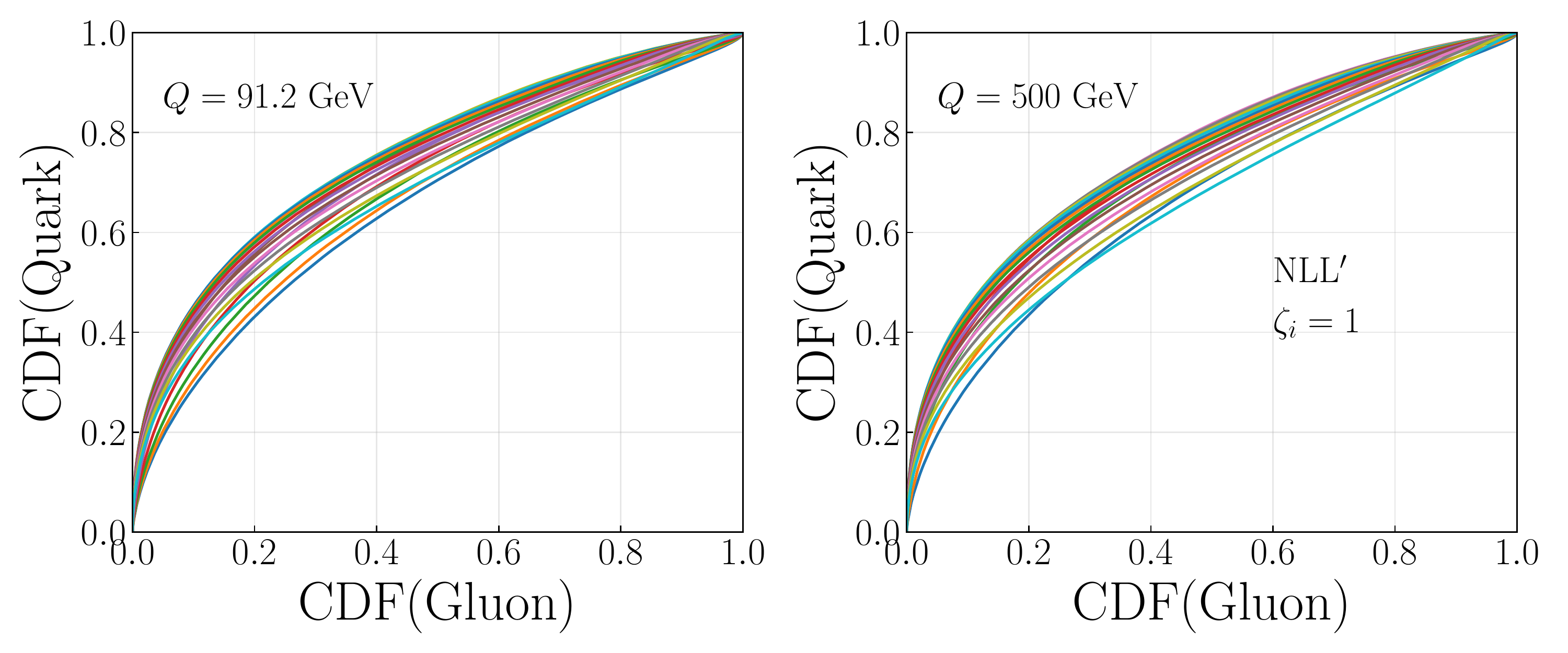}
\caption{ROC curves for quark/gluon jet discrimination with leading (sub)jets for different values of the jet radius $R$ and two different values of the energy scale $Q$.~\label{fig:ROC}}
\end{figure}

Typically, the performance of a classifier is quantified by studying the ROC curve. The ROC curve shows the quark/gluon true positives (cumulative distribution function (CDF)) vs. the false positives rates for a given decision threshold. The result for (sub)leading jets and two values of the energy scale $Q$ and various values of the jet radius $R$ are shown in Fig.~\ref{fig:ROC}. We observe that the discrimination power changes significantly as a function of the jet radius.

\begin{figure}[t]
\vspace*{.7cm}
\centering
\includegraphics[width=.65\textwidth]{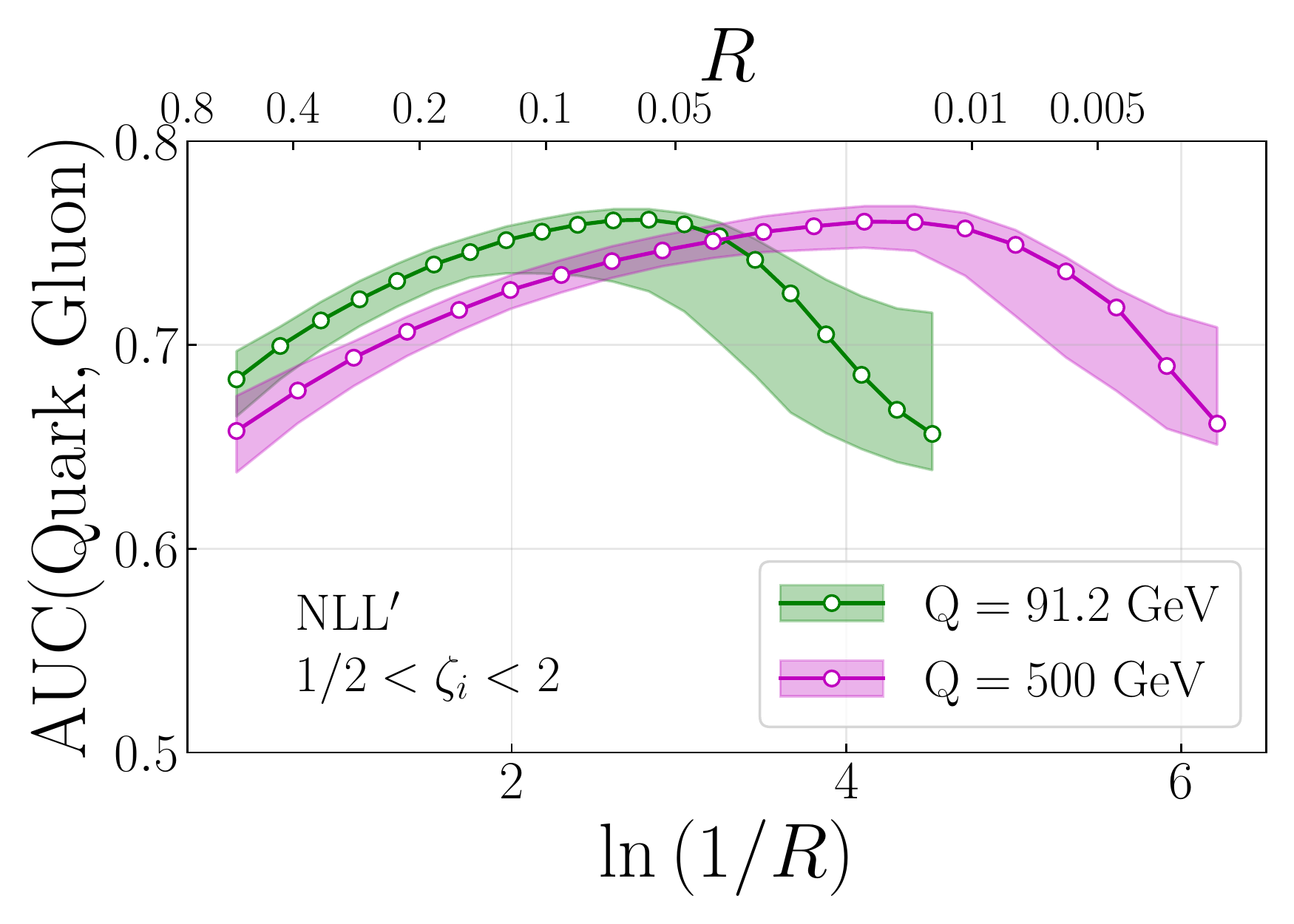}
\caption{Quark/gluon discrimination power of leading (sub)jets (AUC) as a function of the jet radius $R$ and two different values of the energy scale $Q$.~\label{fig:AUC}}
\end{figure}

In order to obtain a single value to quantify the performance of leading (sub)jets as a quark/gluon discriminant, the area under the ROC curve (AUC) is commonly used which is shown in the right panel of Fig.~\ref{fig:AUC}. We note that the location of the peak differs slightly between the AUC and the KL divergence in Fig.~\ref{fig:KL} above. As mentioned above, in both cases, the peak of the distributions is in the perturbative region. Instead, other observables that perform very well are often nonperturbative such as the particle multiplicity. Since leading (sub)jets correspond to the most energetic part of a jet, it may capture relevant information with little overlap with e.g. the particle multiplicity. We plan to explore the information content of leading (sub)jets relative to other observables in future work. See for example Ref.~\cite{Datta:2017rhs}.

\section{Further applications~\label{sec:7}}

In this section we discuss several cross sections involving leading jets and hadrons. We start by considering event-wide leading jets in $e^+e^-$ collisions in section~\ref{sec:epemeventleading} in contrast to hemisphere leading jets. We present numerical results and compare to Pythia~8~\cite{Sjostrand:2014zea} simulations. Similar measurements were performed at LEP. We then discuss leading jets in SIDIS (section~\ref{sec:SIDIS}), photon-jet correlations (section~\ref{sec:photon_jet}) and eventually extend our parton shower framework toward leading hadrons instead of jets (section~\ref{sec:hadrons}).

\subsection{$e^+e^-$ event-wide leading jets~\label{sec:epemeventleading}}

\begin{figure}[t]
\vspace*{.6cm}
\centering
\includegraphics[width=.9\textwidth]{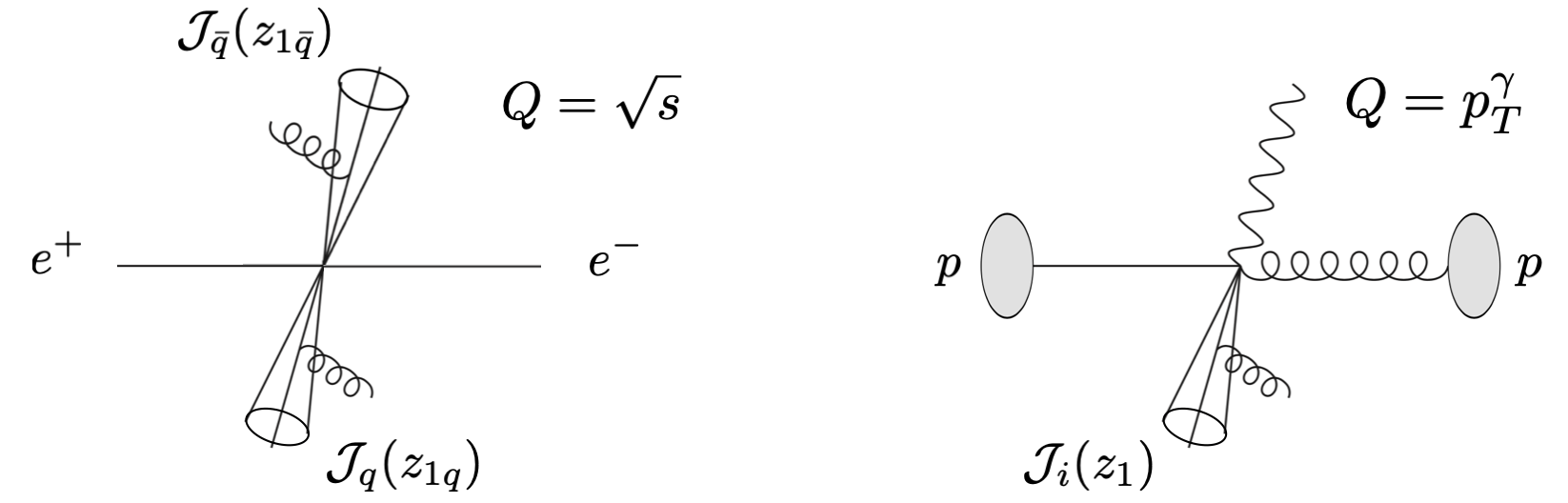}
\caption{Illustration of event-wide leading jets in $e^+e^-$ collisions (left) and the leading jet recoiling against a direct photon in proton-proton collisions (right).~\label{fig:eventwide+photon}}
\end{figure}

Instead of the $e^+e^-$ hemisphere leading jets, which we discussed in the previous sections, we are now going to consider the leading jet in the entire event. The necessary factorization structure here is similar to leading jets in proton-proton collisions which was discussed in section~\ref{sec:lead_factorization}, except that here we have direct access to the initial reference scale $Q=\sqrt{s}$. We consider the cross section differential in the energy fraction of the leading jet $z_1=2E_1/Q$. At LL accuracy, we have
\begin{align}\label{eq:epem_leadingjet4}
\frac{1}{\sigma_{\rm tot}}\frac{{\rm d}\sigma_{e^+e^-\to{\rm jet}_1+X}^{(0)}}{{\rm d}z_{1}}=&\; {\cal H}_{q\bar q}^{(0)}(Q,\mu) \int{\rm d}z_q\,{\rm d}z_{\bar q} \, {\cal J}_q(z_q,QR/2,\mu)\,{\cal J}_{\bar q}(z_{\bar q},QR/2,\mu)
\nonumber\\ &\times\,
\delta(z_{1}-{\rm max}\{z_q,z_{\bar q}\})\,.
\end{align}
The quark and anti-quark each fragment into a leading jet with energy fraction which we denote by $z_{q,\bar q}$. The event-wide leading jet momentum fraction is then obtained by picking the larger value of $z_{q,\bar q}$ which is taken into account by the delta function Eq.~(\ref{eq:epem_leadingjet4}). The topology of event-wide leading jets in $e^+e^-$ collisions is illustrated in the left panel of Fig.~\ref{fig:eventwide+photon}. As discussed in section~\ref{sec:lead_factorization}, at higher perturbative accuracy we need to take into account additional jet functions as well as integrals over the hard function. We stress again that the factorization structure here is very different compared to inclusive jets. In this case any jet is taken into account independent of their energy fraction which allows for a factorization structure in terms of a convolution integral and which is independent of the perturbative order. We include again the threshold resummation for $e^+e^-$ event-wide leading jets. Working at NLL$'$ accuracy, the threshold resummed jet and hard functions can be obtained analogous to the results in section~\ref{sec:4}. We calculate the cross sections in two different ways. First, we evolve the threshold resummed hard and jet functions with the parton shower, i.e. for one initial parton, and we then compute the integral in Eq.~(\ref{eq:epem_leadingjet4}). Second, we initialize the parton shower with two (back-to-back) partons and determine the event-wide leading jet directly from the output of the parton shower. We find that both methods give the same result as expected. We note that we cannot directly connect event-wide leading jets in $e^+e^-$ collisions to parton energy loss due to the structure of the factorization in Eq.~(\ref{eq:epem_leadingjet4}). The main difference compared to $e^+e^-$ hemisphere leading jets is that now we have two instead of one parton at LO, see criterion 3 in the Introduction.

\begin{figure}[t]
\vspace*{.7cm}
\centering
\includegraphics[width=.99\textwidth]{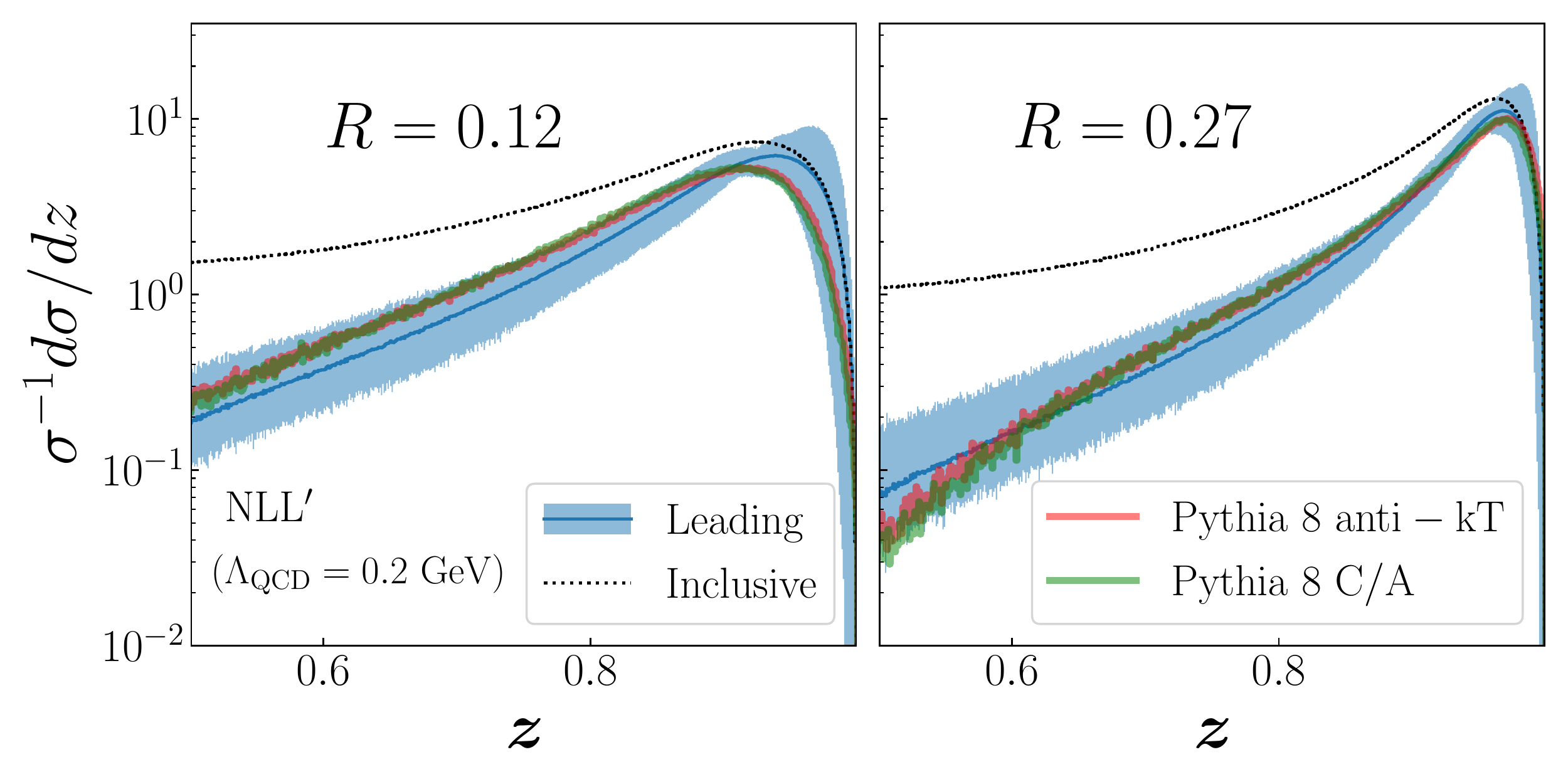}
\caption{Event-wide leading jet energy spectrum in $e^+e^-$ collisions for two values of the jet radius $R$. We compare to the event-wide inclusive jet spectrum and simulations from Pythia~8~\cite{Sjostrand:2014zea}.~\label{fig:eventwide_jets}}
\end{figure}

We show our numerical results for $e^+e^-$ event-wide leading jets in Fig.~\ref{fig:eventwide_jets} for two exemplary values of the jet radius. We observe that the spectrum peaks close to $z\approx 1$ and falls off steeply toward smaller values of $z$. We note that the spectrum looks significantly different compared to $e^+e^-$ hemisphere jets and is more peaked at large values of $z$ since there are now at least two jets produced in each hemisphere and we pick the more energetic one. For comparison, we also show the inclusive jet spectrum in Fig.~\ref{fig:eventwide_jets}. The inclusive spectrum starts to deviate from the leading jet result around $z\approx 0.94$ which differs from the hemisphere jet case where the two spectra only start to differ for $z<0.5$. In addition, we show Pythia~8~\cite{Sjostrand:2014zea} results for the anti-k$_T$~\cite{Cacciari:2008gp} and C/A~\cite{Dokshitzer:1997in,Wobisch:1998wt} algorithm. Pythia is tuned very well to LEP data and can serve as a benchmark for $e^+e^-$ jet observables. We find good agreement with our numerical results where the nonperturbative parameter is chosen as $\Lambda_{\rm QCD}=0.2$~GeV. The two jet algorithms used for the Pythia simulations give identical analytical results in the threshold region at NLL$'$. Indeed, we observe that the Pythia results for the two algorithms are very similar. 

\begin{figure}[t]
\vspace*{.6cm}
\centering
\includegraphics[width=.8\textwidth]{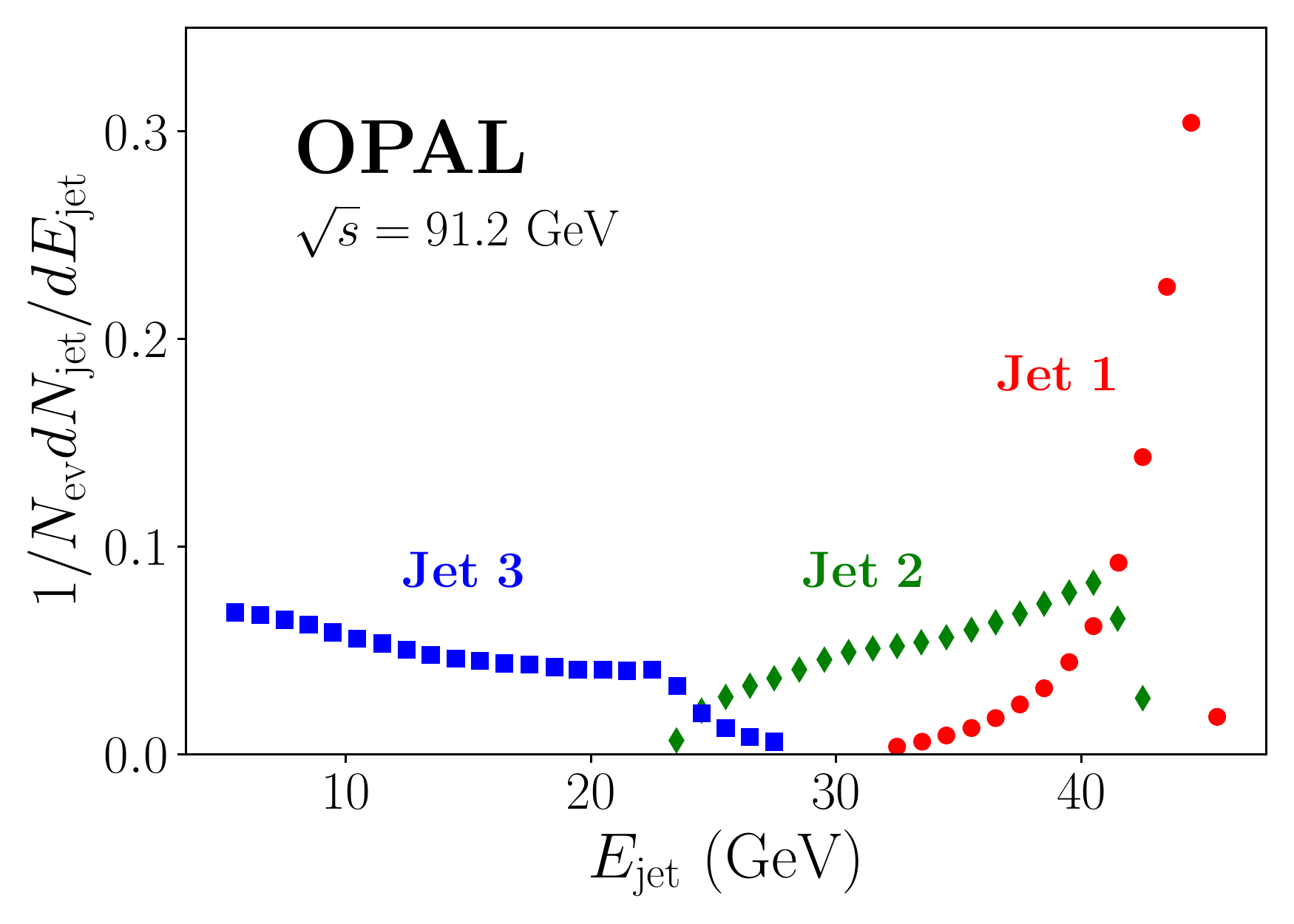}
\caption{OPAL results for the energy spectrum of the leading jet (Jet 1) and the first two subleading jets (Jet 2,3) using 3-jet events. Data taken from~\cite{Abbiendi:2004pr}.~\label{fig:OPAL}}
\end{figure}

Interestingly, the OPAL Collaboration at LEP provided data for the leading and the first two subleading jet cross sections in Ref.~\cite{Abbiendi:2004pr}. OPAL used an algorithm that finds,  by construction, three jets in every recorded event. This procedure does not correspond to our definition of jets making a one-to-one comparison impossible. We show the OPAL data in Fig.~\ref{fig:OPAL}. The OPAL measurement imposed an intra-jet angle which roughly corresponds to the jet radius chosen in the right panel of Fig.~\ref{fig:eventwide_jets}. While the shape of the spectrum is quantitatively different compared to our results in Fig.~\ref{fig:eventwide_jets}, we do find qualitative agreement. We note that for event-wide jets, the second leading jet is not required to carry an energy fraction of $z_2<0.5$ (as for $e^+e^-$ hemisphere jets) since we have at least two jets in the event. In fact, from Fig.~\ref{fig:OPAL} we find that the first subleading jet energy extends up to $z_2\sim 0.94$. The OPAL results demonstrate that it is possible to experimentally measure leading and subleading jet spectra at LEP. We expect that experiments like the LHC and EIC can measure leading jet observables equally well.

\subsection{Semi-Inclusive Deep-Inelastic Scattering~\label{sec:SIDIS}}

SIDIS measurement allow for a clean measurement of jet/parton energy loss since we only have one quark which fragments into the observed leading jet at LL accuracy. We consider the process $e(k)+p(P)\to e'(k')+{\rm jet}_1(P_{\rm jet_1})+X$ where both the final state electron and the leading jet are observed. The reference scale with respect to which the leading jet energy is measured is set by the virtuality $Q^2=-q^2=-(k-k')^2$ of the exchanged photon. We consider the cross section
\begin{equation}
    \frac{{\rm d}\sigma_{ep\to e'+{\rm jet}_1+X}}{{\rm d}x_B\, {\rm d}y\, {\rm d}z_1}\,.
\end{equation}
where we introduced the usual variables for SIDIS with hadrons in the final state: Bjorken $x_B$ and the inelasticity $y$ which are given by
\begin{align} 
x_B  = \frac{Q^{2}}{2 P \cdot q} \,,\quad
y  = \frac{P \cdot q}{P \cdot k} \,.
\end{align}
We have $Q^2=x_Bys$, where $\sqrt{s}$ is the electron-proton CM energy. The momentum fraction of the leading jet $z_1$ is defined as
\begin{equation}
z_1 = \frac{P \cdot P_{\rm jet_1}}{P \cdot q}=\frac{P_{\rm jet_1}^+}{Q}\,.
\end{equation}
Here the last equality holds in the Breit frame and $P^+_{{\rm jet}_1}$ denotes the large lightcone momentum component of the leading jet. In the target rest frame we have $z_1=E_{\rm jet_1}/Q$ which makes this asymmetric process very similar to $e^+e^-$ hemisphere leading jets. Suitable jet clustering algorithms for this type of observable in the Breit frame were discussed in Ref.~\cite{Arratia:2020ssx}. Within collinear factorization, the cross section for inclusive jets can be written in terms of ${\cal F}_{T,L}$, the transverse and longitudinal structure functions~\cite{Altarelli:1979kv,Furmanski:1981cw}
\begin{equation}
\frac{{\rm d}\sigma_{ep\to e'+{\rm jet}+X}}{{\rm d}x\, {\rm d}y\, {\rm d}z} =\frac{4 \pi \alpha^{2}}{Q^{2}}\left[\frac{1+(1-y)^{2}}{2 y} \mathcal{F}_{T}\left(x, z, Q\right)+\frac{1-y}{y} \mathcal{F}_{L}\left(x, z, Q\right)\right]\,,
\end{equation}
where $\alpha$ is the fine structure constant and we dropped the subscript $B$ of Bjorken $x_B$ for simplicity. In the current fragmentation region, the structure functions $a=T,L$ can be written as 
\begin{equation}
    {\cal F}_{a}(x,y,Q)= \sum_{ij} H^a_{ij}(x,z,Q,\mu)\otimes f_i(x,\mu)\otimes J_j(z,QR,\mu) \,.
\end{equation}
Here, $\otimes$ denote convolution integrals in $x$ and $z$, $f_i$ are the PDFs and $J_j$ are the inclusive jet functions. Therefore, also for SIDIS at LL accuracy we can consider the first moment as a direct measure of the average quark energy loss
\begin{equation}
    \int_0^1{\rm d}z_1\, z_1\,\frac{1}{\sigma_{\rm tot}} \frac{{\rm d}\sigma_{ep\to e'+{\rm jet_1}+X}^{(0)}}{{\rm d}x\, {\rm d}y\, {\rm d}z_1} = \langle z_q\rangle\,.
\end{equation}
Here $\sigma_{\rm tot}$ is the inclusive DIS cross section ${\rm d}\sigma/{\rm d}x/{\rm d}y$.

The threshold resummation for jets in SIDIS follows from Refs.~\cite{Cacciari:2001cw,Fleming:2006cd,Anderle:2012rq,Lustermans:2019cau,Arratia:2020ssx} which can also be implemented in the parton shower algorithm discussed in section~\ref{sec:4}. We leave more detailed numerical studies especially for kinematics at the EIC for future work.

\subsection{Photon-jet correlations~\label{sec:photon_jet}}

\begin{figure}[t]
\vspace*{.7cm}
\centering
\includegraphics[width=10cm]{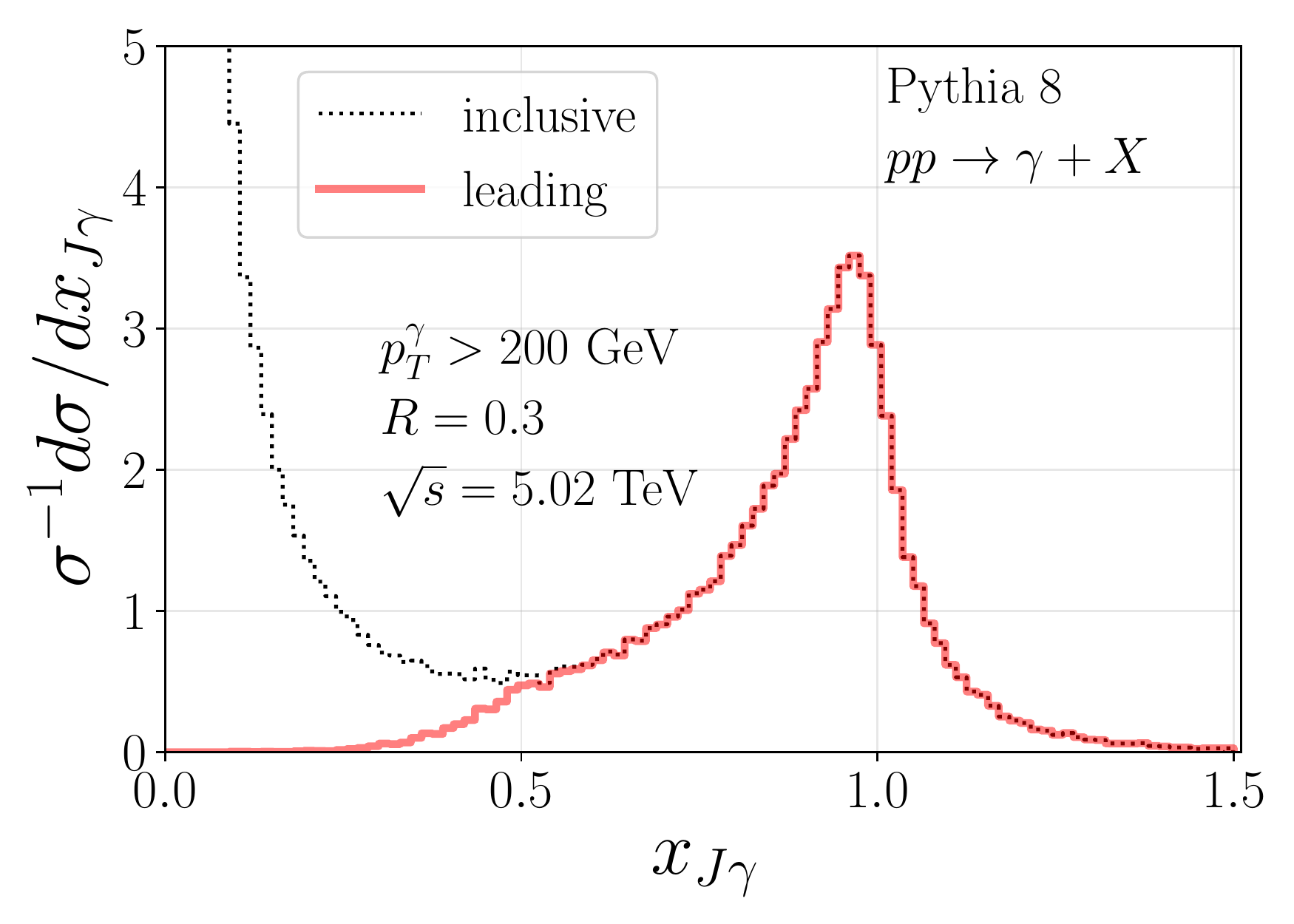}
\caption{The $x_{J\gamma}$ distribution for inclusive and leading jets recoiling against a direct photon with large transverse momentum in proton-proton collisions at $\sqrt{s}=5.02$~TeV. The result shown here is obtained from Pythia~8~\cite{Sjostrand:2014zea}.~\label{fig:photon-jet}}
\end{figure}

In this section we consider the production of a direct photon and measure the momentum of the leading jet in the opposite hemisphere. See Refs.~\cite{Sirunyan:2017qhf,Adamczyk:2017yhe,Aaboud:2018anc,Acharya:2020sxs} for related recent experimental results. We consider the cross section differential in the photon's transverse momentum and rapidity $p_T^\gamma,\,\eta^\gamma$. In addition, we measure the recoiling leading jet's transverse momentum $p_T$ relative to the photon $x_{J1\gamma}=p_T/p_{T}^{\gamma}$. At LO/LL accuracy the factorization of the cross section can be written as~\cite{Owens:1986mp}
\begin{equation}\label{eq:photonjet}
    \frac{{\rm d}\sigma^{(0)}_{pp\to\gamma+{\rm jet_1}+X}}{{\rm d}\eta^{\gamma}\,{\rm d}p_T^\gamma\,{\rm d}x_{J1\gamma}}=\sum_{ijk}f_i(x_i,\mu)\otimes f_j(x_j,\mu)\otimes  H_{ijk}^{(0)}(x_i,x_j,\eta^\gamma,p_T^\gamma,\mu)\otimes{\cal J}_i(x_{J\gamma},p_{T}^\gamma R,\mu)\,.
\end{equation}
Here we consider the transverse momentum of the direct photon $p_T^\gamma$ as the reference scale with respect to which we measure the energy loss of the recoiling leading jet. This observable can also give direct access to the weighted average quark/gluon energy loss
\begin{equation}
    \langle x_{J1\gamma}\rangle=\int_0^1{\rm d}x_{J1\gamma}\,x_{J1\gamma}\frac{1}{\sigma_{\rm tot}}\frac{{\rm d}\sigma^{(0)}_{pp\to\gamma+{\rm jet_1}+X}}{{\rm d}\eta^{\gamma}\,{\rm d}p_T^\gamma\,{\rm d}x_{J1\gamma}}= f_q\,\langle x_{J1\gamma,q}\rangle+f_g\,\langle z_{J1\gamma,g}\rangle\,.
\end{equation}
However, different than the processes considered above, $x_{J\gamma}$ can generally be $>1$ making a clear interpretation in terms of energy loss more difficult. We may nevertheless get a handle on the energy loss of the leading jet recoiling against the photon by considering the difference between inclusive and leading jets. At the jet function level, we find that we can rewrite the average momentum fraction of the leading jet as
\begin{align}\label{eq:energy_loss_leading_inclusive}
    \langle z \rangle = &
    \int_0^1{\rm d}z\, z\, {\cal J}_i(z,p_T^\gamma R,\mu)
    \nn\\=&\,
    \int_0^{1/2}{\rm d}z\, z\, {\cal J}_i(z,p_T^\gamma R,\mu) + \int_{1/2}^1{\rm d}z\, z\, J_i(z,p_T^\gamma R,\mu)
    \nn\\=&\,
    1+\int_0^{1/2}{\rm d}z\, z\,\left[ {\cal J}_i(z,p_T^\gamma R,\mu) - J_i(z,p_T^\gamma R,\mu) \right] \,.
\end{align}
We can therefore also get access to the average jet energy loss by measuring the difference between the inclusive and leading jet spectra recoiling the direct photon. The inclusive and leading spectra from Pythia~8~\cite{Sjostrand:2014zea} are shown in Fig.~\ref{fig:photon-jet}. In practice, jets can only be reconstructed down to a certain transverse momentum. Therefore, we need to introduce a small cutoff $z_{\rm cut}$ in Eq.~(\ref{eq:energy_loss_leading_inclusive}), which gives
\begin{align}
    \langle z \rangle(z_{\rm cut}) = 
    1+\int_{z_{\rm cut}}^{1/2}{\rm d}z\, z\,\left[ {\cal J}_i(z,p_T^\gamma R,\mu) - J_i(z,p_T^\gamma R,\mu) \right] \,.
\end{align}
We leave more detailed numerical studies using our parton shower framework for future work.
 
\subsection{Toward leading hadrons~\label{sec:hadrons}}

\begin{figure}[t]
\vspace*{.7cm}
\centering
\includegraphics[width=\textwidth]{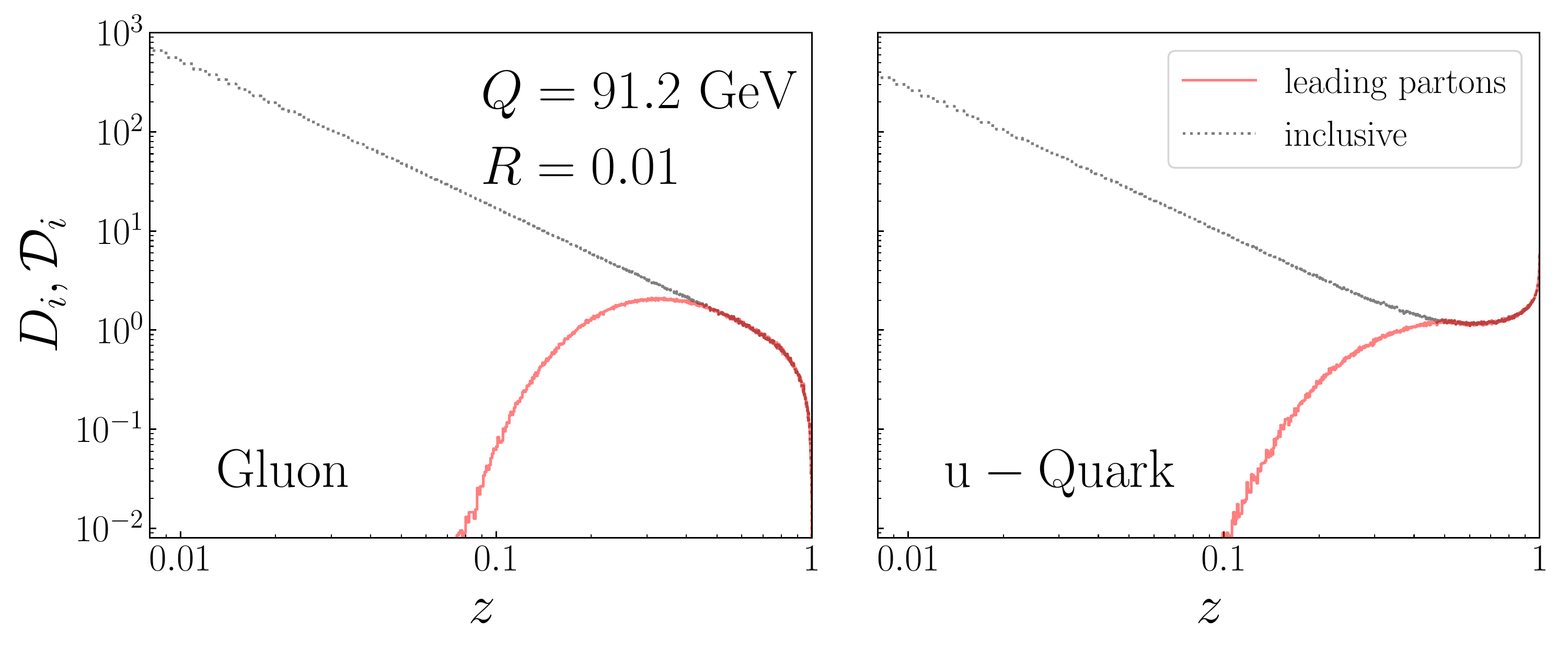}
\caption{Fragmentation spectrum of inclusive and leading partons for gluons (left) and $u$-quarks (right) at the scale $Q=91.2$~GeV.~\label{fig:leading_hadron}}
\end{figure}

In order to extend our calculation from leading jets to leading hadrons, we need to evolve the shower down to the nonperturbative scale $\mu_{\rm NP}={\cal O}(1 \text{GeV})$ and include hard-scattering functions and fragmentation functions in the shower algorithm. The leading and subleading hadron fragmentation functions are currently not known and we are not aware of existing data sets that could constrain them. Similar to jet functions, the leading and inclusive parton-to-hadron fragmentation functions agree for $z>1/2$ but differ for smaller values of $z$. We expect that the necessary experimental measurements are feasible which can provide important new insights into the QCD fragmentation mechanism. We leave more detailed phenomenological studies for future work. Here we only focus on the parton shower evolved to the nonperturbative scale without including fragmentation functions and hard-scattering functions.

Analogous to leading jets discussed above, we can use the Monte Carlo setup introduced in section~\ref{sec:3} to calculate the spectrum of leading partons since the corresponding leading hadron fragmentation functions satisfy the same non-linear evolution equations. Now we evolve the shower down to a scale of order $\mu_{\rm NP}={\cal O}(1~\text{GeV})$ which we consider as the onset of nonperturbative physics. To be specific, the Monte Carlo algorithms terminates when $t>t^{\rm max}=t(Q,1/Q)$ or equivalently $Q{\cal R}={\cal O}(1~\text{GeV})$ in Eq.~(\ref{eq:MCtime}). 

In the upper two panels of Fig.~\ref{fig:leading_hadron}, we show the result for the $z$-spectrum for leading ${\cal D}_i$ and inclusive $D_i$ partons evolved down to $\mu_{\rm NP}$ starting from $Q=91.2$~GeV. For the numerical results shown here we choose $R=0.01$. Even without sampling from the nonperturbative fragmentation function, the gluon distribution (left) already has the typical shape of a hadron fragmentation spectrum which falls off steeply toward $z\to 1$. Instead, the quark spectrum (right) still has a peak at very large values of $z$ similar to the evolved LL jet functions discussed above. See Fig.~\ref{fig:Comparison_MC_Mellin}. The leading and inclusive parton spectra start to deviate around $z\approx 0.4$. We note that all the leading jet cross sections discussed in previous sections can also be calculated for hadrons. In particular, we can also define a nonperturbative version of energy loss in terms of leading hadrons instead of jets. We leave more detailed studies of the leading hadron energy loss for future work.

\section{Conclusions~\label{sec:8}}

In this work we discussed leading and subleading jet cross sections which probe fundamental aspects of the QCD fragmentation process. Different than inclusive jets, the formation and evolution of leading jets is described by jet functions with non-linear DGLAP-type evolution equations. These leading and subleading jet functions constitute normalized probability densities to find a (sub)leading jet with a given longitudinal momentum fraction from an initial quark or gluon. Instead, the jet functions relevant for inclusive jet production are number densities where the total number of jets per event is not fixed and produced dynamically event-by-event. Motivated by results in probability theory~\cite{Derrida_1987}, we established relations between leading/subleading jets and inclusive single-, di- and tri-jet functions which we plan to explore in more detail in future work.

We focused in particular on cross sections where we have access to an additional reference scale $Q$ with respect to which we can define the longitudinal momentum fraction $z_1$ which is contained in the leading jet. We are then able to define the (average) out-of-jet radiation or the energy loss of leading jets as $z_{\rm loss}=1- z_1$ which can be computed order-by-order in perturbation theory and which can be accessed directly through experimental measurements. For observables where we only have one parton at leading order, the experimentally accessible leading jet energy loss can be identified with parton energy loss at leading-logarithmic accuracy. We identified criteria of suitable observables which allow for a direct measurement of the leading jet energy loss which include $e^+e^-$ hemisphere jets, subjets in proton-proton collisions, jets in Semi-Inclusive Deep Inelastic Scattering and photon-jet correlations. From these cross sections we can obtain the average energy loss of leading jets $\langle z_{\rm loss}\rangle$ which can be compared to our theoretical results.

One of the main new developments of our work is a parton shower framework that allows us to compute threshold resummed leading jet cross sections at next-to-leading logarithmic accuracy (NLL$'$). Hard-scattering functions, jet functions and also fragmentation functions can be included directly in the parton shower framework. The results of the parton shower agree exactly with analytical results for inclusive jets and allow for a well-defined extension to leading jet cross sections. The threshold resummation which we include for both the hard and jet functions is phenomenologically important for leading jet observables. We derived the threshold resummation for leading jets in $e^+e^-$ collisions and leading subjets in proton-proton collisions. While the developed framework is a ``few purpose'' parton shower, we expect that it can be extended systematically to other observables.

We presented numerical results for $e^+e^-$ hemisphere and event-wide leading jets as well as leading subjets in proton-proton collisions at NLL$'$ accuracy using the parton shower framework. For $e^+e^-$ event-wide leading jets we compared to Pythia~8 results and found good agreement. Interestingly, the OPAL Collaboration at LEP measured similar leading jet cross sections in Ref.~\cite{Abbiendi:2004pr}. While a one-to-one comparison to the existing data is not possible since their definition of leading jets differs from ours, these measurements demonstrate that leading jet and hadron measurements are generally feasible.

We investigated the differences of the average energy loss between leading quark and gluons jets. We observed that the differences are surprisingly small compared to a leading-order estimate and that they are rather independent of the jet radius $R$. Besides the average energy loss, the mean of the energy loss probability distribution, we also considered for the first time the variance which quantifies event-by-event fluctuations of energy loss. In addition, we computed the Shannon entropy and the KL divergence. The latter quantifies the difference between quark and gluon jet energy loss. We further explored the potential of leading (sub)jets to discriminate between quark and gluon jets. We presented ROC curves and the AUC for different values of the jet radius $R$. Interestingly, the best discrimination power is achieved for a perturbative value of $R$. In the future, we plan to explore the tagging performance of leading (sub)jets and study their complementarity to other typical observables such as particle multiplicity.

In addition, we outlined how our work can be extended to leading hadrons. Similar to leading jets, leading hadron fragmentation functions are normalized probability densities which allow us to establish a well-defined but nonperturbative notion of energy loss. Our results constitute the first quantitative calculation of leading jet energy loss which can compared directly to experimental data. We expect that our results will be particularly useful to study leading jets which traverse hot or cold nuclear matter in heavy-ion collisions or electron-nucleus collisions at the future EIC. Medium induced emissions can generally increase the energy loss of leading jets and the corresponding energy loss spectrum introduced here can provide important information about the Quark-Gluon Plasma/cold nuclear matter and its interaction with hard probes.

\section*{Acknowledgments}

We thank Miguel Arratia, Yi Chen, Peter Jacobs, Yue-Shi Lai, Yen-Jie Lee, Yiannis Makris, James Mulligan, Dennis Perepelitsa, Mateusz Ploskon, Darren Scott and Wouter Waalewijn for helpful discussions. D.N. was supported by the U.S. DOE under Contract DE-AC52-06NA25396 at LANL and through the LANL/LDRD Program. F.R. was supported by LDRD funding from Berkeley Lab provided by the U.S. Department of Energy under Contract No. DE-AC02-05CH11231. N.S was supported through DOE Contract No. DE-AC05-06OR23177 under which JSA operates the Thomas Jefferson National Accelerator Facility.

\medskip

\appendix

\section{Fixed order expressions and anomalous dimensions~\label{app:NLO_expressions}}

Here we summarize the NLO results of the relevant functions that appear in the refactorization of $e^+e^-$ hemisphere jets and subjets at threshold in Eqs.~(\ref{eq:refactorization_epem}) and~(\ref{eq:refactorization_subjets}). In Mellin space for anti-k$_T$ jets, we find\footnote{At one loop, the matching coefficient $H_g$ has no single log contribution, since the squared field strength operator $F^2$ itself has an anomalous dimension. This anomalous dimension is reproduced in the effective theory matching procedure through the jet functions of the factorization theorem for the gluon form factor.}
\begin{align}
    H_q(Q,\mu)&=1+\frac{\alpha_s}{4\pi}C_F\bigg(-2\ln^2\Big(\frac{\mu^2}{Q^2}\Big) - 6 \ln\Big(\frac{\mu^2}{Q^2}\Big) -16+\frac{7\pi^2}{3}\bigg) \,, 
    \nonumber \\
    H_g(Q,\mu)&=1+\frac{\alpha_s}{4\pi}C_A\bigg(-2\ln^2\Big(\frac{\mu^2}{Q^2}\Big) +\frac{7\pi^2}{3} \bigg) \,, 
    \nonumber \\
    \mathscr{J}_q(Q/N,\mu)&=1+\frac{\alpha_s}{4\pi}C_F\bigg(2\ln^2\Big(\frac{\mu^2\bar N}{Q^2}\Big)+3\ln\Big(\frac{\mu^2\bar N}{Q^2}\Big)+7-\frac{2\pi^2}{3}\bigg) \,, 
    \nonumber \\
    \mathscr{J}_g(Q/N,\mu)&=1+\frac{\alpha_s}{4\pi}C_A\bigg(2\ln^2\Big(\frac{\mu^2\bar N}{Q^2}\Big)+\frac{\beta_0}{C_A}\ln\Big(\frac{\mu^2\bar N}{Q^2}\Big)+\frac{67}{9}-\frac{2\pi^2}{3}-\frac{20}{9}\frac{T_FN_f}{C_A}\bigg) 
    \nonumber \\
    S_i(QR/N,\mu)&=1+\frac{\alpha_s}{4\pi}C_i\bigg(-\ln ^{2}\Big(\frac{\mu^2 \bar{N}^2}{Q^2 R^2}\Big)-\frac{\pi^{2}}{2}\bigg) \,, 
    \nonumber \\
    J_q(QR,\mu)&=1+\frac{\alpha_s}{4\pi}C_F\bigg(\ln^2\Big(\frac{\mu^2}{Q^2R^2}\Big) + 3 \ln\Big(\frac{\mu^2}{Q^2R^2}\Big) + 13 + \frac{3\pi^2}{2} \bigg) \,, 
    \nonumber \\
    J_g(QR,\mu)&=1+\frac{\alpha_s}{4\pi}C_A\bigg(\ln^2\Big(\frac{\mu^2}{Q^2R^2}\Big) + \frac{\beta_0}{C_A} \ln\Big(\frac{\mu^2}{Q^2R^2}\Big)+ 2\left(\frac{67}{9}-\frac{3 \pi^{2}}{4}\right)-2 \frac{23}{9} \frac{T_{F} N_{f}}{C_A} \bigg) \,, \nonumber \\
    {\cal S}_i(QR/N,\mu)&=1+\frac{\alpha_s}{4\pi}C_i\bigg(\ln ^{2}\Big(\frac{\mu^2 \bar{N}^2}{Q^2 R^2}\Big)+\frac{\pi^{2}}{2}\bigg) \,.
\end{align}
The hard functions ${\cal H}_{ij}$ of the refactorized subjet cross section can be found in Refs.~\cite{Kang:2017mda,Kang:2017glf}. The corresponding anomalous dimensions (defined via ${\rm d}/{\rm d}\ln\mu$) are given by
\begin{align}
    \gamma^H_q(Q,\mu)&=\frac{\alpha_s}{\pi}C_F\bigg(-2\ln\Big(\frac{\mu^2}{Q^2}\Big) - 3\bigg) \,, 
    \nonumber \\
    \gamma^H_g(Q,\mu)&=\frac{\alpha_s}{\pi}C_A\bigg(-2\ln\Big(\frac{\mu^2}{Q^2}\Big) \bigg) \,, 
    \nonumber \\
    \gamma^\mathscr{J}_q(Q/N,\mu)&=\frac{\alpha_s}{\pi}C_F\bigg(2\ln\Big(\frac{\mu^2\bar N}{Q^2}\Big)+\frac{3}{2}\bigg) \,, 
    \nonumber \\
    \gamma^\mathscr{J}_g(Q/N,\mu)&=\frac{\alpha_s}{\pi}C_A\bigg(2\ln\Big(\frac{\mu^2\bar N}{Q^2}\Big)+\frac{\beta_0}{2C_A}\bigg) 
    \nonumber \\
    \gamma^S_i(QR/N,\mu)&=-\frac{\alpha_s}{\pi}C_i\ln \Big(\frac{\mu^2 \bar{N}^2}{Q^2 R^2}\Big) \,, 
    \nonumber \\
    \gamma^J_q(QR,\mu)&=\frac{\alpha_s}{\pi}C_F\bigg(\ln\Big(\frac{\mu^2}{Q^2R^2}\Big) + \frac{3}{2} \bigg) \,, 
    \nonumber \\
    \gamma^J_g(QR,\mu)&=\frac{\alpha_s}{\pi}C_A\bigg(\ln\Big(\frac{\mu^2}{Q^2R^2}\Big) + \frac{\beta_0}{2C_A}\bigg) \,, 
    \nonumber \\
    \gamma^H_{qq}(QR,N,\mu)&=\frac{\alpha_s}{\pi}C_F\bigg(-\ln\Big(\frac{\mu^2}{Q^2R^2}\Big) - \frac32\bigg) + P_{qq}(N) \,, 
    \nonumber \\
    \gamma^H_{gg}(QR,N,\mu)&=\frac{\alpha_s}{\pi}C_A\bigg(-\ln\Big(\frac{\mu^2}{Q^2R^2}\Big) -\frac{\beta_0}{2C_A} \bigg) + P_{gg}(N) \,, 
    \nonumber \\    
    \gamma^{\cal S}_i(QR/N,\mu)&=\frac{\alpha_s}{\pi}C_i\ln \Big(\frac{\mu^2 \bar{N}^2}{Q^2 R^2}\Big) \,,
\end{align}
where $P_{ji}(N)$ denote the Altarelli-Parisi splitting functions in Mellin space. In addition, we have $\gamma^H_{ji}=\alpha_s/\pi P_{ji}(N)$ for $i\neq j$.

\bibliographystyle{JHEP}
\bibliography{bibliography}

\end{document}